\documentclass[sigconf]{acmart}

\usepackage{times}
\usepackage{color}
\usepackage{subfigure}

\usepackage[linesnumbered, ruled, vlined]{algorithm2e}

\newtheorem{example}{Example}

\usepackage{balance}

\usepackage{color}
\definecolor{Xiang}{rgb}{1,0,0}
\definecolor{Weilong}{rgb}{0,0,1}

\newtheorem{theorem}{theorem}[section]
\newtheorem{lemma}{lemma}[section]

\newcommand{\nop}[1]{}

\newtheorem{definition}{Definition}

\AtBeginDocument{%
  \providecommand\BibTeX{{%
    \normalfont B\kern-0.5em{\scshape i\kern-0.25em b}\kern-0.8em\TeX}}}

\copyrightyear{2021}
\acmYear{2021}
\setcopyright{acmcopyright}\acmConference[SIGMOD '21]{Proceedings of the 2021 International Conference on Management of Data}{June 20--25, 2021}{Virtual Event, China}
\acmBooktitle{Proceedings of the 2021 International Conference on Management of Data (SIGMOD '21), June 20--25, 2021, Virtual Event, China}
\acmPrice{15.00}
\acmDOI{10.1145/3448016.3457238}
\acmISBN{978-1-4503-8343-1/21/06}

\begin{document}
\fancyhead{}

\title{Online Topic-Aware Entity Resolution Over Incomplete Data Streams (Technical Report)}

\author{Weilong Ren}
\affiliation{%
  \institution{Department of Computer Science, Kent State University}
  \streetaddress{}
  \city{Kent}
  \country{U.S.A.}}
\email{wren3@Kent.edu}

\author{Xiang Lian}
\affiliation{%
  \institution{Department of Computer Science, Kent State University}
  \streetaddress{}
  \city{Kent}
  \country{U.S.A.}}
\email{xlian@Kent.edu}

\author{Kambiz Ghazinour}
\affiliation{%
  \institution{ Center for Criminal Justice, Intelligence and Cybersecurity, State University of New York}
  \streetaddress{}
  \city{Canton}
  \country{U.S.A.}}
\email{ghazinourk@canton.edu}

\begin{abstract}

In many real applications such as the data integration, social network analysis, and the Semantic Web,
the \textit{entity resolution} (ER) is an important and fundamental problem, which identifies and links the same
real-world entities from various data sources. While prior works usually consider ER over static and complete data, in practice, application data are usually collected in a streaming fashion, and often incur missing attributes (due to the inaccuracy of data extraction techniques). Therefore, in this paper, we will formulate and tackle a novel problem, \textit{topic-aware entity resolution over incomplete data streams} (TER-iDS), which online imputes incomplete tuples and detects pairs of topic-related matching entities from incomplete data streams. In order to effectively and efficiently tackle the TER-iDS problem, we propose an effective imputation strategy, carefully design effective pruning strategies, as well as indexes/synopsis, and develop an efficient TER-iDS algorithm via index joins. Extensive experiments have been conducted to evaluate the effectiveness and efficiency of our proposed TER-iDS approach over real data sets.

\end{abstract}


\maketitle

\section{Introduction}
In many real applications such as the data fusion \cite{dong2009data}, social
network analysis \cite{bartunov2012joint}, and the Semantic Web \cite{gangemi2013comparison}, one
important and fundamental problem is to identify and link the same
real-world entities from data sources, also known as the \textit{entity resolution}
(ER) problem \cite{papadakis2019survey} (or record linkage \cite{winkler2006overview}). Specifically,
an ER problem retrieves from data sources the matching pairs of records or profiles that represent the same entities, which can be
inferred by their similar or the same attribute values. In the era of big data, new data records often arrive very fast in a streaming fashion, thus, the ER problem becomes more challenging over dynamic data sources (e.g., data streams). An
efficient solution to such an online ER problem can be used as a critical step
during the process of the data integration.

Although there are many existing works on the ER problem over static data (e.g., \cite{papadakis2014meta,li2015linking,shen2014probabilistic,papadakis2019survey,ebraheem2018distributed}) or data streams (e.g., \cite{firmani2016online,dragut2015query}), they often assume that the underlying data are complete and
accurate. However, in practice, data can be missing due to the unreliability of data sources. For example, on social networks (e.g., Twitter) or health-related forums, users often post daily textual comments/messages such as tweets/retweets (i.e., entities) about different topics/events. Since users may not fully describe their opinions or \textit{information extraction} (IE) \cite{poon2007joint} techniques are sometimes not very accurate, some extracted attributes from the unstructured comment/message texts may be missing and incomplete. In this case, it is rather challenging to conduct online ER operator over such incomplete comment/message streams.

Below, we give a motivation example in the application of \textit{online health community support}.\vspace{-1ex}

\setlength{\textfloatsep}{1pt}
\begin{figure}[t!]
\centering\vspace{-1ex}
\hspace{-1ex}\includegraphics[scale=0.135]{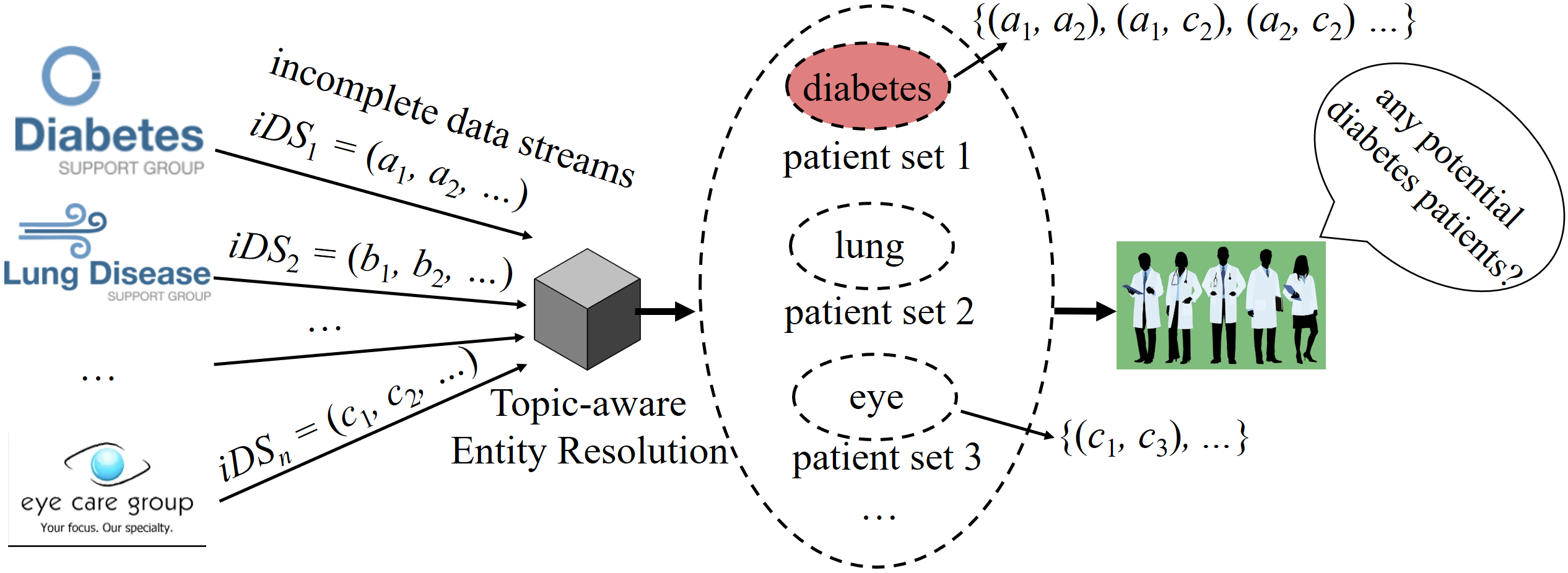}\vspace{-1ex}
\caption{\small Online Topic-Aware Entity Resolution for Online Health Community Support.}
\label{fig:motivation}
\end{figure}

\begin{table}[t!]
\centering\scriptsize
\caption{\small The extracted tuples from posts of health groups in Figure \ref{fig:motivation}.}\vspace{-2ex}
\begin{tabular}{|c||c|l|l|l|}
\hline
\textbf{ID} & \textbf{Gender} & \textbf{Symptom} & \textbf{Diagnosis} & \textbf{Treatment} \\
\hline
\hline
$a_1$ & \textit{male}  & \textit{loss of weight} & \textit{diabetes} & \textit{dietary therapy, drug therapy}  \\ \hline
$a_2$ & \textit{male}  & \textit{loss of weight, blurred vision} & {\bf $-$ } & {\bf $-$ }\\ \hline
...  & ... & ... & ... & ... \\ \hline\hline
$b_1$ & \textit{female}  & \textit{fever, low spirit, cough} & \textit{pneumonia} & {\bf $-$ } \\ \hline
$b_2$ & \textit{male}  & \textit{fever, poor appetite, cough} & \textit{flu} & \textit{drink more, sleep more}  \\ \hline
...  & ... & ... & ... & ... \\ \hline\hline
$c_1$ & \textit{female}  & \textit{red eye, eye itchy, shed tears} & \textit{conjunctivitis} & \textit{eye drop}  \\ \hline
$c_2$ & \textit{male}  & \textit{blurred vision} & \textit{diabetes} & \textit{drug therapy}  \\ \hline
...  & ... & ... & ... & ... \\ \hline
\end{tabular}
\label{table:motivation}
\end{table}

\begin{example} {\bf (Online Health Community Support)}
In online health communities \cite{chen2020linguistic} such as WebMD \cite{huh2016lessons}, PatientsLikeMe \cite{brubaker2010patientslikeme}, or social networks (e.g., Twitter and Facebook), patients often post messages or comments about their symptoms, (self-)diagnosis, and/or current treatment on health-related groups/forums. However, patients may not be able to describe their symptoms in the right disease groups. For example, the symptom ``blurred vision'' may not be an eye-related disease, but may be a sign of diabetes. Therefore, it is quite important to timely detect and group posts with similar symptoms/diagnoses/treatment, and alert medical professionals who are ready to provide online health support/help. 

In this case, each medical professional needs to specify one's expertise or disease topics (e.g., diabetes-related keywords). Then, we can perform an online entity resolution over streams of patients' posts with relevant topics, and identify patient sets with similar symptoms/diagnoses/treatment from different health groups/forums, which will be sent to the medical professional for support. 

As illustrated in Figure  \ref{fig:motivation} and Table \ref{table:motivation}, we can extract attributes, (ID, Gender, Symptom, Diagnosis, Treatment), from new posts on health groups in a streaming fashion. Due to incomplete data entered by users or the inaccuracy of IE techniques \cite{poon2007joint}, some extracted attributes in Table \ref{table:motivation} are missing and denoted as ``$-$'' (e.g., attributes \textit{Diagnosis} and \textit{Treatment} of post $a_2$). Therefore, we should perform online  topic-aware ER over such incomplete (textual) post streams and obtain similar post pairs (e.g., pair $(a_1, c_2)$ related to diabetes) for medical professionals.
\label{example:motivation}
\end{example}

Inspired by the example above, in this paper, we will formulate and tackle a novel problem of \textit{topic-aware entity resolution over incomplete data streams} (TER-iDS), which online obtains the matching pairs of incomplete tuples that represent the same entities, extracted from (sliding windows  \cite{tao2006maintaining,ananthakrishna2003efficient} of) incomplete data streams.

The TER-iDS problem has many other real applications, such as the \textit{recommendation system}, \textit{public opinion analysis}, and \textit{data fusion}, which identify the same entities (e.g., products, public opinions, or data records, resp.) from various (incomplete) data sources (e.g., e-commerce websites, social networks, or data sets) for timely decision/policy making. For example, a customer may want to conduct the TER-iDS operator over (incomplete) descriptions/features (i.e., extracted attributes) of a product type (i.e., topics) from the crawled e-commerce websites in a streaming manner, and obtain groups of the latest products with similar features to choose from.

\noindent {\bf Challenges.} To tackle the TER-iDS problem, there are three major challenges. First, it is non-trivial how to effectively impute the missing textual attributes of data records from incomplete data streams, since we need to explore accurate dependence relationships between complete (non-missing) and missing textual attributes. Moreover, under the streaming environment, it is rather challenging to efficiently detect and retrieve matching pairs of topic-related entities over sliding windows of incomplete data streams, since stream data usually arrive and then expire very fast. Furthermore, it is not trivial how to perform the data imputation and topic-aware ER over incomplete data streams at the same time, which requires both high accuracy and ER efficiency, respectively. 

\noindent {\bf State-of-the-art Approaches}. Previous works usually focused on one-time ER task (i.e., resolving all data records that refer to the same entities) on static data \cite{shen2014probabilistic,papadakis2014meta,ebraheem2018distributed} or streaming data \cite{dragut2015query,li2015linking,firmani2016online,simonini2018schema,wang2019efficient} within a fixed-size window. However, such an ER task has several drawbacks. First, we are often interested in finding topic-related entities only (e.g., diabetes-related posts in Example \ref{example:motivation}), thus, it is neither necessary nor efficient to retrieve the matching entities (posts) of \textit{all} topics from data sources. Second, in the streaming environment with an unlimited-size window, it is not always practical to consider ER over \textit{all} historical data records (e.g., patients' concerns 5 years ago), due to the features of stream data (i.e., high volume and velocity). Finally, most previous works \cite{papadakis2014meta,shen2014probabilistic,dragut2015query,li2015linking,firmani2016online,ebraheem2018distributed,wang2019efficient} focused on complete data and cannot support ER over incomplete data (e.g., some attributes extracted from posts might be unavailable, or missing due to the extraction inaccuracy).

To our best knowledge, most prior works  \cite{papadakis2014meta,shen2014probabilistic,dragut2015query,li2015linking,firmani2016online,ebraheem2018distributed,simonini2018schema,wang2019efficient} did not fully consider topic-aware ER over incomplete and streaming data, which requires high ER accuracy and efficiency. In contrast, our TER-iDS problem obtains the matching entities of ad-hoc topics (rather than all topics), considers the most recent data in \textit{sliding windows} of streams \cite{tao2006maintaining,ananthakrishna2003efficient} (instead of all historical data), and allows ER over incomplete data (but not complete data).

\noindent {\bf Our Proposed Approach.} To solve our TER-iDS problem, in this paper, we propose effective rule-based data imputation techniques (i.e., imputation based on \textit{conditional differential dependency} (CDD) \cite{kwashie2015conditional,wang2017discovering}) for dealing with missing attributes, design effective pruning/indexing schemes to filter out false alarms of matching pairs during the ER process, and develop a novel algorithm to enable efficient and effective TER-iDS processing via index joins over incomplete data streams.

Specifically, in this paper, we make the following contributions:\vspace{-1ex}

\begin{itemize}
    \item We formalize a novel problem of \textit{topic-aware entity resolution over
    incomplete data streams} (TER-iDS) in Section \ref{sec:prob_def}, which considers both ad-hoc topics and data incompleteness in the ER process.
    \item We design an effective ad-hoc data imputation approach
    to infer incomplete data in Section \ref{sec:data_imputation} (i.e., achieving 94.62\%$\sim$ 97.34\% topic-related ER accuracy on our tested real data sets in Section \ref{sec:exper}).
    \item We propose effective pruning strategies to reduce the search space of the TER-iDS problem in Section \ref{sec:pruning_strategy} (i.e., pruning 98.32\%$\sim$ 99.43\% entity pairs on our tested real data sets in Section \ref{sec:exper}). 
    \item We devise novel indexing mechanisms, and propose an efficient TER-iDS algorithm to allow the data imputation and ER processing at the same time in Section \ref{sec:TER_iDS} (i.e., faster than a straightforward method without index/synopsis by 3 order of magnitude on our tested real data sets in Section \ref{sec:exper}).
    \item We demonstrate the efficiency and effectiveness of our
    proposed TER-iDS processing approaches in Section \ref{sec:exper}, which outperforms the state-of-the-art approaches with higher topic-related ER accuracy and better efficiency (by 1-4 orders of magnitude).\vspace{-1ex}
\end{itemize}

Section \ref{sec:related} reviews previous works on the ER problem,
and Section \ref{sec:conclusion} concludes this paper.

\vspace{-1ex}
\section{Problem Definition}
\label{sec:prob_def}

In this section, we formally define the problem of
\textit{topic-aware entity resolution over incomplete data streams}
(TER-iDS), which takes into account topics and missing data during the streaming ER
process.

\subsection{Incomplete Data Streams}

\begin{definition} \textbf{(Incomplete Data Stream, iDS)} 
An \textit{incomplete
data stream}, iDS, contains \textit{an ordered sequence} of records (tuples), $(r_1, r_2, ..., r_t, ...)$, where each record $r_i$ arrives at time $i$. Each record $r_i$ consists of a
unique profile identifier, $rid_i$, and a set of $d$ extracted
attribute-value pairs in the form $(A_j, v_j)$ ($1\leq j\leq d$), where $A_j$ is the
attribute name, and $v_j$ is the value of attribute $A_j$.
\end{definition}

We also use $r_i[A_j]$ to represent the value, $v_j$, of attribute $A_j$ in record
$r_i$. In practice, if the attribute value,
$r_i[A_j]$, is missing (i.e., incomplete or null), we denote it as
$r_i[A_j]$ = ``$-$''. 

Following the literature of stream processing, in this paper, we adopt the \textit{sliding window} model \cite{ananthakrishna2003efficient} over incomplete data streams. 

\begin{definition} \textbf{(Sliding Window, $W_t$)} 
Given an incomplete data stream $iDS$, a current timestamp $t$, and an integer $w$, a \textit{sliding window}, $W_t$, contains w most recent tuples, $r_i$, from $iDS$.
\label{def:sw}
\end{definition}

In Definition \ref{def:sw}, at timestamp $t$, the sliding window $W_t$ contains a set of $w$ most recent tuples, $(r_{t-w+1}, r_{t-w+2}, ..., r_t)$, from incomplete data stream $iDS$. At timestamp $(t+1)$, the oldest tuple $r_{t-w+1}$ will expire and be evicted from $W_t$; meanwhile, a new tuple $r_{t+1}$ will arrive and be added to $W_t$, forming a new sliding window $W_{t+1} = (r_{t-w+2}, ..., r_t, r_{t+1})$.

Note that, there are two models of the \textit{sliding window}: count-based \cite{ananthakrishna2003efficient} and time-based \cite{tao2006maintaining}. In this paper, we adopt the count-based \textit{sliding window} \cite{ananthakrishna2003efficient}, however, our proposed solution can be easily extended to the time-based one \cite{tao2006maintaining}, by assuming that more than one tuple arrives in $iDS$ at a new timestamp, which we would like to leave as our future work.

\subsection{Imputation Over Incomplete Data Stream}
\label{subsec:imputation}

In this paper, we consider \textit{conditional differential dependency} (CDD) \cite{kwashie2015conditional,wang2017discovering}, an imputation method that extends \textit{differential dependency} (DD) \cite{song2011differential}, as an imputation tool to estimate possible values of missing attributes.

\noindent {\bf Rules for Data Imputation:} In this paper, we will use a set of imputation rules (i.e., CDD rules) w.r.t. attributes, to impute values of missing attributes. We will first use an example to illustrate the basic idea of CDD rules \cite{kwashie2015conditional,wang2017discovering}.

\begin{example}
As depicted in Table \ref{table:R}, assume that we have a data repository $R$ with three attributes $A$, $B$, and $C$. From samples in $R$, we can obtain a rule: for any two samples (e.g., $s_1$ and $s_2$), if their values on attribute $A$ equal to $a_1$ (i.e., $s_1[A] = s_2[A] = a_1$) and their distance difference on attribute $B$ is within $0.1$ (i.e., $|s_1[B] - s_2[B]| = 0.1 \in [0, 0.1]$), then 
their distance difference on attribute $C$ must be within $0.1$ (i.e., $|s_1[C]-s_2[C]| = 0.1 \in [0, 0.1]$). This way, we can obtain a so-called CDD rule, in the form $CDD_1: AB \to C,\{a_1, [0, 0.1],$ $[0, 0.1]\}$.
\label{example:CDD}
\end{example}

Next, we give formal definition of the CDD rule as follows. 

\begin{definition} {\bf (Conditional Differential Dependency, CDD)}
A \textit{conditional differential dependency} (CDD) is in the form, $(X\rightarrow A_j, \phi[XA])$, where $X$ is a set of determinant attributes, $A_j$ is a dependent attribute ($A_j \notin X$), and $\phi[Y]$ is a constraint function on attributes $Y$ ($=X$ or $A_j$), where $\phi[A_x]$ is either a distance constraint $A_x.I$ ($=[\epsilon_{A_x}.min, \epsilon_{A_x}.max]$) or a specific value $v$ ($\in dom(A_x)$) on determinant attribute $A_x\in X$, and $\phi[A_j]$ is a distance constraint $A_j.I$ on dependent attribute $A_j$.
\label{def:CDD}
\end{definition}

Given two records (tuples), $r_1$ and $r_2$, and a CDD rule, $(X\to A_j,$ $\phi[X A_j])$, the CDD rule requires that $r_1$ and $r_2$ have similar values on the dependent attribute $A_j$ (i.e., the difference between $r_1$ and $r_2$ on attribute $A_j$ must be within the interval $A_j.I$), if these two records satisfy any of the two following requirements: the differences between $r_1$ and $r_2$ on determinant attributes $A_x\in X$ are within the distance constraint $[\epsilon_{A_x}.min,\epsilon_{A_x}.max]$ (i.e., $\epsilon_{A_x}.min \le |r_1[A_x]$ $-r_2[A_x]|\le \epsilon_{A_x}.max$); or $r_1[A_x]$ and $r_2[A_x]$ are equal to a value $v$ (i.e., $r_1[A_x] = r_2[A_x] = v$). Specifically, we use $(r_1, r_2)\asymp \phi[Y]$ ($Y = X$ or $A_j$) to represent that tuples $r_1$ and $r_2$ satisfy the constraints of a CDD rule on attributes $Y$. 

Note that, instead of setting $\epsilon_Y.min$ to 0 in \cite{kwashie2015conditional,wang2017discovering}, in this paper, we relax this limitation to let the $\epsilon_Y.min$ be any non-negative value less than $\epsilon_Y.max$ (i.e., $0\le \epsilon_Y.min < \epsilon_Y.max$), such that the CDD rule can have tighter intervals for distance constraints.

\begin{table}[t!]
\centering\scriptsize
\caption{\small An example of a complete data repository $R$.}
\begin{tabular}{|c||c|c|c|}
\hline
\textbf{sample} & $A$ & $B$ & $C$ \\
\hline
\hline
$s_1$ & $a_1$ & 0.2 & 0.1   \\ \hline
$s_2$ & $a_1$ & 0.3 & 0.2  \\ \hline
$s_3$ & $a_1$ & 0.5 & 0.35   \\ \hline
$s_4$ & $a_2$ & 0.7 & 0.7  \\ \hline
\end{tabular}
\label{table:R}
\end{table}

\noindent {\bf CDD Rule Detection:} We assume that a static data repository $R$ is available, which can be collected/inferred by historical stream data \cite{mayfield2010eracer,song2015enriching,song2015screen,zhang2017time}. Following the literature \cite{kwashie2015conditional,wang2017discovering}, to infer a CDD rule in the form $X\to A_j$ from $R$, we first obtain determinant attributes $X$ from ($d$-1) attributes (other than $A_j$), where attributes $X$ are correlated with $A_j$ in $R$. Then, for each determinant attribute $A_x\in X$, we obtain a \textit{differential dependency} \cite{song2011differential} in the form $A_x\to A_j$ from data repository $R$. Specifically, if any determinant attributes $A_x$ cannot accurately impute $A_j$ with an acceptable interval (i.e., large $A_j.I$), we will adopt \textit{editing rule} \cite{fan2010towards} for $A_x$ to impute $A_j$, by considering constant values of $A_x$. This way, we can divide attributes $X$ into two parts, which take intervals and specific constant values as constraints in CDD rules, respectively. Please refer to \cite{kwashie2015conditional,wang2017discovering} for more details of CDD detection, and Appendix \ref{subsec:CDD_detect_eva} for the time cost of detecting (creating) CDD rules on our tested real data sets.

\noindent {\bf Imputing Missing Attributes:} Assume that there is a static data repository, $R$, consisting of complete data records $s$, that can be used to impute missing data. Given an incomplete tuple $r_i\in iDS$ with missing attribute $A_j$, we can utilize CDD rules in the form $X\to A_j$ (detected from $R$) to find some samples $s\in R$ to fill the missing attribute $r_i[A_j]$ with $s[A_j]$. 

In our previous example of Figure \ref{fig:motivation}, assume that we have a CDD rule $(Gender, Symptom \to Diagnosis, \{male, [0, 0.3], [0, 0.2]\})$. Then, given a complete tuple \textit{($p_1$, ``male'', ``weight loss, blurred vision'', ``diabetes'', ``drug therapy'')} in $R$, we find that tuples $p_1$ and $a_2$ (as depicted in Table \ref{table:motivation}) have the same or similar attributes, $Gender$ and $Symptom$ (i.e., satisfying distance constraints of the CDD). Thus, we can use the diagnosis result ``diabetes'' in $p_1$ to impute the missing \textit{Diagnosis} attribute of tuple $a_2$.

We will discuss more details later in Section \ref{sec:data_imputation} on how to impute the missing attribute values of $r_i\in iDS$, based on CDD rules and data repository $R$. This way, we can turn all incomplete records $r_i\in iDS$ into complete (imputed) ones, and obtain an imputed data stream, which is defined as follows.

\begin{definition} {\bf (Imputed Data Stream, $pDS$)}
Given an incomplete data stream $iDS=(r_1, r_2, ..., r_t, ...)$, an imputed data stream, $pDS$, contains an ordered sequence of imputed records (tuples), $(r_1^p, r_2^p, ..., r_t^p, ...)$. Each tuple $r_i^p\in pDS$ is the imputed version of $r_i\in iDS$, and contains some mutually exclusive instances (samples), $r_{i,m}$ (for $m\ge 1$), each of which is associated with an existence probability $r_{i,m}.p$, where $\sum_{\forall r_{i,m}} r_{i,m}.p \le 1$.
\end{definition}

\subsection{Topic-Aware Entity Resolution Over Incomplete Data Streams}
\label{subsec:TER-iDS}

\noindent {\bf The Similarity Function for ER:} In this paper, we assume that attributes in tuples are of textual data types (e.g., the extracted topic/attribute strings). Given two tuples $r$ and $r'$, a key problem of the ER process is how to measure the similarity between $r$ and $r'$. Specifically, we consider the \textit{Jaccard similarity} between two token sets (from two tuples, resp.) for each attribute, and define the similarity function for ER as the summation of similarities on all the $d$ attributes as follows.

\begin{definition} (The Similarity Function, $sim(r, r')$)
Given two $d$-dimensional complete tuples $r$ and $r'$, their similarity can be measured by:
\begin{equation}
\hspace{-1ex}sim(r, r') = \sum_{j=1}^d sim(r[A_j], r'[A_j]) = \sum_{j=1}^{d}\frac{|T(r[A_j])\cap T(r'[A_j])|}{|T(r[A_j])\cup T(r'[A_j])|}
\label{eq:SF}
\end{equation}
where $T(r[A_j])$ is a set of tokens in attribute $r[A_j]$.
\label{def:SF}
\end{definition}

For simplicity, in this paper, we consider data streams with homogeneous data schema. For the similarity function over data sets with heterogeneous schema \cite{papadakis2019survey}, we can take into account the \textit{Jaccard similarity} between two token sets $T(r)$ and $T(r')$ (from all attributes of two tuples, respectively), that is, $\frac{|T(r)\cap T(r')|}{|T(r)\cup T(r')|}$, which we would like to leave as our future work.

Next, we will formally define the TER-iDS problem as follows.

\noindent {\bf Problem Statement (Topic-Aware Entity Resolution Over Incomplete Data Streams).} Given $n$ ($\ge 2$) incomplete data streams, $\{iDS_1$, $iDS_2$, ..., $iDS_n\}$, a set, $\mathcal{K}$, of query topic keywords, a current timestamp $t$, an integer $w$, a similarity threshold $\gamma$ ($\in(0, 1)$), and a probabilistic threshold
$\alpha$ ($\in[0, 1)$), the problem of the \textit{topic-aware
entity resolution over incomplete data streams} (TER-iDS) is to retrieve
matching pairs, $(r_i, r_j)$, of (incomplete) tuples, $r_i$ and $r_j$, from two of $n$ data streams (sliding windows $W_t$ of size $w$), respectively, such that either $r_i$ or $r_j$ contains at least one query topic $k \in \mathcal{K}$ and represent the same entity with
probability, $Pr_{TER\text{-}iDS}(r_i, r_j)$, greater than threshold
$\alpha$, that is:
\begin{eqnarray}
&&\hspace{-4ex}Pr_{TER\text{-}iDS}(r_i, r_j) = \sum_{\forall r_{i,m}} \sum_{\forall r_{j,m'}} r_{i,m}.p \cdot r_{j,m'}.p  \label{eq:eq2}\\
&&\hspace{-4ex} \cdot  \chi((\varpi(r_{i,m},\mathcal{K}) \vee \varpi(r_{j,m'},\mathcal{K})) \wedge sim(r_{i,m}, r_{j,m'}) > \gamma) > \alpha \notag
\end{eqnarray}
\noindent where $\varpi(r_{i,m},\mathcal{K})$ is a Boolean function that indicates whether the token set of $r_{i,m}$ contains at least one keyword $k\in \mathcal{K}$; $sim(., .)$ is the Jaccard similarity function given by Equation~(\ref{eq:SF}); $\chi(z) = 1$, if $z$ is $true$ (otherwise, $\chi(z) = 0$); and $r_{i,m}$ is an instance of tuple $r_i$ with an existence probability $r_{i,m}.p$.

From our problem statement, the TER-iDS problem aims to monitor pairs of incomplete tuples, $(r_i, r_j)$, from sliding windows of any two streams, which are related to specific topics in $\mathcal{K}$ and represent the same entity with high entity resolution (ER) probability (as given in Inequality~(\ref{eq:eq2})). Specifically, in the TER-iDS problem, query keywords can be online specified by users, that is, we do not need to know query keywords in advance. Moreover, the TER-iDS approach can also support ER without any constraint of topics/keywords by setting the set, $\mathcal{K}$, of query topic keywords as the domain of all possible keywords.

\vspace{0.5ex}\noindent{\bf A Straightforward Method:}  A straightforward method to solve the TER-iDS problem is as follows. For each newly arriving tuple $r$ (with a missing attribute $r[A_j]$), we first obtain all CDD rules, $X\to A_j$. Then, we use these CDDs to retrieve samples $s$ (satisfying CDD constraints) in a data repository $R$, which can be used for imputing $r[A_j]$. Finally, we can search for tuples $r'$ from other data streams that represent the same entity as $r$ satisfying the topic and ER requirements (as given in our problem statement).

However, in practice, there are many CDD rules (e.g., 2,500 detected CDD rules over only 600 tuples, each with 7 attributes, on real data set, \textit{Cora} \cite{wang2017discovering}), and it is not efficient to obtain all CDDs with $A_j$ as a dependent attribute. Moreover, due to the large scale of the data repository $R$, it is rather time-consuming to retrieve all samples $s\in R$ to fill the missing attribute $r[A_j]$. Furthermore, it is not trivial how to efficiently obtain the matching tuples $r'$ from other $(d-1)$ data streams, since the computation of the ER probability (i.e., $Pr_{TER\text{-}iDS}(r, r')$ in Inequality (\ref{eq:eq2})) is very costly. Therefore, the straightforward method is rather inefficient.

\vspace{0.5ex}\noindent{\bf Challenges:} There are three major challenges to solve the TER-iDS problem. First, many previous ER works (e.g., \cite{papadakis2014meta,li2015linking,shen2014probabilistic,papadakis2019survey,ebraheem2018distributed,firmani2016online,dragut2015query}) usually assume that the underlying data are complete, or simply discard data records with missing attributes. However, in reality, application data often incur incompleteness (e.g., packet losses in sensor networks), and the strategy of ignoring incomplete data may cause inaccurate or erroneous ER results. Thus, previous ER techniques cannot be directly applied to solve the TER-iDS problem over incomplete data streams, and we need propose effective and efficient imputation approach to impute the missing attribute values of incomplete objects. 

Second, due to the intrinsic quadratic time complexity of the ER task, it is very challenging to efficiently and dynamically maintain all the topic-related entity pairs in the stream environment. Therefore, we need to devise some effective pruning strategies to reduce the search space of the TER-iDS problem.

Third, it is non-trivial how to efficiently and effectively impute the missing attribute values and conduct topic-aware ER analyses at the same time. Therefore, we need carefully design some effective index techniques to enable an efficient TER-iDS processing algorithm. 

Inspired by the challenges above, in this paper, we will propose an efficient framework for TER-iDS processing, as will be discussed in the next subsection.

\vspace{0.5ex}\noindent{\bf Discussions on The TER-iDS Problem:} Different from existing approaches \cite{papadakis2014meta,shen2014probabilistic,dragut2015query,li2015linking,firmani2016online,ebraheem2018distributed,simonini2018schema,wang2019efficient} that are not topic-aware, our TER-iDS problem only reports ER results with topics that users are interested in (i.e., related to one or multiple topics/keywords in a query keyword set $\mathcal{K}$). Moreover, our TER-iDS problem considers the \textit{sliding window} model (Definition \ref{def:sw}) for online ER \cite{li2015linking,firmani2016online,simonini2018schema,wang2019efficient}, since users are usually interested in the most recent data (in the sliding window), instead of old data (e.g., data from years ago). Thus, TER-iDS will return ER results that users are interested in; in other words, those ER results with irrelevant topics or expired entities will not be outputted by our TER-iDS approach. Based on our TER-iDS problem statement, we are not losing any information that users are not interested in. Nevertheless, our problem can be easily extended to consider arbitrary topics and all stream data by setting the query keyword set, $\mathcal{K}$, to the domain of keywords, and the size of sliding window to be infinite.

\begin{algorithm}[t!]\scriptsize
\KwIn{$n$ incomplete data stream $iDS_i$ ($1\le i\le n$), a data repository $R$, a timestamp $t$, and two thresholds $\gamma$ and $\alpha$} 
\KwOut{a TER result entity set, $ES$, over sliding windows $W_{i,t}$ ($1\le i\le d$)}

\tcp{Pre-Computation Phase}

offline select pivot tuples from the data repository $R$

offline compute CDD rules from $R$

offline construct CDD-indexes, $I_j$ ($1\leq j\leq d$), over CDD rules

offline construct a DR-index, $I_R$, over $R$

create a data synopsis, \textit{ER-grid}, over $n$ data streams $iDS_i$

$ES\leftarrow$ entity resolution result set within time interval $[t-w, t-1]$

\tcp{Imputation and TER-iDS Pruning Phase}
\For{expired objects $r_{t\text{-}w}$ in each data stream $iDS_i$}{
    remove the $r_{t\text{-}w}$ from \textit{ER-grid} 
    
    remove from entity result set $ES$ entity pairs involving $r_{t\text{-}w}$
}

\For{newly arriving (incomplete) tuple $r_t$ in each data stream $iDS_i$
}{
    simultaneously traverse index $I_j$ over CDDs, index $I_R$ over $R$, and data synopsis \textit{ER-grid} over sliding windows $W_t$ to enable the data imputation and ER query processing at the same time 
  
    add candidate matching entities of $r_t$ to the $r_t.ES$  
    
    insert $r_t^p$ into the \textit{ER-grid}
    
}

\tcp{TER-iDS Refinement Phase}
\For{newly arriving tuple $r_t$ in each data stream $iDS_i$}{
    \For{each candidates $r_c\in r_t.ES$}{
        \If{$Pr_{TER\text{-}iDS}(r_c, r_t)\le \alpha$}{
            remove $r_c$ from the $r_t.ES$
        }
    }

    add $r_t.ES$ to $ES$
}

return the entity set, $ES$, as the result of the TER-iDS problem
\caption{The TER-iDS Processing Framework}
\label{alg:framework}
\end{algorithm}

\subsection{The TER-iDS Framework}
\label{subsec:framework}

Algorithm \ref{alg:framework} illustrates a general framework for our TER-iDS solution, which consists of three phases: \textit{pre-computation phase},  \textit{imputation and TER\text{-}iDS pruning phase}, and \textit{TER-iDS refinement phase}.

In the first \textit{pre-computation phase}, we offline select pivot tuples from the data repository $R$ (Section \ref{subsec:metric_space}) (line 1), which will be used for constructing imputation indexes and ER synopsis. Then, we offline compute CDD rules from $R$, and construct indexes, $I_j$ and $I_R$ (Section \ref{subsec:indexes_CDD_and_R}), over CDD rules and data repository $R$, respectively (lines 2-4). Moreover, we create a data synopsis, \textit{ER-grid} (Section \ref{subsec:synopsis_Wt}), over $n$ data streams $iDS_i$ (line 5). We also use an entity result set, $ES$, to maintain all ER results from $n$ data streams $iDS_i$ at timestamp $(t-1)$ (line 6). 

In the \textit{imputation and TER\text{-}iDS pruning phase}, we online maintain the data synopsis \textit{ER-grid} over $d$ data streams $iDS_i$. 
Specifically, at timestamp $t$, we remove the expired tuple $r_{t\text{-}w}$ in each data stream $iDS_i$ from the \textit{ER-grid}, as well as all entity pairs containing $r_{t\text{-}w}$ from $ES$  (lines 7-9). For each newly arriving (incomplete) tuple $r_t$ in each stream $iDS_i$, we simultaneously traverse index $I_j$ (over CDDs), index $I_R$ (over data repository $R$), and \textit{ER-grid} (over streams), and obtain ER results $r_t.ES$ w.r.t. tuple $r_t$ (lines 10-12). In addition, we also insert tuple $r^p$ into \textit{ER-grid} (line 13).

In the \textit{TER-iDS refinement phase}, for each newly arriving tuple $r_t$ from any data stream $iDS_i$, we calculate actual TER-iDS probabilities, $Pr_{TER\text{-}iDS}(r_c, r_t)$ (as given in Equation~(\ref{eq:eq2})), of its candidate pairs $(r_t, r_c)$ (for  $r_c\in r_t.ES$),  and add final entity set $r_t.ES$ of $r_t$ to the result set $ES$ (lines 14-18). Finally, we return $ES$ as the result of the TER-iDS problem (line 19).

Table \ref{symbols_and_descriptions} depicts the commonly-used symbols and their descriptions in this paper.

\section{Incomplete Data Imputation}
\label{sec:data_imputation}
In the sequel, we will first illustrate how to leverage a single CDD \cite{kwashie2015conditional,wang2017discovering}, $X\rightarrow A_j$, to impute incomplete objects $r$ with missing attributes $A_j$. Then, we discuss how to impute the missing attribute values $r[A_j]$ via multiple available CDDs. 

\begin{table}\hspace{-2ex}
\centering
{\small\scriptsize
    \caption{\small Notation.}
    \label{symbols_and_descriptions}\vspace{-2ex}
    \begin{tabular}{l|l} \hline
    {\bf Symbol} & \qquad\qquad\qquad\qquad{\bf Description} \\ \hline \hline
    $iDS$ & an incomplete data stream \\ \hline 
    $pDS$ & an imputed (probabilistic) data stream \\ \hline 
    $r_i$ & an incomplete tuple from $iDS$ \\ \hline 
    $r_i^p$ & the imputed (probabilistic) tuple of an incomplete tuple $r_i$ \\ \hline 
    $W_t$ & a sliding window containing $w$ most recent objects from $iDS$ \\ \hline 
    $R$ & a static complete data repository for assisting the data imputation \\ \hline 
    $CDD$ & conditional differential dependency rule \\ \hline
    $sim(r, r')$ & a similarity function measuring objects $r$ and $r'$ \\ \hline 
    \end{tabular}\vspace{2ex}  
}
\end{table}

\noindent {\bf Data Imputation via a Single CDD:} Assume that we have a static data repository $R$, which stores (historical) complete data records (tuples) for data imputation. Denote $dom(A_j)$ as the domain of attribute $A_j$, which contains all possible values of attribute $A_j$ in repository $R$.

Given a single CDD rule $CDD: X\to A_j$ and an incomplete tuple $r$ with a missing attribute $r[A_j]$, we can retrieve all sample tuples $s$ from data repository $R$ that satisfy distance constraints on attributes $X$. Then, for each sample $s \in R$, we can obtain a candidate set, $cand(s[A_j])$, of possible imputed values, $val \in dom(A_j)$, for missing attribute $r[A_j]$, such that the Jaccard distance, $dist(s[A_j], val)$, between (token sets of) $s[A_j]$ and $val$ is within the interval $A_j.I$. This way, we can compute all possible values to impute missing attribute $r[A_j]$, by taking a union of candidate sets $cand(s[A_j])$ for all sample tuples $s$, that is, $\bigcup_{\forall s} cand(s[A_j])$.

Let $F(\cdot)$ be a frequency distribution of all possible values $v$ ($\in \cup_{\forall s} cand(s[A_j])$) of missing attribute $r[A_j]$, where the frequency, $F(v)$, of each value $v$ is given by the times that $v$ appears in $cand(s[A_j])$ for all samples $s$. In order to obtain the probability confidence, $v.p$, of each imputed value $v$, we normalize the frequency distribution and calculate the probability $v.p$ as follows:
\begin{equation}
v.p= \frac{F(v)}{\sum_{\forall val} F(val)}.
\label{eq:eq1}
\end{equation}

\begin{example}
Consider a data schema with 3 attributes $\{A, B,$ $C\}$, a CDD rule $CDD_1:AB \to C,\{a_1, [0, 0.1], [0, 0.1]\}$, and a data repository $R=\{s_1, s_2, s_3, s_4\}$ (as depicted in Table \ref{table:R}). We have the domain of attribute $C$ inferred from $R$, that is, $dom(C)=\{0.1, 0.2, 0.35, 0.7\}$. For an incomplete tuple $r=(a_1, 0.3,$ $-)$, we can obtain two samples, $s_1$ and $s_2$, from $R$ satisfying distance constraints on attributes $AB$ w.r.t. $r$ (i.e., $\phi[AB]=\{a_1, [0, 0.1]\}$). For example, for sample $s_1$, we have $s_1[A]=r[A_j]=a_1$ and $|s_1[B]-r[B]|=0.1\in [0, 0.1]$.

With samples $s_1$ and $s_2$, we can obtain two candidate sets for imputing $r[C]$, that is, $cand(s_1[C]) = \{ 0.1, 0.2\}$ and  $cand(s_2[C]) = \{ 0.1, 0.2 \}$, respectively. This way, we can compute a frequency distribution with 2 possible imputed values $\{0.1, 0.2\}$ and their frequencies $\{2, 2\}$, respectively.  Therefore, each of the 2 imputed values (i.e., $0.1$ and $0.2$) has the existence probability $\frac{2}{4}$. 
\label{example:single_CDD}
\end{example}

\noindent{\bf Data Imputation via Multiple CDDs.} Given a data repository $R$, there may exist more than one CDD rule, $X_1\to A_j$, $X_2\to A_j$, ..., and $X_l\to A_j$, which can be used for imputing the missing attribute $A_j$, where attributes in $X_i$ are non-missing, for $1\leq i\leq l$. In this case, we have two imputation strategies to impute an incomplete object $r$ with missing attribute $A_j$. That is, we can either choose one suitable CDD rule or use all the $l$ CDDs for imputation. In this paper, we will consider the latter strategy (i.e., all CDDs) and leave the former one as our future work.

Specifically, for each of $l$ CDDs $X_i \to A_j$ (for $1\leq i\leq l$), we can impute a missing attribute $r[A_j]$ of an incomplete tuple $r$ with a set of candidate values $v$, each with its frequency, denoted as $F_i(v)$. Instead of considering one single CDD rule (Equation~(\ref{eq:eq1})), with $l$ CDDs, we can obtain the existence probability, $v.p$, of each possible imputed value $v$ as follows.
\begin{equation}
v.p= \frac{\sum_{i=1}^l F_i(v)}{\sum_{i=1}^l \sum_{\forall val} F_i(val)},
\label{eq:eq4}
\end{equation}
\noindent where $F_i(v)$ is the frequency of the imputed value $v$ suggested by $CDD_i$ ($1\le i\le l$).

Intuitively, for candidate values $v$ to fill the missing attribute $r[A_j]$, we give more weights (existence probabilities) $v.p$ (as given in Equation~(\ref{eq:eq4})), if values $v$ are suggested by more CDD rules (or with higher frequencies). 

\begin{example}
Continue with Example \ref{example:single_CDD}. For the data repository $R$ in Table \ref{table:R}, assume that we have two CDD rules $CDD_1:AB \to C,\{a_1, [0, 0.1], [0, 0.1]\}$ and $CDD_2:AB \to C,\{a_1,$ $(0.1, 0.2], [0, 0.2]\}$. As mentioned in Example \ref{example:single_CDD}, by $CDD_1$, we can obtain the frequency distribution $F_1(v)$, where possible values $v$ to impute $r[C]$ are $\{0.1, 0.2\}$ with frequencies $\{2, 2\}$, respectively. Similarly, by $CDD_2$, we can obtain another frequency distribution $F_2(v)$, with possible values $\{0.2, 0.35\}$ and their frequencies $\{1, 1\}$, respectively.

By combining $F_1(\cdot)$ with $F_2(\cdot)$, we can obtain a set of possible imputed values $\{0.1, 0.2, 0.35\}$ for attribute $C$, with frequencies $\{2, 3, 1\}$, respectively. Correspondingly, their existence probabilities can be calculated as $\left\{\frac{2}{4+2}, \frac{3}{4+2}, \frac{1}{4+2}\right\}$, respectively. 
\label{example:mul_CDD}
\end{example}

\section{Pruning Strategies}
\label{sec:pruning_strategy}

As discussed in Section \ref{subsec:TER-iDS}, it is rather challenging to efficiently and effectively tackle the TER-iDS problem (as given in our problem statement in Section \ref{subsec:TER-iDS}) in the streaming environment. In order to reduce the problem search space, in this section, we will propose effective pruning strategies to significantly filter out false alarms. For proofs of all the theorems/lemmas below, please refer to Appendix \ref{sec:all_proofs}.

\vspace{0.5ex}\noindent{\bf Pruning with Topic Keywords.} We first present an effective pruning method, \textit{topic keyword pruning}, with respect to the constraint of topic keywords. Intuitively, given two (incomplete) tuples $r_i$ and $r_j$, if neither $r_i$ nor $r_j$ contains any topic keywords $k\in \mathcal{K}$, based on Inequality~(\ref{eq:eq2}) (in our problem statement in Section \ref{subsec:TER-iDS}), we do not need to further check whether they refer to the same entities. 

Formally, we have the following pruning theorem.

\begin{theorem} {\bf (Topic Keyword Pruning)}
Given two (incomplete) tuples $r_i$ and $r_j$, the tuple pair $(r_i, r_j)$ can be safely pruned, if  $\varpi(r_{i,m}, \mathcal{K}) = false$ and $\varpi(r_{j,m'}, \mathcal{K}) = false$ hold, for all possible instances, $r_{i,m}$ and $r_{j,m'}$, of the imputed (probabilistic) tuples $r_i^p$ and $r_j^p$, respectively.
\label{lem:lem1}
\end{theorem}

By Theorem \ref{lem:lem1}, we can filter out false alarms of pairs $(r_i, r_j)$ that do not contain any keywords in $\mathcal{K}$. Specifically, for incomplete tuples $r_i$ and $r_j$, in the process of the data imputation, we can prune a tuple pair $(r_i, r_j)$, if we can ensure that the imputed tuples $r_i^p$ and $r_j^p$ have no chance to contain any keywords in $\mathcal{K}$. 

\vspace{0.5ex}\noindent{\bf Pruning via Similarity Upper Bound.} We next present the second pruning strategy, namely \textit{similarity upper bound pruning}, which filters out tuple pairs with low similarity scores $sim(r_i, r_j)$ (as given by Equation~(\ref{eq:SF})). 

Denote $ub\_sim(r_i, r_j)$ as the upper bound of similarity scores $sim(r_{i,m}, r_{j,m'})$, for all possible instance pairs $(r_{i,m}, r_{j,m'})$ of imputed tuples $r_i^p$ and $r_j^p$, respectively. Then, we have the following theorem.
\begin{theorem} {\bf (Similarity Upper Bound Pruning)}
Given two (incomplete) tuples $r_i$ and $r_j$, the tuple pair $(r_i, r_j)$ can be safely pruned, if $ub\_sim(r_{i,m}, r_{j,m'})\le \gamma$ holds for all possible instance pairs, $(r_{i,m}, r_{j,m'})$, of imputed tuples $r_i^p$ and $r_j^p$, respectively.
\label{lem:lem2}
\end{theorem}

In Theorem \ref{lem:lem2}, if the similarity upper bound $ub\_sim(r_i, r_j)$ is less than or equal to threshold $\gamma$ (i.e.,  $ub\_sim(r_i, r_j) \le \gamma$), then we can safely prune this tuple pair $(r_i, r_j)$ (due to Inequality (\ref{eq:eq2})). 

Below, we will discuss how to calculate the similarity upper bound $ub\_sim(r_i, r_j)$, by either the token set size or pivot.

\underline{\it Similarity upper bound via token set size.} Given (incomplete) objects $r_i$ and $r_j$, we can obtain their similarity upper bound based on the token set sizes of possible attribute values as follows.

\begin{lemma} 
A similarity upper bound, $ub\_sim(r_i, r_j)$, of (incomplete) tuples $r_i$ and $r_j$ can be given by summing up the similarity upper bounds, $ub\_sim(r_i[A_k], r_j[A_k])$, for all attributes $A_k$, that is, $ub\_sim(r_i, r_j) = \sum_{k=1}^d ub\_sim(r_i[A_k], r_j[A_k])$.

Here, we have:
\begin{eqnarray}
&&ub\_sim(r_i[A_k], r_j[A_k])\notag\\
&=&\begin{cases} \frac{|T^+(r_j^p[A_k])|}{|T^-(r_i^p[A_k])|} &\hspace{2ex} \text{if } |T^-(r_i^p[A_k])| > |T^+(r_j^p[A_k])|\notag\\  
\frac{|T^+(r_i^p[A_k])|}{|T^-(r_j^p[A_k])|} &\hspace{2ex} \text{if } |T^+(r_i^p[A_k])| < |T^-(r_j^p[A_k])|\notag\\ 
1 &\hspace{2ex} \text{otherwise} \notag
\end{cases},
\label{eq:Aj_ub}
\end{eqnarray}
\label{cor:cor1}
\noindent where $|T^-(r_x^p[A_k])|$ and $|T^+(r_x^p[A_k])|$ are the minimum and maximum sizes of token sets $T(r_{x,m}[A_k])$ for all instances $r_{x,m}$ of imputed objects $r_x^p$ ($x=i$ or $j$) on attributes $A_k$, respectively.
\end{lemma}

\begin{example}
Given a data schema with three textual attributes, $\{A, B, C\}$, and two incomplete tuples, $r_1=(a_1, b_1, -)$ and $r_2=(a_2, b_2, -)$, with missing attribute $C$, assume that the attribute token sets of imputed tuples $r_1^p$ and $r_2^p$ have sizes (intervals) as follows: $|T(r_1[A])| = 10$, $|T(r_2[A])| = 8$, $|T(r_1[B])| = 7$, $|T(r_2[B])| = 10$, $|T(r_1[C])| \in [5, 7]$, and $|T(r_2[C])| \in [10, 12]$. Therefore, we can obtain the similarity upper bounds of tuples $r_1^p$ and $r_2^p$ on attributes as: $ub\_sim(r_1[A], r_2[A])=\frac{8}{10}$, $ub\_sim(r_1[B],\\ r_2[B])=\frac{7}{10}$, and $ub\_sim(r_1[C], r_2[C])=\frac{7}{10}$. Finally, we can obtain the similarity upper bound of $r_1$ and $r_2$ as $ub\_sim(r_1, r_2)=0.8+0.7+0.7=2.2$.
\label{ex:cor1}
\end{example}

Given imputed tuples $r_x^p$ of incomplete tuples $r_x$ ($x=i$ or $j$), we can obtain the size intervals, $[|T^-(r_x^p[A_k])|, |T^+(r_x^p[A_k])|]$, of the token sets $T(r_x^p[A_k])$ of imputed tuples $r_x^p$ on attributes $A_k$, based on the possible values $r_x^p[A_k]$ of $r_x^p$ on attributes $A_k$. Thus, we can quickly obtain similarity upper bounds of tuple pairs in Lemma \ref{cor:cor1}.

\underline{\it Similarity upper bound via a pivot tuple.} Next, we will derive another similarity upper bound, based on the property of Jaccard similarity function. Specifically, the Jaccard similarity, $sim(r_i[A_k],$ $r_j[A_k])$, is given by $1-dist(r_i[A_k], r_j[A_k])$, where $dist(r_i[A_k],$ $r_j[A_k])$ is called \textit{Jaccard distance} which is a metric function, following the triangle inequality. 

According to the property of Jaccard similarity mentioned above, we can transform the matching condition $sim(r_i, r_j) > \gamma$ (given in Equation (\ref{eq:eq2})) to its equivalent form $d- dist(r_i, r_j) > \gamma$, where $dist(r_i, r_j) = \sum_{k=1}^d dist(r_i[A_k], r_j[A_k])$.

Denote $min\_dist(r_i[A_k], r_j[A_k)$ as the minimum possible Jaccard distance, $dist(r_i[A_k], r_j[A_k])$, between (token sets of) attributes $r_i^p[A_k]$ and $r_j^p[A_k]$, which can be computed with respect to a pivot tuple $piv$. Then, we can obtain a similarity upper bound: $ub\_sim(r_i, r_j)$ $=d$ $-\sum_{k=1}^d$ $min\_dist(r_i[A_k], r_j[A_k])$.

\begin{lemma}
Given a pivot tuple $piv$, a similarly upper bound, $ub\_sim(r_i, r_j)$, of (incomplete) tuples $r_i$ and $r_j$ is given by:
$ub\_sim(r_i,$ $r_j) = d - \sum_{k=1}^d min\_dist(r_i[A_k],$ $r_j[A_k])$.

Let $X_k = dist(r_i[A_k], piv[A_k])$  and $Y_k = dist(r_j[A_k],$ $piv[A_k])$, where $dist(\cdot, \cdot)$ is a Jaccard distance function. Assume that $X_k \in [lb\_X_k, ub\_X_k]$ and $Y_k \in [lb\_Y_k, ub\_Y_k]$. Then, we can obtain:
\begin{eqnarray}
min\_dist(r_i[A_k], r_j[A_k])= \begin{cases}lb\_X_k - ub\_Y_k & \text{if } lb\_X_k > ub\_Y_k\notag\\  
lb\_Y_k - ub\_X_k & \text{if } lb\_Y_k > ub\_X_k\notag\\  
0 & otherwise \notag
\end{cases}.
\end{eqnarray}
\label{cor:cor2}
\end{lemma}

In Lemma \ref{cor:cor2}, the bounds, say $[lb\_X_k, ub\_X_k]$, of Jaccard distance $dist(r_i[A_k], piv[A_k])$ (i.e., $X_k$) can be computed as follows. First, based on the data repository $R$ and CDD rules, we can infer possible imputed values (texts or token sets), $val$, of the missing attribute $r_i[A_k]$ (as discussed in Section \ref{sec:data_imputation}). Then, the interval bounds, $[lb\_X_k, ub\_X_k]$, can be obtained by taking the minimum and maximum Jaccard distances, $dist(val, piv[A_k])$, for all possible values $val$, respectively.

We will discuss how to obtain a good pivot tuple later in Section \ref{subsec:metric_space}.

\begin{example}
Consider two (incomplete) tuples $r_1$ and $r_2$ with 3 attributes $\{A, B, C\}$, and a pivot tuple $piv$. Assume that tuples $r_1$ and $r_2$ have Jaccard distances (or distance bounds) with pivot $piv$ on 3 attributes as $\{0.3, 0.3, [0.1, 0.2]\}$ and $\{0.7, 0.8, [0.7, 0.9]\}$, respectively. From Lemma \ref{cor:cor2}, we can compute the similarity upper bound as $ub\_sim(r_1, r_2)= d - \sum_{\forall A_k} min\_dist(r_i[A_k],$ $r_j[A_k])$ $= 3 - ((0.7-0.3)+(0.8-0.3)+(0.7-0.2)) = 3 - 1.4 = 1.6$.
\label{ex:cor2}
\end{example}

In this paper, we will quickly compute similarity upper bounds (via token set size and/or pivot), and utilize them to enable effective similarity upper bound pruning (as given in Theorem \ref{lem:lem2}). 

\vspace{0.5ex}\noindent{\bf Pruning via Probability Upper Bound.} Theorem \ref{lem:lem2} can prune false alarms of tuple pairs with $ub\_sim(r_i, r_j) \leq \gamma$ (in other words, $Pr_{TER\text{-}iDS}(r_i, r_j)=0$). Next, we will present an effective pruning strategy, namely \textit{probability upper bound pruning}, which can filter out false alarms of tuple pairs with low TER-iDS probability (i.e., $Pr_{TER\text{-}iDS}(r_i, r_j)\leq \alpha$).

Specifically, assume that we can quickly compute an upper bound, $UB\_Pr_{TER\text{-}iDS}(r_i, r_j)$, of the TER-iDS probability $Pr_{TER\text{-}iDS}(r_i,$ $r_j)$ (in Inequality~(\ref{eq:eq2})). If it holds that $UB\_Pr_{TER\text{-}iDS}(r_i, r_j)$ $\le \alpha$, then we can safely prune the tuple pair $(r_i, r_j)$. Formally, we have:

\begin{theorem} {\bf (Probability Upper Bound Pruning)}
Given two (incomplete) tuples $r_i$ and $r_j$, the tuple pair $(r_i, r_j)$ can be safely pruned, if $UB\_Pr_{TER\text{-}iDS}(r_i, r_j) \le \alpha$ holds.
\label{lem:lem3}
\end{theorem}

To obtain the probability upper bound $UB\_Pr_{TER\text{-}iDS}(r_i, r_j)$ in Theorem \ref{lem:lem3}, we have the following lemma.

\begin{lemma}(Paley-Zygmund Based Probability Upper Bound)
Given two (incomplete) tuples $r_i$ and $r_j$, and a pivot tuple $piv$, based on Paley-Zygmund Inequality \cite{paley1932some}, we can obtain a probability upper bound: 
\begin{eqnarray}
&&UB\_Pr_{TER\text{-}iDS}(r_i, r_j)\notag\\
&=&\begin{cases}
1 - (1 - \frac{d-\gamma}{E(X) - E(Y)})^2 & \text{if } 0\leq\frac{d-\gamma}{E(X) - E(Y)}\leq 1 \\
\qquad \cdot \frac{E(X) - E(Y)}{ub\_X - lb\_Y}, & \text{\quad and } lb\_X\geq ub\_Y\notag\\
1 - (1 - \frac{d-\gamma}{E(Y) - E(X)})^2 & \text{if } 0\leq\frac{d-\gamma}{E(Y) - E(X)}\leq 1\\
\qquad \cdot \frac{E(Y) - E(X)}{ub\_Y - lb\_X}, & \text{\quad and } lb\_Y\geq ub\_X \notag\\
1, & otherwise \notag\\
\end{cases},
\end{eqnarray}
\noindent where $X$ and $Y$ denote the Jaccard distances, $dist(r_i, piv)$ and $dist(r_j, piv)$, from the imputed tuples $r_i^p$ and $r_j^p$, respectively, to pivot $piv$, $E(Z)$ is the expectation of variable $Z$, and $lb\_X = \sum_{k=1}^d lb\_X_k$ and $ub\_X = \sum_{k=1}^d ub\_X_k$ are the minimal and maximal values of variable $X$, respectively ($lb\_Y$ and $ub\_Y$ are the same). 
\label{cor:corPZ}
\end{lemma}

To obtain $E(X)$ in Lemma \ref{cor:corPZ}, we first calculate the Jaccard distances, $dist(val,$ $piv[A_k])$, from all possible (textual) values $val\in r_i^p[A_k]$ of the imputed tuple $r_i^p$ to the pivot attribute $piv[A_k]$. Then, we can obtain: $$E(X) = \sum_{k=1}^d E(X_k) = \sum_{k=1}^d \sum_{\forall val\in r_i^p[A_k]} dist(val, piv[A_k])\cdot val.p.$$ 

Moreover, lower/upper bounds, $lb\_X$ and $ub\_X$, of variable $X$ ($=dist(r_i, piv)$) can be given by $lb\_X=\sum_{k=1}^d lb\_X_k$ and $ub\_X$ $=\sum_{k=1}^d ub\_X_k$, respectively, where $lb\_X_k$ and $ub\_X_k$ are lower and upper bounds of $dist(r_i[A_k], piv[A_k])$, respectively. 

The case of $E(Y)$, $lb\_Y$, and $ub\_Y$ is similar and thus omitted here.

\begin{example}
Given two incomplete tuples $r_1$ and $r_2$ with 3 attributes $\{A, B, C\}$, a pivot tuple $piv$, and a similarity threshold $\gamma=2.8$, assume that the imputed tuples $r_1^p$ and $r_2^p$ have Jaccard distances to $piv$ on the three attributes as: $\{0.1, 0.1, \{0.1,$ $0.5, 0.9\}\}$ and $\{0.2, 0.2, \{0.7, 0.9\}\}$, respectively. Note that, there are multiple possible distance values from $piv$ to $r_1^p$ ($r_2^p$) on attribute $C$, and we consider their existence probabilities as equal. Denote $X$ and $Y$ as the distances from pivot $piv$ to tuples $r_1$ and $r_2$, respectively. Therefore, we can obtain: 

$E(X)=\sum_{k=1}^3 E(X_k) = 0.1 + 0.1 + \frac{0.1+0.5+0.9}{3} = 0.7,$

$E(Y)=\sum_{k=1}^3 E(Y_k)=0.2+0.2+\frac{0.7+0.9}{2}=1.2,$

$lb\_X = 0.1 + 0.1 + 0.1 = 0.3$,

$ub\_X = 0.1 + 0.1 + 0.9 = 1.1$,

$lb\_Y= 0.2 + 0.2 + 0.7 = 1.1$, and

$ub\_Y = 0.2 + 0.2 + 0.9 = 1.3$. 

Based on Lemma \ref{cor:corPZ}, since $lb\_Y (1.1)\geq ub\_X (1.1)$ and $0\leq \frac{d-\gamma}{E(Y)-E(X)} = \frac{3-2.8}{1.2 - 0.7} \leq 1$ hold, we can obtain a probability upper bound: $UB\_Pr_{TER\text{-}iDS}(r_i, r_j) =$ $1 - (1 - \frac{3-2.8}{1.2-0.7})^2\times \frac{1.2-0.7}{1.3 - 0.3} = 0.82$.
\end{example}

\vspace{0.5ex}\noindent{\bf Instance-Pair-Level Pruning.} Next, we present an effective \textit{instance-pair-level pruning method}, which filters out those tuple pairs, $(r_i, r_j)$, when most of their instance pairs $(r_{i,m}, r_{j,m'})$ do not match with high probability. Intuitively, in Inequality~(\ref{eq:eq2}), it is costly to compute the TER-iDS probability, $Pr_{TER\text{-}iDS}(r_i, r_j)$, by enumerating different possible combinations of instance pairs $(r_{i,m}, r_{j,m'})$. Thus, our instance-pair-level pruning method will overestimate the matching probability for those instance pairs that have not been calculated, and prune false alarms with low TER-iDS probability.

\begin{theorem}{\bf (Instance-Pair-Level Pruning)} Given two imputed tuples $r_i^p$ and $r_j^p$, assume that we have computed the TER-iDS probability for a set, $S$, of instance pairs $(r_{i,m}, r_{j,m'})$. Then, the tuple pair $(r_i, r_j)$ can be safely pruned, if it holds that: 
\begin{eqnarray}
&&\hspace{-4ex}\sum_{\forall (r_{i,m}, r_{j,m'})\in S} \hspace{-4ex}Pr(r_{i,m}, r_{j,m'})  + \left(1-\sum_{\forall (r_{i,m}, r_{j,m'})\in S} \hspace{-4ex}r_{i,m}.p \cdot r_{j,m'}.p\right)\notag\\
&&\hspace{-4ex}\leq \alpha,\notag
\end{eqnarray}
where $Pr(r_{i,m}, r_{j,m'}) = r_{i,m}.p \cdot r_{j,m'}.p \cdot  \chi((\varpi(r_{i,m},\mathcal{K}) \vee \varpi(r_{j,m'},$ $\mathcal{K})) \wedge sim(r_{i,m}, r_{j,m'})> \gamma)$.
\label{lem:lem5}
\end{theorem}

Theorem \ref{lem:lem5} uses partially checked instance pairs of the tuple pair to prune false alarms, and overestimates the probability for those instance pairs that have not been processed, which saves the computation cost of checking the remaining instance pairs. Note that, the instance-pair-level pruning in Theorem \ref{lem:lem5} can be used on the instance level, when we want to calculate the actual TER-iDS probability for the refinement. 

To reduce the TER-iDS search space, we will apply the 4 pruning strategies in the order of topic keyword pruning (Theorem \ref{lem:lem1}), similarity upper bound pruning (Theorem \ref{lem:lem2}), probability upper bound pruning (Theorem \ref{lem:lem3}), and instance-pair-level pruning (Theorem \ref{lem:lem5}). As we will discuss later in Section \ref{subsec:pruning_evaluation}, these 4 pruning strategies can together achieve high pruning power (98.32\%$\sim$99.43\%) over the tested real data sets.

\section{Topic-aware Entity Resolution over Incomplete Data Streams}
\label{sec:TER_iDS}
Section \ref{subsec:indexes_CDD_and_R} presents two types of indexes over CDD rules and a data repository $R$, respectively, to facilitate the data imputation. Section \ref{subsec:synopsis_Wt} proposes a data synopsis, namely \textit{ER-grid}, over tuples in sliding window $W_t$ of incomplete data streams. Section \ref{subsec:Algorithm_TER_iDS} provides an efficient TER-iDS processing algorithm, which essentially traverses and joins indexes/synopses over CDDs, data repository, and stream data. Section \ref{subsec:metric_space} discusses our proposed cost model to select ``good'' \textit{pivot tuples} over textual tuples, which can enable fast pruning for the TER-iDS processing. Finally, Section \ref{subsec:dynamic_discussion} illustrates the basic idea of extending our TER-iDS method with dynamic data repository $R$.

\subsection{Imputation Indexes Over CDDs and Data Repository $R$}
\label{subsec:indexes_CDD_and_R}

In order to facilitate efficient and effective imputations for missing attribute data, we will propose two types of indexes, namely CDD-indexes and DR-index, over CDD rules and data repository $R$, respectively. 

\vspace{1ex}\noindent{\bf CDD-Indexes, $I_j$, Over CDD Rules.} Assume that we obtain all the CDD rules from a data repository $R$. Then, for each attribute $A_j$ ($1\le j\le d$), we can build a CDD-index, denoted as $I_j$, over those CDD rules in the form of $X_f\to A_j$ (for $1\leq f\leq L$). For any incomplete tuple $r$ with a missing attribute $r[A_j]$, we can utilize CDD-index $I_j$ to quickly select some suitable CDD rules and impute attribute $r[A_j]$.

\begin{figure}[t!]
\centering\vspace{-3ex}                                     
\hspace{0ex}\scalebox{0.18}[0.18]{\includegraphics{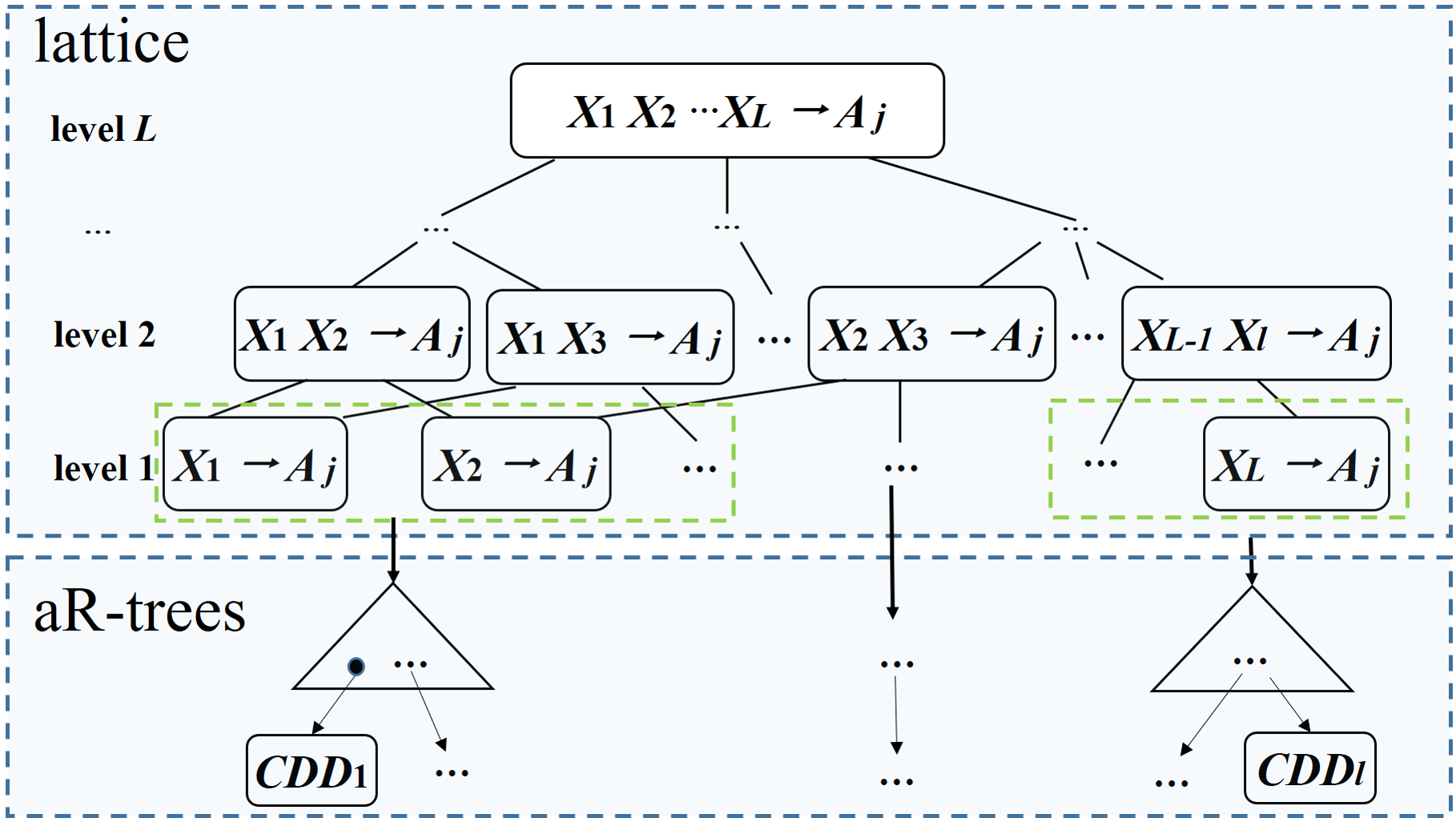}}\vspace{-2ex}
\caption{\small Illustration of the CDD-index $I_j$.} \label{fig:index_CDD}
\end{figure}

As illustrated in Figure \ref{fig:index_CDD}, the CDD-index $I_j$ is a hierarchical structure, which contains two portions, lattice and aR-trees \cite{lazaridis2001progressive}.

\underline{Lattice.} As shown in Figure \ref{fig:index_CDD}, the lattice part is composed of multiple levels, where each level contains some CDD rule(s), $X_f\to A_j$ (without the constraint function $\phi[\cdot]$). Specifically, at the bottom of the lattice (Level 1), we have $L$ nodes, each corresponding to CDD rules, in the same form of $X_i\to A_j$ ($1\leq i\leq L$), but with different constraint function $\phi[\cdot]$; on Level 2, we have the combined CDD rules in the form $X_aX_b\to A_j$ ($1\leq a\ne b\leq L$), each of which is a combination of two CDD rules, $X_a\to A_j$ and $X_b\to A_j$, on Level 1; and so on. Finally, on the top level (Level $L$), we have one combined CDD rule in the form of $X_1X_2...X_L\to A_j$. 

As an example in Figure \ref{fig:index_CDD}, the CDD rule $X_1X_2\to A_j$ on Level 2 is a combined rule of 2 CDDs, $X_1\to A_j$ and $X_2\to A_j$, on Level 1. 

\underline{aR-Trees.} We can cluster CDDs on Level 1 of the lattice into $g$ groups of combined CDD rules, $X_m\to A_j$ ($1\leq m\leq g$), each of which contains CDDs $X_f\to A_j$ such that $X_f\subseteq X_m$.

For each of the $g$ CDD groups, we construct an aR-tree \cite{lazaridis2001progressive} over CDD rules, $X_f\to A_j$, in the group. Essentially, we build the aR-tree on constraint functions, $\phi[X_m]$, of determinant attributes $A_x \in X_m$, which can be one of three types: (1) constant values (e.g., $A_x = v$), (2) intervals (e.g., $A_x.I = [0, 0.1]$), and, (3) ``-1'' (i.e., $A_x.I = [-1, -1]$, indicating that $A_x$ is missing). 

In this paper, we assume that attributes of tuples are textual data. Therefore, instead of directly indexing textual constants $v$ (for attributes $A_x$) in CDDs, we introduce some pre-selected pivot tuples, convert textual constants $v$ into numeric values via pivot tuples, and index the converted values in aR-trees. In particular, for each attribute $A_x$ with constant constraints $v$ in CDDs, we offline pre-select $n_x$ pivots, $piv_1$, $piv_2$, ..., and $piv_{n_x}$. Then, we convert text $v$ into a numeric value, which is defined as the Jaccard distance, $dist(v, piv_1[A_x])$, between $v$ and the \textit{main pivot} $piv_1[A_x]$, and will be indexed in aR-tree. The Jaccard distances from $v$ to the remaining $(n_x-1)$ \textit{auxiliary pivots}, $piv_a$ ($2\leq a\leq n_x$), will be used as aggregates in the aR-tree (discussed later).

This way, we can use the aR-tree \cite{lazaridis2001progressive} to index constraints (i.e., converted Jaccard distance via pivot tuples, intervals, and ``-1'') on different determinant attributes (dimensions) in $X_f$ for CDDs, via standard ``insert'' function.

{\it Aggregates in aR-trees:} In leaf nodes of the aR-tree, each CDD rule $X_f\to A_j$ is associated with two types of aggregate values, that is, (1) the distance constraint, $A_j.I$, of dependent attribute $A_j$, and; (2) distances, $dist(v, piv_a[A_x])$, from constants $v$ specified on attributes $A_x\in X_f$ to auxiliary pivot attributes $piv_a[A_x]$. Note that, we do not include the missing attribute $A_x$ ($A_x.I=[-1,1]$) for aggregates, since we only index on non-missing constraint attributes.

Moreover, each entry, $e_{CDD}$, of non-leaf nodes contains aggregates for CDD rules under $e_{CDD}$ as follows.
\begin{itemize}
\item a minimum interval, $A_j.I_e$, that bounds constraint intervals, $A_j.I$, for all CDDs under node $e_{CDD}$, and;
\item intervals, $I_{x,a}$, that minimally bound the distances, $dist(v,$ $piv_a[A_x])$, between constant constraints $v$ and pivot attributes $piv_a[A_x]$, for all CDD rules under node $e_{CDD}$.
\end{itemize}
where $piv_a$ are \textit{auxiliary pivots} (excluding the main pivot $piv_1$). 

\begin{figure}[t!]
\centering\vspace{-3ex}                                     
\hspace{0ex}\scalebox{0.16}[0.16]{\includegraphics{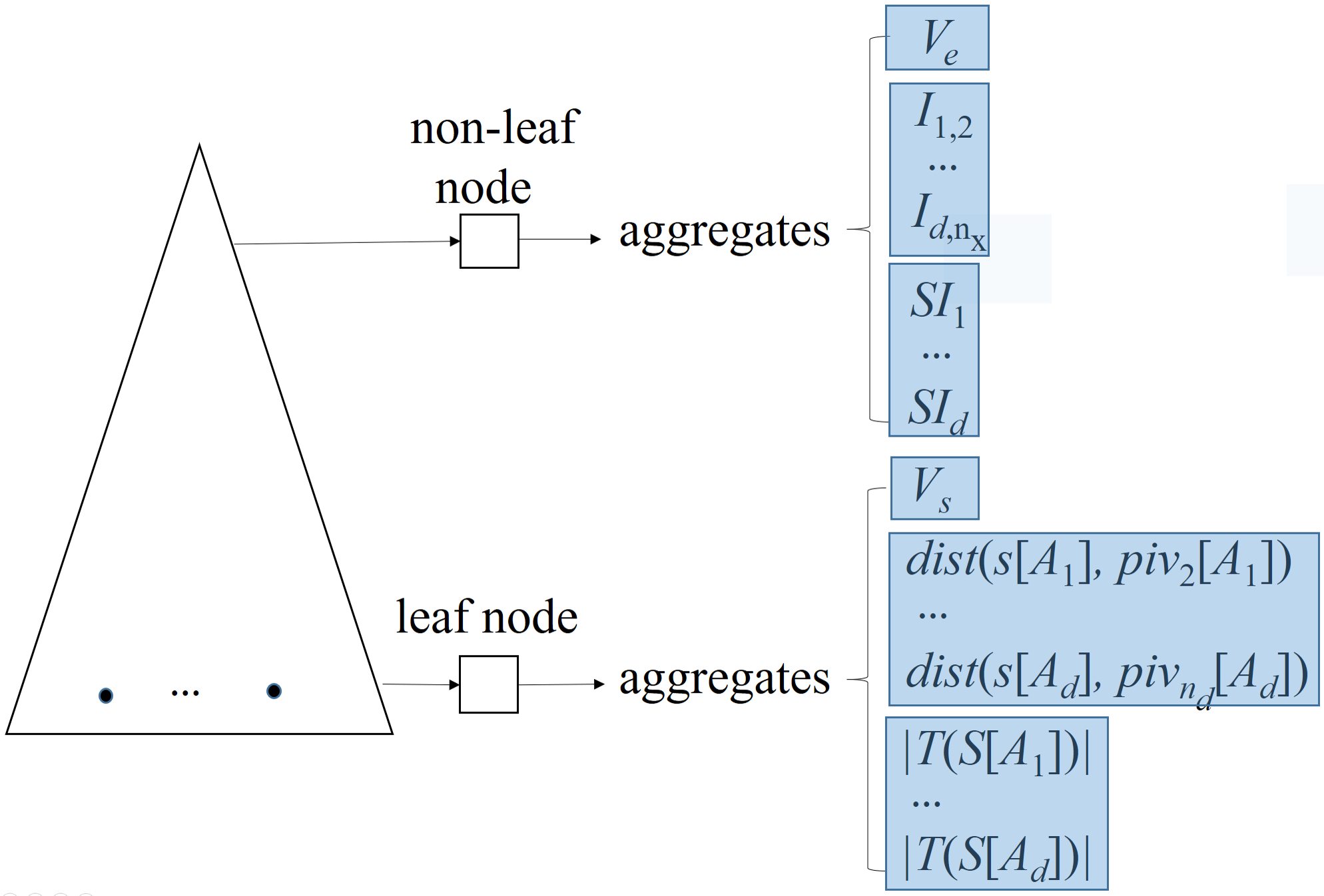}}\vspace{-2ex}
\caption{\small  Illustration of the DR-index $I_R$.} \label{fig:index_R}
\end{figure}

\vspace{1ex}\noindent{\bf DR-Index, $I_R$, Over Data Repository $R$.} Given a data repository $R$ and a pre-selected main pivot $piv_1$, we can convert each tuple (sample) $s\in R$ into a $d$-dimensional data point in the metric space, where the $x$-th attribute $A_x$ ($1\leq x\leq d$) has the converted value: $dist(s[A_x], piv_1[A_x])$. Then, as shown in Figure \ref{fig:index_R}, we can establish an index over $R$, denoted as $I_R$, by inserting the converted $d$-dimensional data points into an aR-tree \cite{lazaridis2001progressive}. 

In leaf nodes of the aR-tree, each tuple (sample) $s\in R$ is associated with three types of aggregate values: (1) a boolean vector, $V_s$, in which each element corresponds to a keyword/topic (i.e., bits ``1'' or ``0'' indicating the existence or non-existence of the keyword/topic in $s$, resp.); (2) distances, $dist(s[A_x], piv_a[A_x])$, between sample $s$ and auxiliary pivots $piv_a$, for attributes $A_x$ ($1\leq x\leq d$), and; (3) sizes, $|T(s[A_x])|$, of token sets, $T(s[A_x])$, for attributes $A_x$ ($1\leq x\leq d$) in sample $s$.

Moreover, each non-leaf node, $e_{R}$, in the index $I_R$ maintains aggregates as follows.
\begin{itemize}
\item a boolean vector, $V_e$, indicating the (non-)existence of some keywords/topics under node $e_R$;
\item intervals, $I_{x,a}$, that minimally bound distances, $dist(s[A_x],$ $piv_a[A_x])$, for all tuples $s\in e_R$ ($1\leq x\leq d$ and $2\leq a \leq n_x$), and;
\item size intervals, $SI_x$, that minimally bounds token set sizes, $|T(s[A_x])|$, for all tuples $s\in e_R$ ($1\leq x\leq d$). 
\end{itemize}

\noindent{\bf Complexity Analysis.} Given a newly arriving incomplete tuple, the worst-case time complexity of obtaining suitable CDD rules via the CDD-index is given by $O(N_{leaf}\cdot N_{CDD})$, where $N_{leaf}$ is the number of non-pruned leaf nodes of the aR-trees in CDD-index that may contain suitable CDD rules for imputation, and $N_{CDD}$ is the number of CDD rules inside leaf nodes. Moreover, the worst-case time complexity of retrieving samples $s$ from data repository $R$ via the DR-index for imputation is given by $O(M_{leaf}\cdot N_s)$, where $M_{leaf}$ is the number of non-pruned leaf nodes of the aR-trees in DR-index, and $N_s$ is the number of suggested samples $s$ inside leaf nodes.

\vspace{1ex}\noindent{\bf Index Join for Imputation.} To impute a missing attribute $A_j$, we can access both CDD rules and data repository $R$ at the same time, by performing an index join over CDD-indexes $I_j$ and DR-index $I_R$, respectively. Specifically, we can
simultaneously traverse indexes $I_j$ and $I_R$, obtain candidate nodes $e_{CDD} \in I_j$ that may contain CDDs to impute attribute $A_j$, and meanwhile retrieve candidate nodes $e_R\in I_R$ that contain samples for imputing $A_j$. When we traverse both indexes in a top-down manner, we apply our proposed pruning methods to rule out unpromising nodes in $I_j$ and $I_R$, until we finally obtain relevant CDDs and matching samples $s\in R$ to impute missing attribute $A_j$.

In this paper, we will perform online data imputation and ER processing at the same time. Therefore, instead of the index join on CDD-indexes $I_j$ and DR-index $I_R$, we will conduct the index join over 3 indexes/synopsis over $I_j$, $I_R$, and data synopses over incomplete data streams (as will be discussed in Section \ref{subsec:Algorithm_TER_iDS}).

\subsection{Data Synopsis for Sliding Window}
\label{subsec:synopsis_Wt}

In this subsection, based on the pre-selected pivot tuples, we design a data synopsis, namely \textit{entity resolution grid} (\textit{ER-grid}) $G_{ER}$, over $n$ incomplete data streams $iDS_y$ (for $1\leq y\leq n$), which can be used for identifying matching (incomplete) tuples from incomplete data streams. 

\vspace{1ex}\noindent{\bf ER-Grid Over Data Streams.} An \textit{ER-grid}, denoted as $G_{ER}$, is a $d$-dimensional grid file, which divides the data space into cells $c$ of the same size. Each cell $c$ in the ER-grid $G_{ER}$ contains tuples $r$ from data streams $iDS_y$ that intersect with cell $c$.

To online construct the ER-grid, for each tuple $r$ from a data stream $iDS_y$, we first convert it into a $d$-dimensional data point, whose $x$-th dimension ($1\leq x\leq d$) is given by the Jaccard distance between attributes $r[A_x]$ and $piv_1[A_x]$, where $piv_1$ is the main pivot. Then, we insert the converted data point of $r$ into cells $c\in G_{ER}$, such that the imputed tuples $r^p$ of $r$ fall into cells $c$.

Each cell $c$ in $G_{ER}$ is associated with aggregates as follows.
\begin{itemize}
\item a boolean vector, $V_c$, indicating the (non-)existence of some keywords/topics under cell $c$;
\item intervals, $I_{x,a}$, that minimally bound distances, $dist(r^p[A_x],$ $piv_a[A_x])$, for all tuples $r\in c$ ($1\leq x\leq d$ and $2\leq a\leq n_x$), and;
\item size intervals, $SI_x$, that minimally bounds token set sizes, $|T(r^p[A_x])|$, for all imputed tuples $r^p\in c$ ($1\leq x\leq d$).
\end{itemize}

In each cell $c\in G_{ER}$, each (imputed) tuple $r^p$ is associated with 4 types of aggregate values: (1) a boolean vector, $V_r$, in which each element corresponds to a probabilistic keyword/topic (i.e., bits ``1'' or ``0'' indicating the existence or non-existence of the keyword/topic in $r^p$, resp.); (2) intervals, $r.SI_x$, that minimally bound token set sizes, $|T(r^p[A_x])|$, of all instances of imputed tuple $r^p$, for attributes $A_x$ ($1\leq x\leq d$); (3) intervals, $r.I_{x,a}$, that minimally bound distances, $dist(r^p[A_x], piv_a[A_x])$, between instances of imputed tuple $r^p$ and pivot tuples $piv_a$, on attributes $A_x$ ($1\leq x\leq d$ and $1\leq a\leq n_x$), and; (4) expectations, $E(dist(r^p[A_x], piv_a[A_x]))$, of distances, $dist(r[A_x], piv[A_x])$, between tuple instances and pivot tuples $piv_a$, for attributes $A_x$ ($1\leq x\leq d$ and $1\leq a\leq n_x$).

\vspace{1ex}\noindent{\bf Dynamic Maintenance of $G_{ER}$.} Since we consider the model of sliding window $W_t$ (given in Definition \ref{def:sw}), we need to incrementally maintain the {\it ER-grid} $G_{ER}$. Specifically, at timestamp $t$, we will evict the expired tuples $r_{t-w}$ from $G_{ER}$, and update the aggregate information of cells $c_a\in G_{ER}$ that store tuple $r_{t-w}$. Moreover, we insert into cells $c_b\in G_{ER}$ those newly arriving tuples $r_t$, where $r_t^p$ intersects with $c_b$, and update the aggregates in $c_b$ with $r_t^p$.

\noindent{\bf Complexity Analysis.} Given an expired tuple $r_{exp}$, the worst-case time complexity of updating $G_{ER}$ is given by $O(N_c\cdot N_r)$, where $N_c$ is the number of cells intersecting with $r_{exp}$, and $N_r$ is the number of tuples inside the intersecting cells. Moreover, given a newly arriving tuple $r_{new}$, the worst-case time complexities of updating $G_{ER}$ and obtaining its matched tuples via $G_{ER}$ are given by $O(M_c)$ and $O(U_c\cdot U_r)$, respectively, where $M_c$ is the number of cells in $G_{ER}$ intersecting with $r_{new}$, $U_c$ is the number of non-pruned cells in $G_{ER}$, and $U_r$ is the number of tuples inside the non-pruned cells.

\begin{algorithm}[t!]\scriptsize
\KwIn{a keyword set $\mathcal{K}$, a data repository $R$, $n$ data streams $iDS_i$ ($1\le i\le n$), pivot tuples $piv_a$ for attributes $A_x$ ($1\le x\le d$, $1\le a\le n_x$), and a timestamp $t$} 
\KwOut{A TER-iDS result set $ES$}

$ES\leftarrow$ entity resolution result set at timestamp $(t-1)$

\For{each expired tuple $r_{t\text{-}w}$ in each data stream $iDS_i$}{
    remove $r_{t\text{-}w}$ from the \textit{ER-grid} $G_{ER}$
    
    \For{each unexpired imputed tuple $r_u\in$ $G_{ER}$ matched with $r_{t\text{-}w}$}{
        remove the pair $(r_{t\text{-}w}, r_u)$ from the $ES$
    }
    
    \For{each cell $c\in$ $G_{ER}$ containing the $r_{t\text{-}w}$}{
        update the aggregate information in the cell $c$
    }
}

\For{each newly arriving (incomplete) tuple $r_t$ (with $r_t[A_j]=$``$-$'') in each data stream $iDS_i$}{
    impute $r_t[A_j]$ and obtain a set, $r_t.ES$, of potentially matching tuples of $r_t$ by joining 3 indexes: $I_j$, $I_R$, and $G_{ER}$
    
    calculate and obtain the aggregate information of imputed object $r_t^p$ on attributes $A_x$ ($1\leq x\leq d$)
    
    insert $r_t^p$ into the $G_{ER}$
    
    \For{each cell $c\in G_{ER}$ containing the $r_t^p$}{
        update the aggregate information in the $c$
    }
    
    \If{$r_t^p$ contains any keyword in $\mathcal{K}$}{
        \For{each $r_u\in r_t.ES$}{
            \If{the pair $(r_t, r_u)$ can be pruned by Theorems \ref{lem:lem2}-\ref{lem:lem5}}{
                remove $r_u$ from $r_t.ES$
            }\ElseIf{$Pr_{TER\text{-}iDS}(r_u, r_t) \le \alpha$}{
                remove $r_u$ from $r_t.ES$
            }
        }
    }\Else{
        \For{each tuple $r_u\in r_t.ES$}{
            \If{$r_u$ does not contain any keyword in $\mathcal{K}$}{
                remove $r_u$ from $r_t.ES$\tcp{Theorem \ref{lem:lem1}}
            }\Else{
                conduct operations in lines 16-19
            }
        }
    }
    
    add $r_t.ES$ to $ES$
    
}

return the TER-iDS result set $ES$

\caption{The TER-iDS Algorithm}
\label{alg:TER-iDS}
\end{algorithm}

\subsection{The TER-iDS Processing Algorithm}
\label{subsec:Algorithm_TER_iDS}

Next, we illustrate our algorithm to solve the TER-iDS problem, which leverages the indexes, $I_j$ and $I_R$, over CDD rules and data repository $R$, respectively, and the data synopsis (i.e., \textit{ER-grid}) over incomplete data streams. 

\vspace{1ex}\noindent{\bf The TER-iDS Algorithm.} Algorithm \ref{alg:TER-iDS} illustrates the basic idea of our TER-iDS processing algorithm. At each timestamp $t$, we first initialize an entity set, $ES$, that contains all matching tuple pairs at timestamp $(t-1)$ (line 1). Then, in lines 2-7, for each expired tuple $r_{t-w}$ in data stream $iDS_i$, we remove $r_{t-w}$ from the \textit{ER-grid} $G_{ER}$, remove tuple pairs $(r_{t-w}, r_u)$ (involving the expired tuple $r_{t-w}$) from $ES$, and update the aggregates in cells $c$ (e.g., boolean vector $V_c$ and intervals $I_{x,a}$ and $SI_x$, as discussed in Section \ref{subsec:synopsis_Wt}). 

Moreover, for each newly arriving (incomplete) tuple $r_t$ in each data stream $iDS_i$, we will impute $r_t[A_j]$ and obtain a set, $r_t.ES$, of potential matching tuples of $r_t$ by joining the indexes/synopsis over CDD rules, data repository $R$, and data streams $iDS_i$ at the same time (lines 8-9). Then, we calculate and maintain the aggregate information of the imputed tuples $r_t^p$ on attributes $A_x$ ($1\leq x\leq d$), which contains the boolean vector $V_r$, expectations $E(dist(r^p[A_x], piv_a[A_x]))$, and intervals $r.SI_x$ and $r.I_{x,a}$ (line 10). Next, in lines 11-13, we insert the imputed tuple $r_t^p$ into the \textit{ER-grid} $G_{ER}$, and update aggregates in cells $c$ intersecting with $r_t^p$ (e.g., the boolean vector $V_c$ and intervals $I_{x,a}$ and $SI_x$). For each potentially matching tuple $r_u\in r_t.ES$ of $r_t$, we will check and remove the non-matching ones from $r_t.ES$, by leveraging the pruning strategies in Section \ref{sec:pruning_strategy} (lines 14-25). Thus, we can obtain the final matching set $r_t.ES$ of $r_t$, and add all pairs in $r_t.ES$ to the entity set $ES$ (line 26). Finally, we return $ES$ as our TER-iDS result set (line 27).

\underline{Index join over $I_j$, $I_R$, and $G_{ER}$}. In line 9 of Algorithm \ref{alg:TER-iDS}, given an incomplete tuple $r_t$ with $r_t[A_j]=$``$-$'', the 3-way index join (i.e., $I_j$, $I_R$, and $G_{ER}$) imputes $r_t[A_j]$ and obtains the entity set $r_t.ES$ (containing the potentially matching tuples of $r_t$) at the same time. The basic idea is as follows. Given an incomplete tuple $r_t$, we will access the CDD-index $I_j$ to obtain entries $e_{CDD}$ from the root, which represent some combined (coarse) CDD rules. Meanwhile, with these CDD-index entries, we can find initial query ranges over DR-index $I_R$, in which samples in $R$ may be used for imputing tuple $r_t$ (with false positives). Similarly, we can obtain a coarse query range (w.r.t. the imputed tuple $r_t^p$) over \textit{ER-grid} $G_{ER}$, in which an initial entity superset, $r_t.ES$, of $r_t$ is retrieved. Next, we will iteratively access children nodes $e_{CDD}$ of CDD-index $I_j$, and examine more accurate (combined) CDDs, which in turn lead to more precise query ranges on lower levels of DR-index $I_R$ and narrower query range over \textit{ER-grid} $G_{ER}$ for the topic-based ER process. This process repeats until we reach the bottom levels of indexes of CDD- and DR-indexes, and obtain final imputed tuple $r_t^p$ and matching entity set $r_t.ES$ of $r_t$. 

\subsection{Cost-Model-Based Pivot Tuple Selection}
\label{subsec:metric_space}
As mentioned in Section \ref{subsec:indexes_CDD_and_R}, we need to select $n_x$ pivot tuples, $piv_a$ ($1\leq a\leq n_x$), and use their attributes $piv_a[A_x]$ to convert textual attribute $A_x$ of data tuple $r_i$ into numeric values $dist(r_i[A_x], piv_a[A_x])$. In this subsection, we will discuss how to select ``good'' pivot tuples.

\noindent{\bf Pivot Attribute Selection.} We select textual attributes, $piv_a[A_x]$, of pivot tuples $piv_a$ from the domains of attributes $A_x$ in data repository $R$. Given samples $s$ in $R$, a good pivot of attribute $A_x$, denoted as $piv_a[A_x]$ ($1\leq a\leq n_x$), should be able to distribute the converted values, $dist(s[A_x], piv_a[A_x])$, of attributes $s[A_x]$ as evenly as possible in the converted space on $A_x$. We evenly divide its converted space (i.e., $[0, 1]$) into $P$ buckets, $p_b$ ($1\leq b\leq P$), of equal length.
Then, we measure the converting quality of a pivot attribute $piv_a[A_x]$ by the Shannon entropy \cite{zhang2013reducing} as follows.
\begin{equation}
H(piv_a[A_x]) = -\sum_{b=1}^P pdf[p_b] \cdot log(pdf[p_b])
\label{eq:entropy}
\end{equation}
\noindent where $pdf[p_b]$ is the ratio of the converted values $dist(s[A_x], piv_a[A_x])$ of $s[A_x]$ falling into the $b$-th bucket.

As given in Equation (\ref{eq:entropy}), larger Shannon entropy indicates better converting quality of the pivot, i.e., evenly distributing the converted attribute values in the converted space. Therefore, for each attribute $A_x$, we design a cost-model-based algorithm to select the best pivot attributes $piv_a[A_x]$ (from the attribute domain in $R$) that maximize the entropy $H(piv_a[A_x])$. Please refer to Appendixes~\ref{sec:cost_model_for_pivots} and \ref{subsec:cost_model_verification} for the cost-model-based algorithm and its evaluation, respectively.

Note that, we utilize pivot tuples to build indexes (i.e., $I_j$, $I_R$, and $G_{ER}$), whose selection is guided by our designed cost model (i.e., Equation (\ref{eq:entropy})), and aim to achieve the best query performance. Since the targeting goal of our cost model is based on heuristics, our Algorithm (i.e., Algorithm \ref{alg:TER-iDS}) is expected to obtain high, but sub-optimal, performance via index in the stream environment.

\subsection{Discussions on TER-iDS with Dynamic Data Repository $R$}
\label{subsec:dynamic_discussion}
In this paper, we assume that data repository $R$ is static, however, our solution can be extended to dynamically updated $R$, where $R$ is periodically updated with a batch of new (complete) data from streams. In this case, we need to incrementally update the DR-index, CDD rules, and CDD-indexes, while our proposed approach still works for the ER-grid (since it is proposed for handling dynamic data streams). 

Specifically, given a batch of new complete data, $s_{new}$, from streams, we will discuss dynamic updates of DR-index, CDD rules, and CDD-indexes. We first update the DR-index by inserting the converted $d$-dimensional data points of $s_{new}$ into the aR-tree \cite{lazaridis2001progressive}. Then, we will update the CDD rules. Given the previously detected $CDDs$, $X\to A_j$, from data repository $R$, we check whether $s_{new}$ meets the constraints of any existing $CDDs$ on determinant attributes (i.e., $\phi[X]$). If $s_{new}$ does not meet the constraint $\phi[A_j]$, then we will delete relevant expired CDD rules, and add new CDD rules that work for $s_{new}$ with larger distance interval on dependent attribute $A_j$. Finally, if we have some newly detected CDD rules, similar to DR-index, we will insert new CDD rules into CDD-indexes; meanwhile, we will remove expired/deleted CDD rules from CDD-indexes. We would like to leave dynamic updates of data repository $R$ for missing data imputation as our future work.

\section{Experimental Evaluation}
\label{sec:exper}

\subsection{Experimental Settings}

\vspace{1ex}\noindent {\bf Real Data Sets.} We evaluate our TER-iDS approach on 5 real-world data sets, $Citations$, $Anime$, $Bikes$, $EBooks$, and $Songs$, as depicted in Table \ref{table:data_sets}. $Citations$ \cite{kopcke2010evaluation} describes citations between DBLP and ACM. $Anime$, $Bikes$, and $EBooks$ data sets were created for entity matching purpose by CS students at UW-Madison in a Fall 2015 data science class \cite{magellandata}. Specifically, $Anime$ was collected from My Anime List and Anime Planet, $Bikes$ came from Bikedekho and Bikewale, and $EBooks$ was extracted from iTunes and eBooks. $Songs$ \cite{das2017falcon} is a self-join set of a million songs.

Based on the missing rate $\xi$, we randomly select $(100 \cdot \xi)$ percent of tuples from streams as incomplete objects, and then mark $m$ random attribute(s) as missing. $Citations$ and $Songs$ data sets include actual groundtruth; for $Anime$, $Bikes$, and $EBooks$ data sets, we obtain the groundtruth of matching pairs based on Equation (\ref{eq:eq2}).

\vspace{1ex}\noindent {\bf State-of-the-art Approaches.} We compare our TER-iDS approach with five competitors, namely $I_j+G_{ER}$, $CDD+ER$, $DD+ER$, $mul+ER$, and $con+ER$. The details of the five baseline methods are as follows (please refer to \cite{fan2010towards,kwashie2015conditional,song2011differential,zhang2016sequential,wang2017discovering} for more implementation details).
\begin{itemize}
\item[$\bullet$]$I_j+G_{ER}$: this baseline first imputes the missing attribute values via the CDD rules \cite{kwashie2015conditional,wang2017discovering} (with help of the CDD-index $I_j$), and then performs the ER query via the ER-Grid $E_{ER}$ over incompelte data streams;
\item[$\bullet$]$CDD+ER$: this baseline first imputes the missing attribute values via the CDD rules \cite{kwashie2015conditional,wang2017discovering} (without help of the CDD-index $I_j$), and then conducts the ER query over incomplete data streams (without help of the ER-Grid $G_{ER}$);
\item[$\bullet$]$DD+ER$: this baseline first imputes the missing attribute values via the DD rules \cite{song2011differential}, and then proceeds the ER query over incomplete data streams;
\item[$\bullet$]$er+ER$: this baseline first imputes the missing attribute values via the \textit{editing rule} \cite{fan2010towards} method, and then performs the ER query over incomplete data streams; 
\item[$\bullet$]$con+ER$: this baseline first imputes the missing attribute values via the \textit{constraint-based imputation method} \cite{zhang2016sequential} method, and then conducts the ER query over incomplete data streams. 
\end{itemize}

\vspace{1ex}\noindent {\bf Measures.} In our experiments, we report the $F\text{-}score$ (effectiveness) of our approach against baselines. Here, the $F\text{-}score$ is defined as:\vspace{-3ex}

\begin{eqnarray}\small
F\text{-}score &=& 2 \times \frac{recall \times precision}{recall + precision},
\label{eq:Fscore}\vspace{-3ex}
\end{eqnarray}

\noindent where the \textit{recall} is given by the number of actual matching pairs in the returned TER-iDS results divided by the size of groundtruth, and the \textit{precision} is given by the number of actual matching pairs in the returned TER-iDS results divided by the total number of the returned pairs. 

We also report the average \textit{wall clock time} (i.e., CPU time) of our proposed TER-iDS approach, for each new timestamp, to impute incomplete data and conduct the topic-based ER.

\begin{table}[t!]
\scriptsize
\caption{\small The tested real data sets.} \label{table:data_sets}\vspace{-2ex}
\begin{tabular}{|c|c|c|c|}
\hline
\textbf{Data Sets} & \textbf{Source A (No. of Tuples)} & \textbf{Source B (No. of Tuples)} & \textbf{No. of Correct Matches} \\
\hline
\hline
Citations & 2,614 & 2,294 & 2,224 \\\hline
Anime & 4,000 & 4,000 & 10,704 \\\hline
Bikes & 4,786 & 9,003 & 13,815\\\hline
EBooks & 6,500 & 14,112 & 16,719 \\\hline
Songs & 1,000,000 & 1,000,000 & 1,292,023 \\\hline
\end{tabular}
\end{table}

\begin{table}[t!]
\scriptsize\vspace{-2ex}
\caption{\small The parameter settings.} \label{table:exp_parameter_setting}\vspace{-2ex}
\begin{tabular}{|l|c|}
\hline
\qquad\qquad\qquad\qquad\textbf{Parameters} & \textbf{Values}\\
\hline
\hline
probabilistic threshold $\alpha$ & 0.1, 0.2, \textbf{0.5}, 0.8, 0.9 \\\hline
the ratio, $\rho$, of similarity threshold $\gamma$ w.r.t. dimensionality &  0.3, 0.4, \textbf{0.5}, 0.6, 0.7 \\\hline
the missing rate, $\xi$, of incomplete tuples in $iDS$ & \textbf{0,1}, 0.2, 0.3, 0.4, 0.5, 0.8 \\\hline
the size, $w$, of the sliding window $W_t$ & 500, 800, \textbf{1000}, 2000, 3000 \\\hline
the size ratio, $\eta$, of data repository $R$ w.r.t. data stream $iDS$ & 0.1, 0.2, \textbf{0.3}, 0.4, 0.5 \\\hline
the number, $m$, of missing attributes & \textbf{1}, 2, 3 \\\hline
\end{tabular}
\end{table}

\begin{figure}[t!]\vspace{-1ex}
\scalebox{0.25}[0.25]{\hspace{0ex}\includegraphics{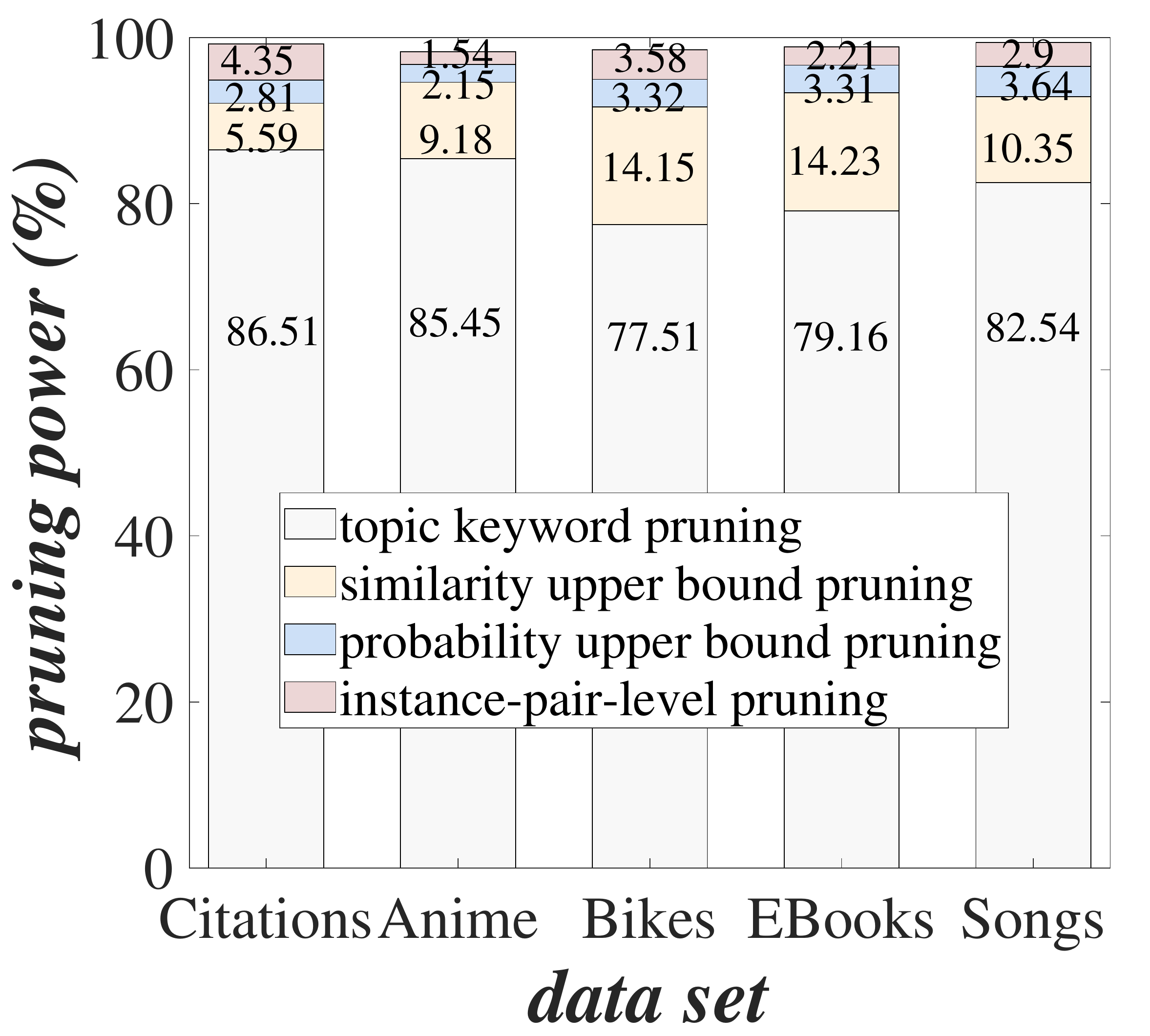}}\vspace{-2ex}
\caption{\small Pruning power evaluation over real data sets.}
\label{fig:pruning_power}
\end{figure}

\begin{figure}[t!]
\centering
\subfigure[][{\small F-score}]{\hspace{-3ex}                  
\scalebox{0.19}[0.18]{\includegraphics{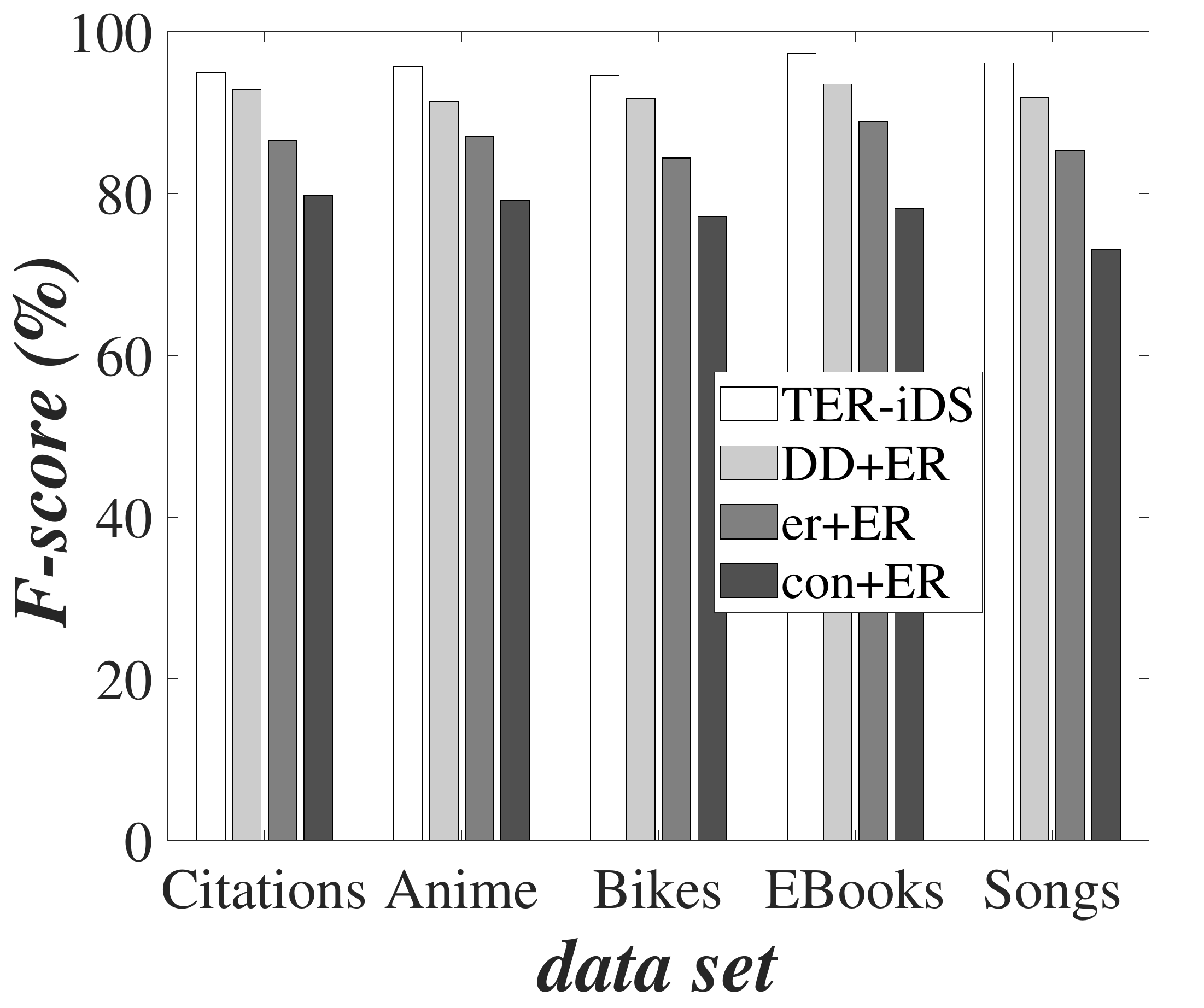}}
\label{subfig:Fscore}
}\quad
\subfigure[][{\small wall clock time}]{\hspace{-3ex}                  
\scalebox{0.19}[0.185]{\includegraphics{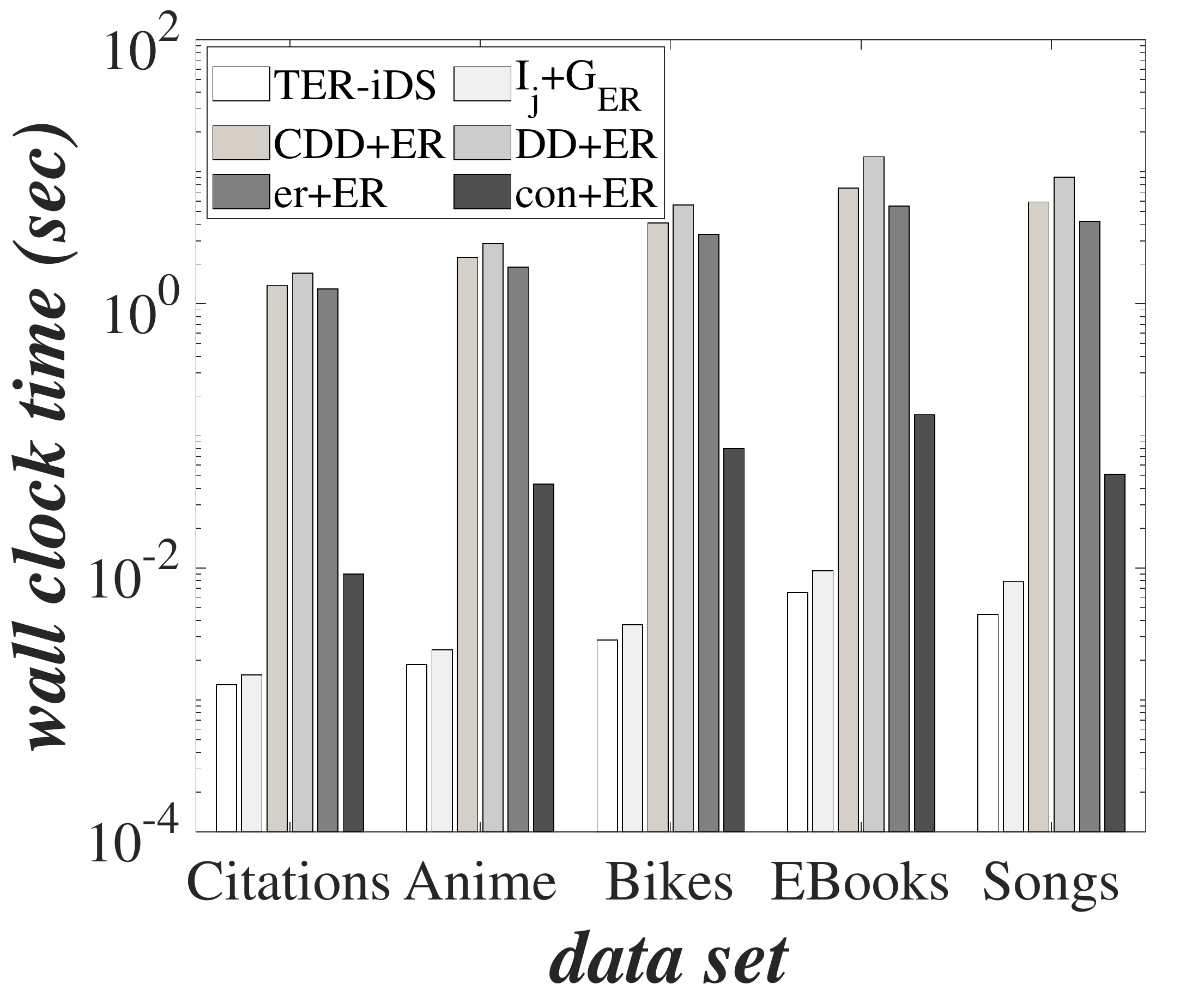}}
\label{subfig:efficiency}
}\vspace{-3ex}
\caption{\small The effectiveness and efficiency vs. real data sets.}
\label{fig:performance_vs_datasets}\vspace{0ex}
\end{figure}

\begin{figure}[t!]\vspace{-1ex}
\scalebox{0.2}[0.2]{\hspace{0ex}\includegraphics{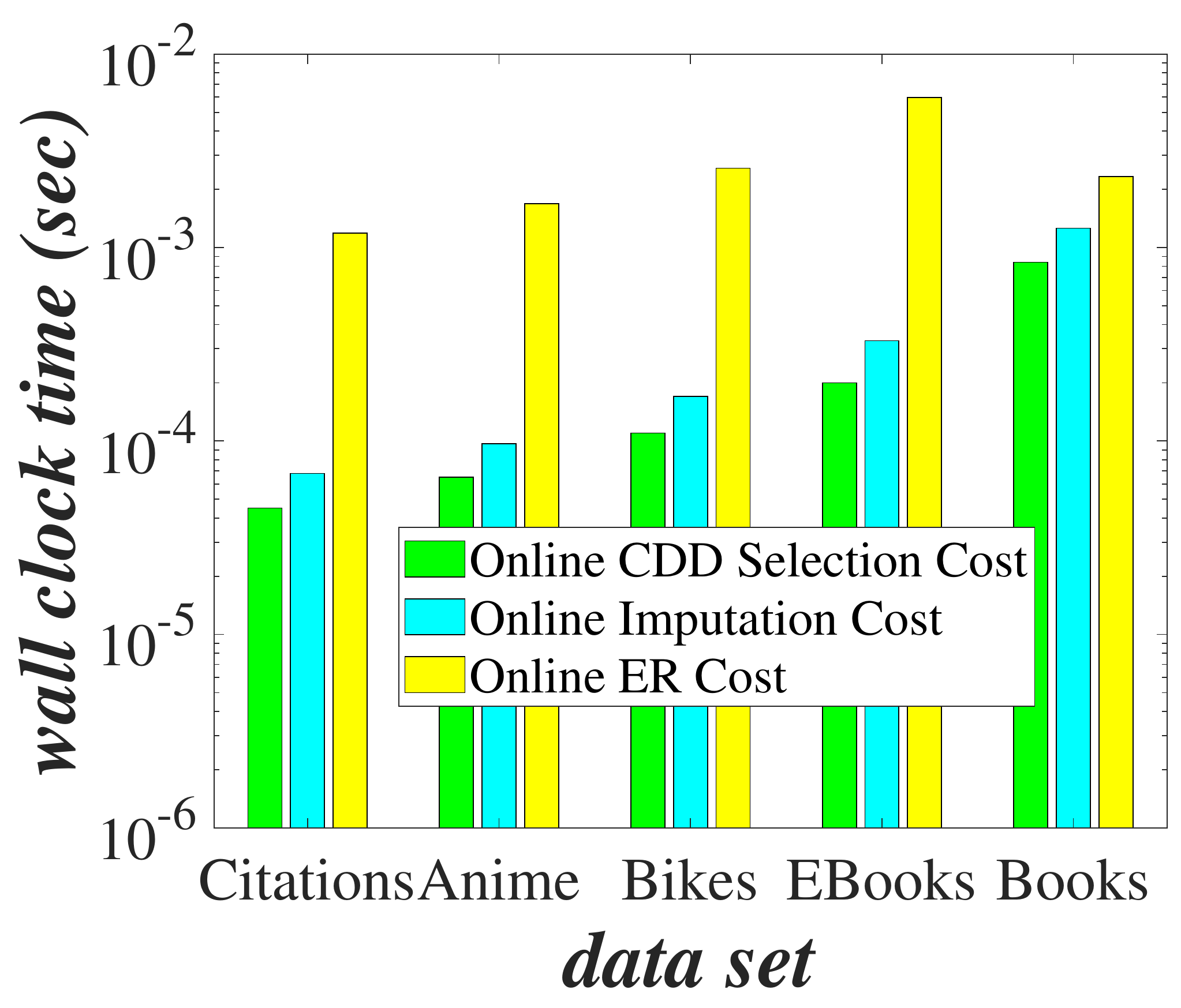}}\vspace{-2ex}
\caption{\small A break-up cost of TER-iDS in Figure \ref{subfig:efficiency}.}
\label{fig:break_up_cost}
\end{figure}

\begin{figure*}[ht]
\centering 
\subfigure[][{\small $Citations$}]{\hspace{-2ex}                  
\scalebox{0.17}[0.15]{\includegraphics{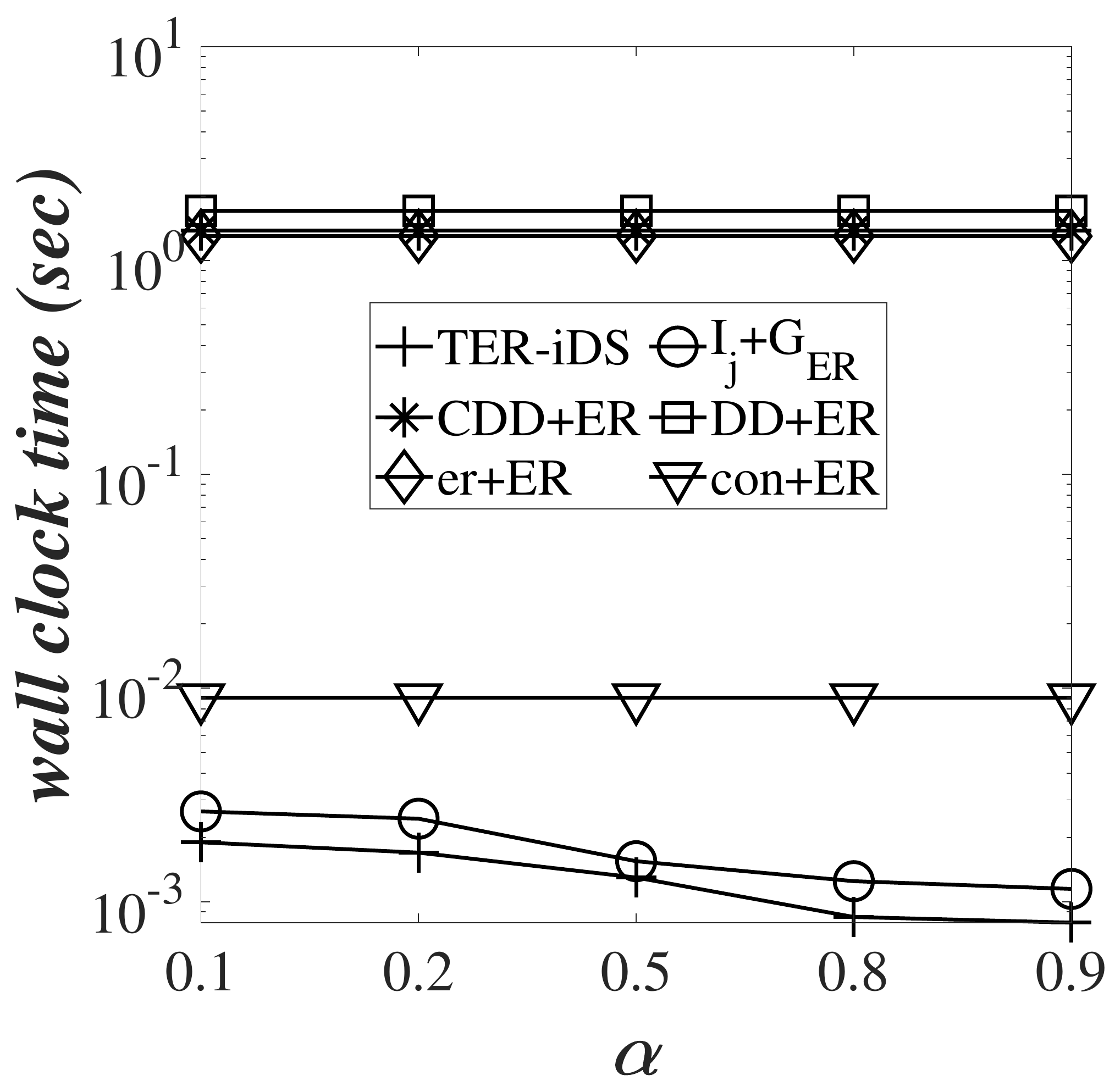}}
\label{subfig:alpha_citations}
}
\subfigure[][$Anime$]{
\scalebox{0.17}[0.15]{\includegraphics{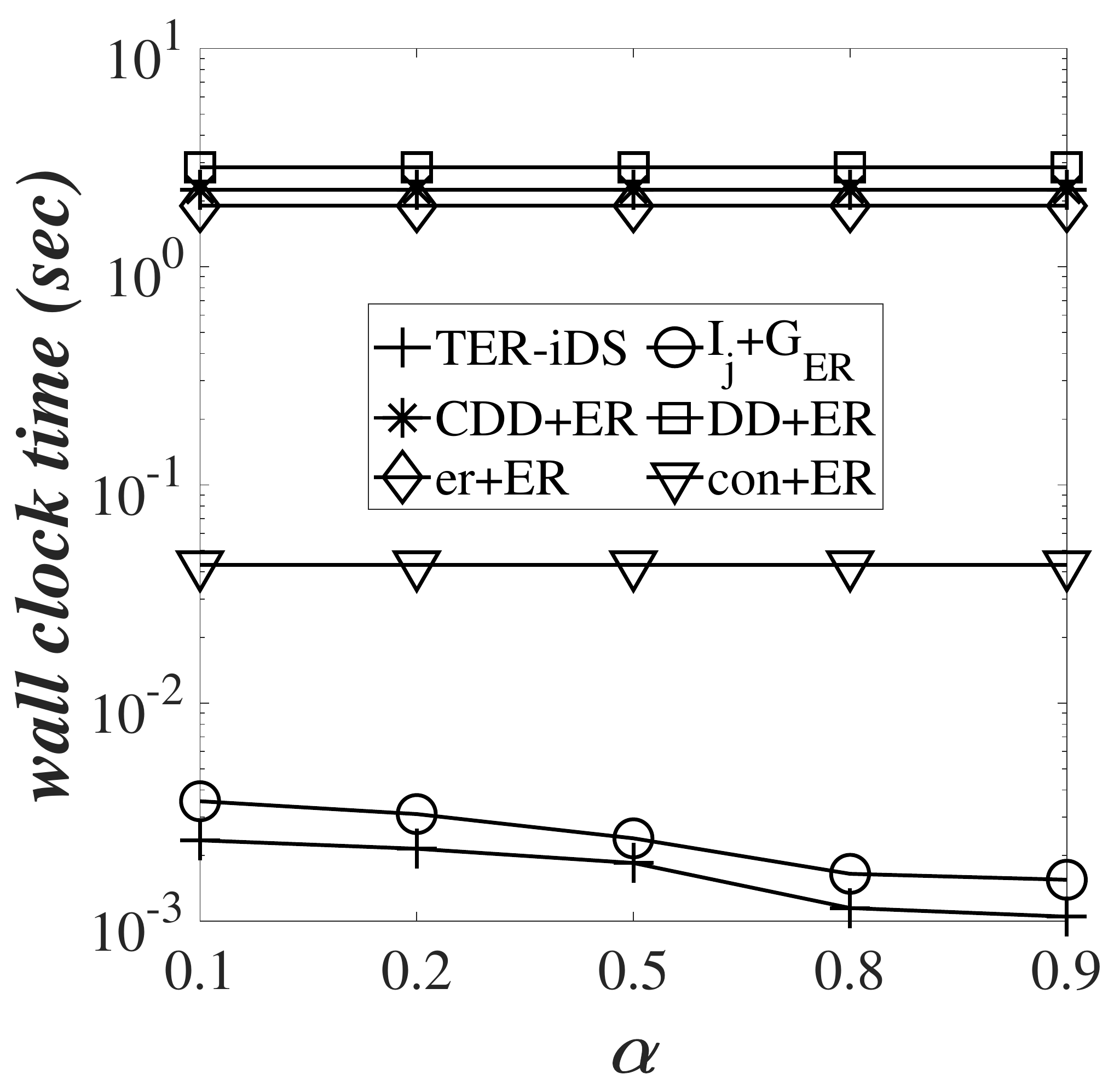}}
\label{subfig:alpha_Anime}
}
\subfigure[][$Bikes$]{
\scalebox{0.17}[0.15]{\includegraphics{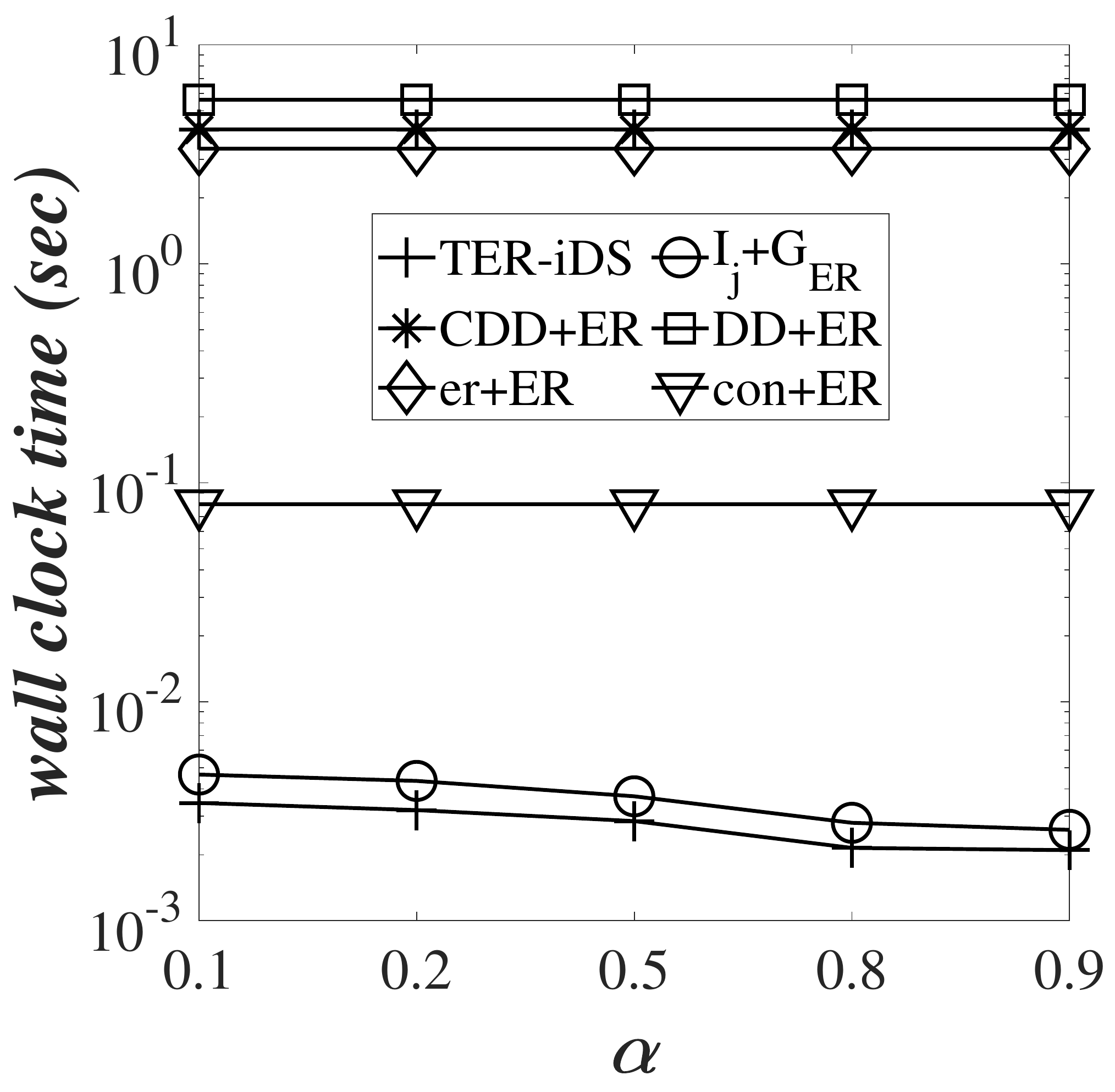}}
\label{subfig:alpha_Bikes}
}
\subfigure[][$EBooks$]{
\scalebox{0.17}[0.15]{\includegraphics{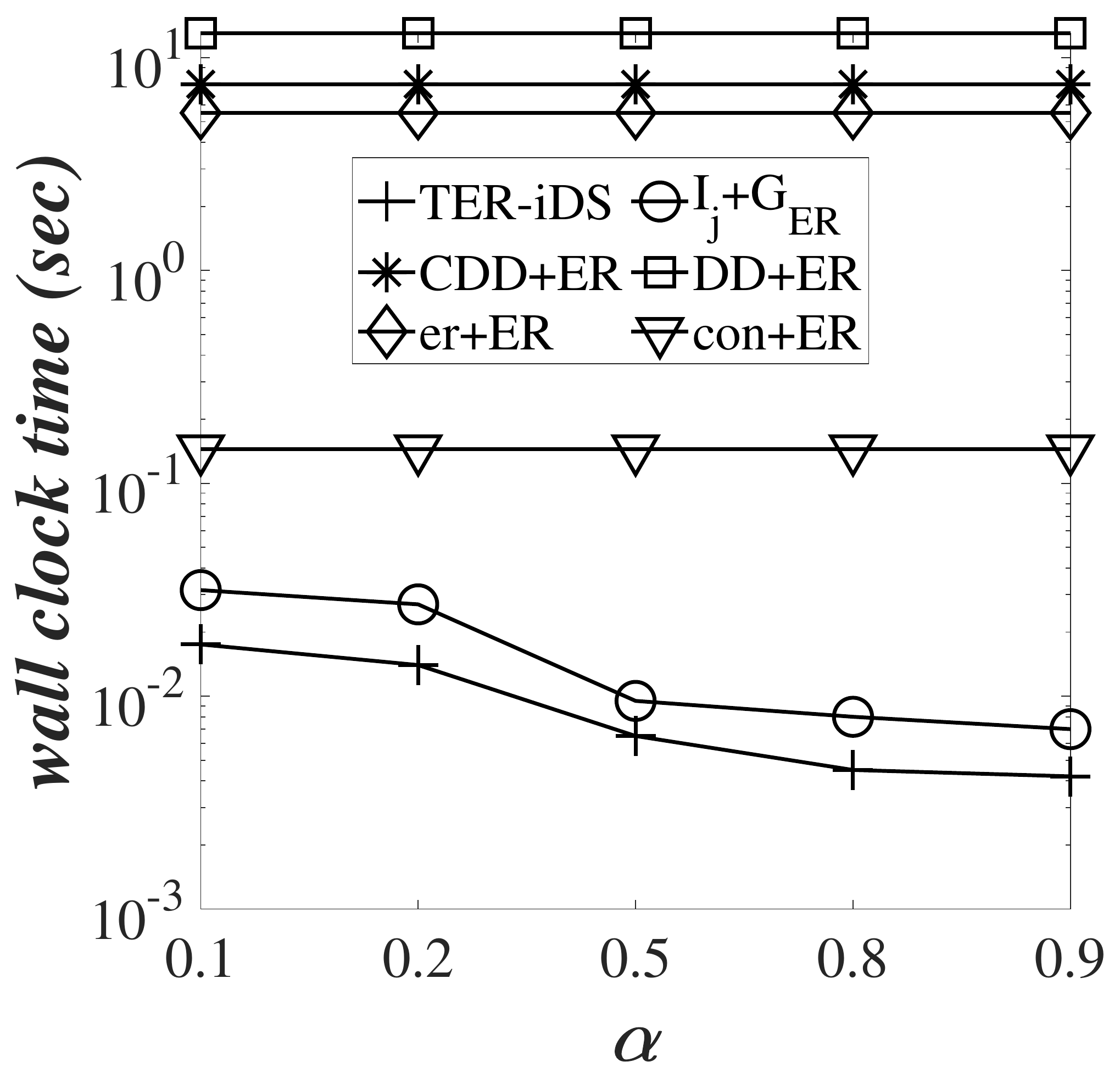}}
\label{subfig:alpha_EBooks}
}
\subfigure[][$Songs$]{\hspace{-2ex}
\scalebox{0.17}[0.15]{\includegraphics{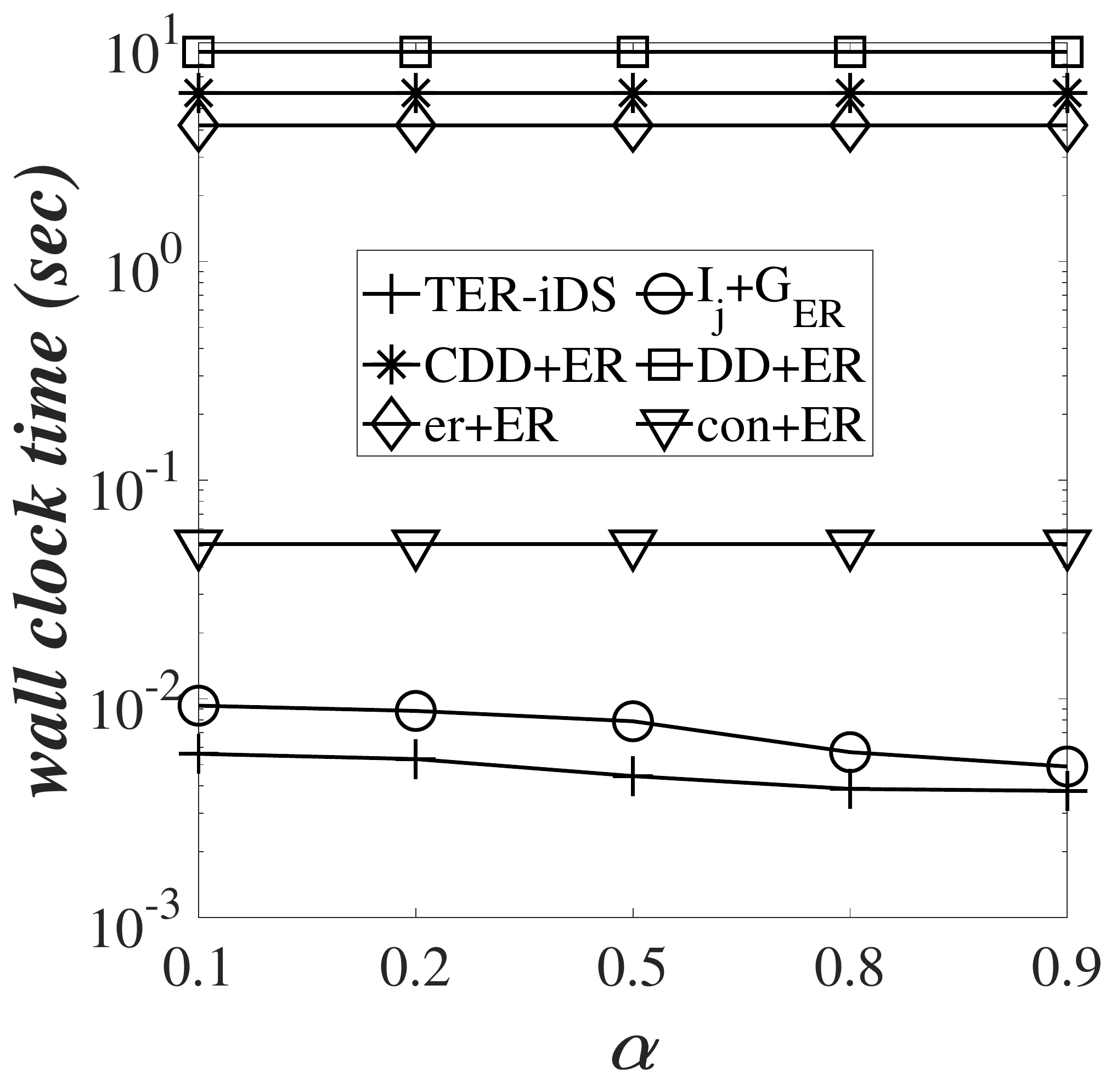}}
\label{subfig:alpha_Songs}
}\vspace{-4ex}
\caption{\small The TER-iDS efficiency vs. probabilistic threshold $\alpha$.} 
\label{exper:alpha} \vspace{-2ex}
\end{figure*} 

\begin{figure*}[ht]
\centering \vspace{-1ex}
\subfigure[][{\small $Citations$}]{\hspace{-2ex}                  
\scalebox{0.17}[0.15]{\includegraphics{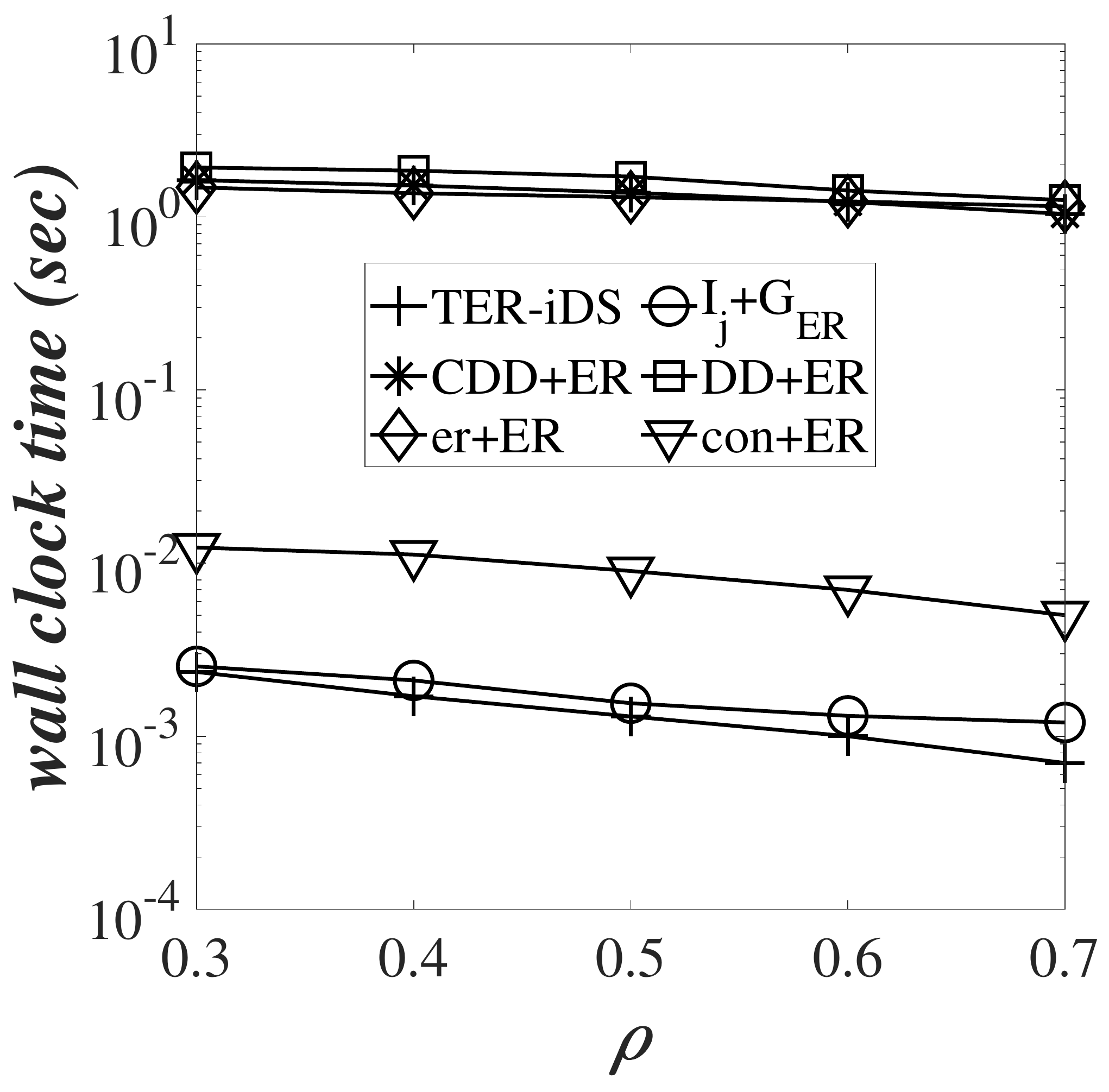}}
\label{subfig:rho_citations}
}
\subfigure[][$Anime$]{
\scalebox{0.17}[0.15]{\includegraphics{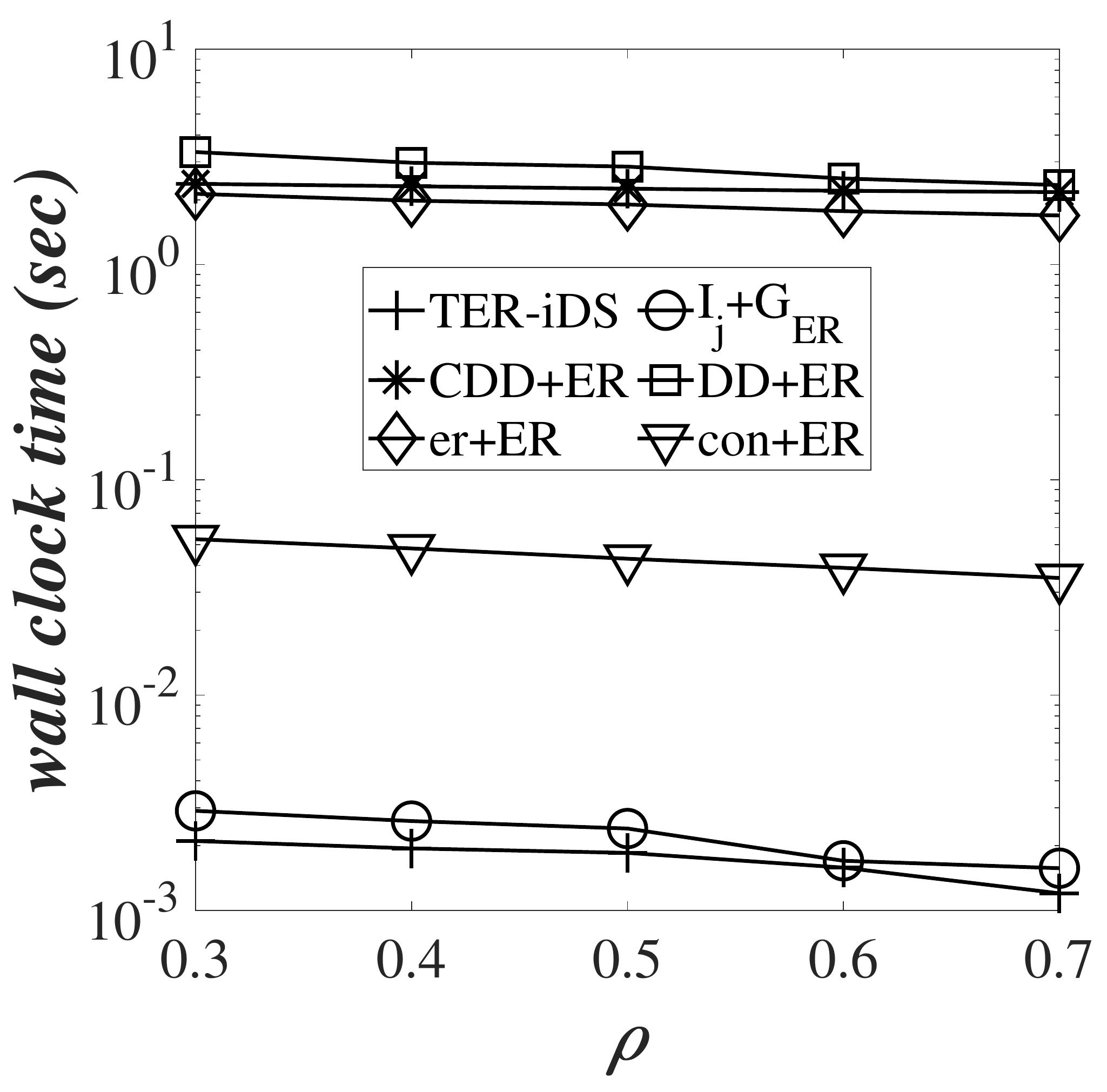}}
\label{subfig:rho_Anime}
}
\subfigure[][$Bikes$]{
\scalebox{0.17}[0.15]{\includegraphics{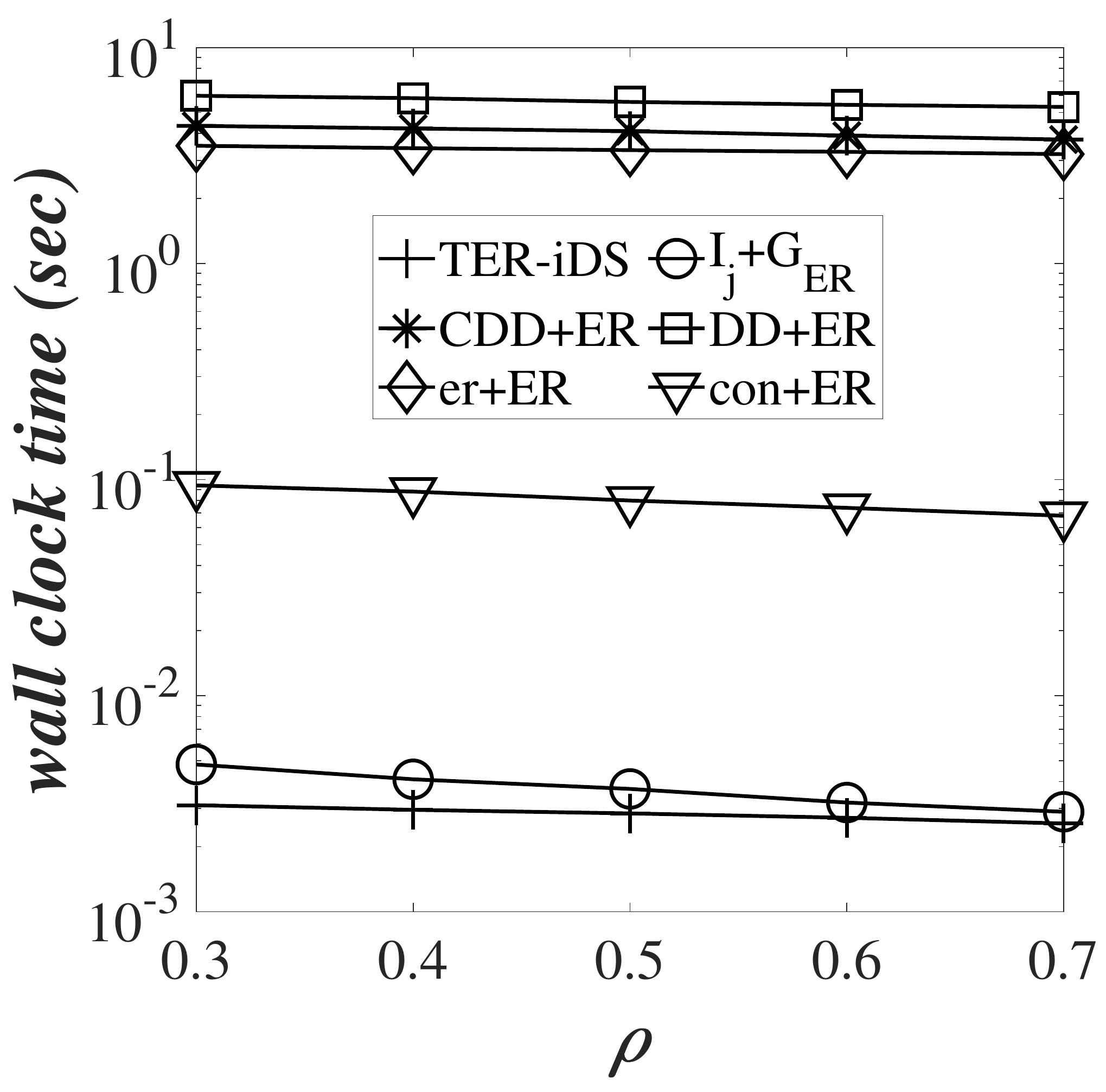}}
\label{subfig:rho_Bikes}
}
\subfigure[][$EBooks$]{
\scalebox{0.17}[0.15]{\includegraphics{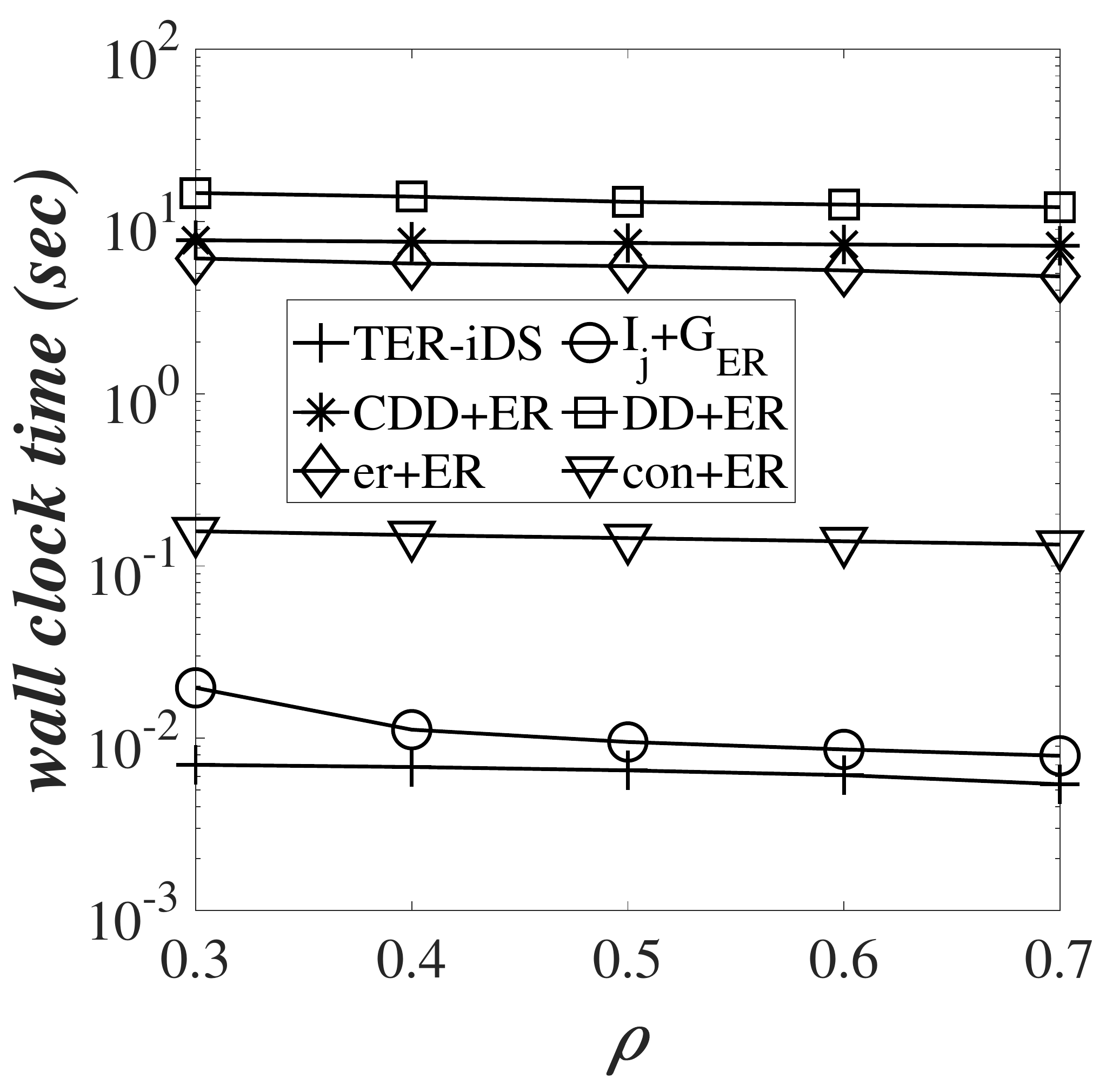}}
\label{subfig:rho_EBooks}
}
\subfigure[][$Songs$]{\hspace{-1ex}
\scalebox{0.17}[0.15]{\includegraphics{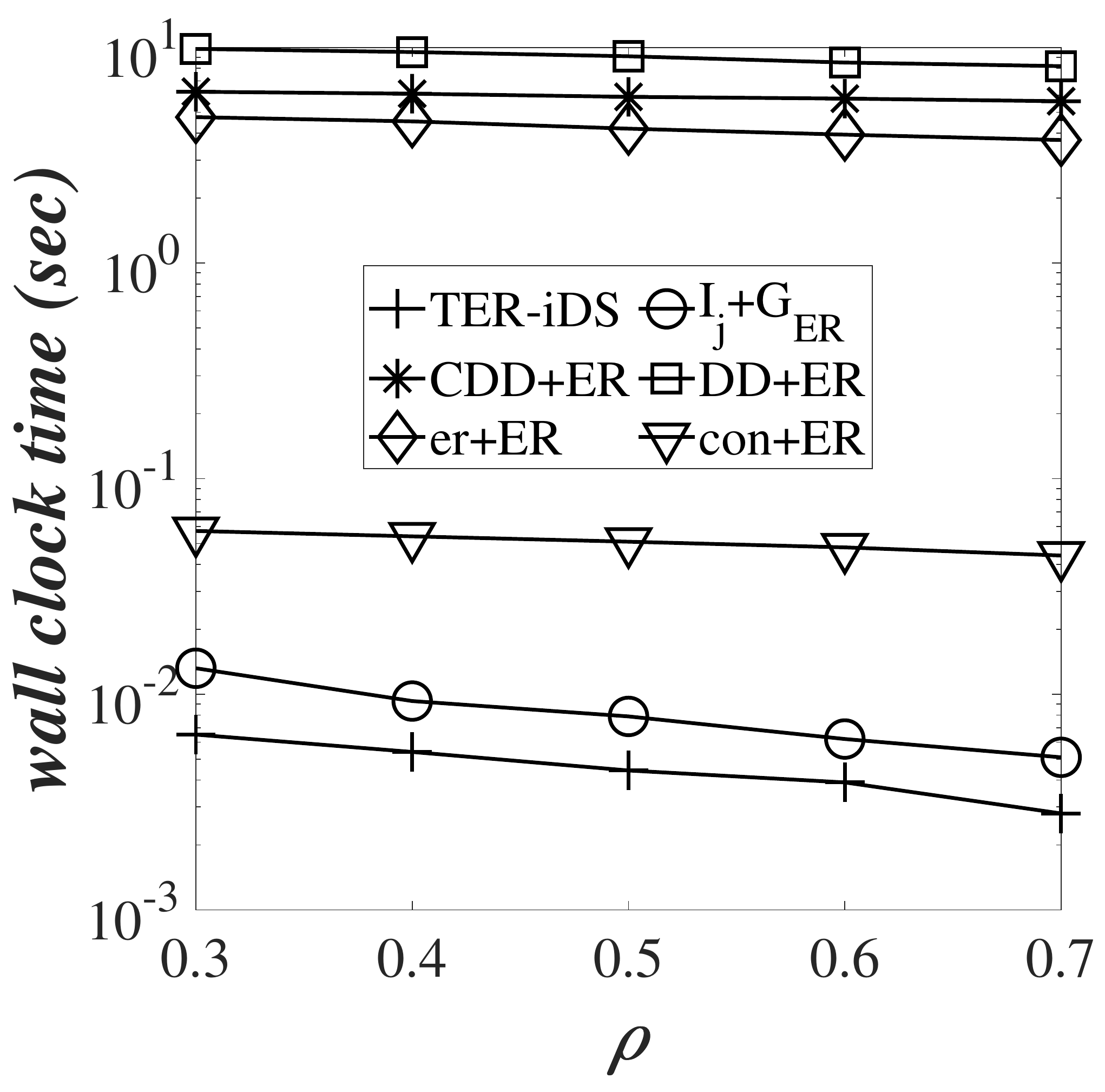}}
\label{subfig:rho_Songs}
}\vspace{-4ex}
\caption{\small The TER-iDS efficiency vs. the ratio, $\rho$, of similarity threshold $\gamma$ w.r.t. dimensionality $d$.} 
\label{exper:rho} \vspace{-2ex}
\end{figure*} 

\begin{figure*}[ht]
\centering \vspace{-1ex}
\subfigure[][{\small $Citations$}]{\hspace{-3ex}   

\scalebox{0.165}[0.13]{\includegraphics{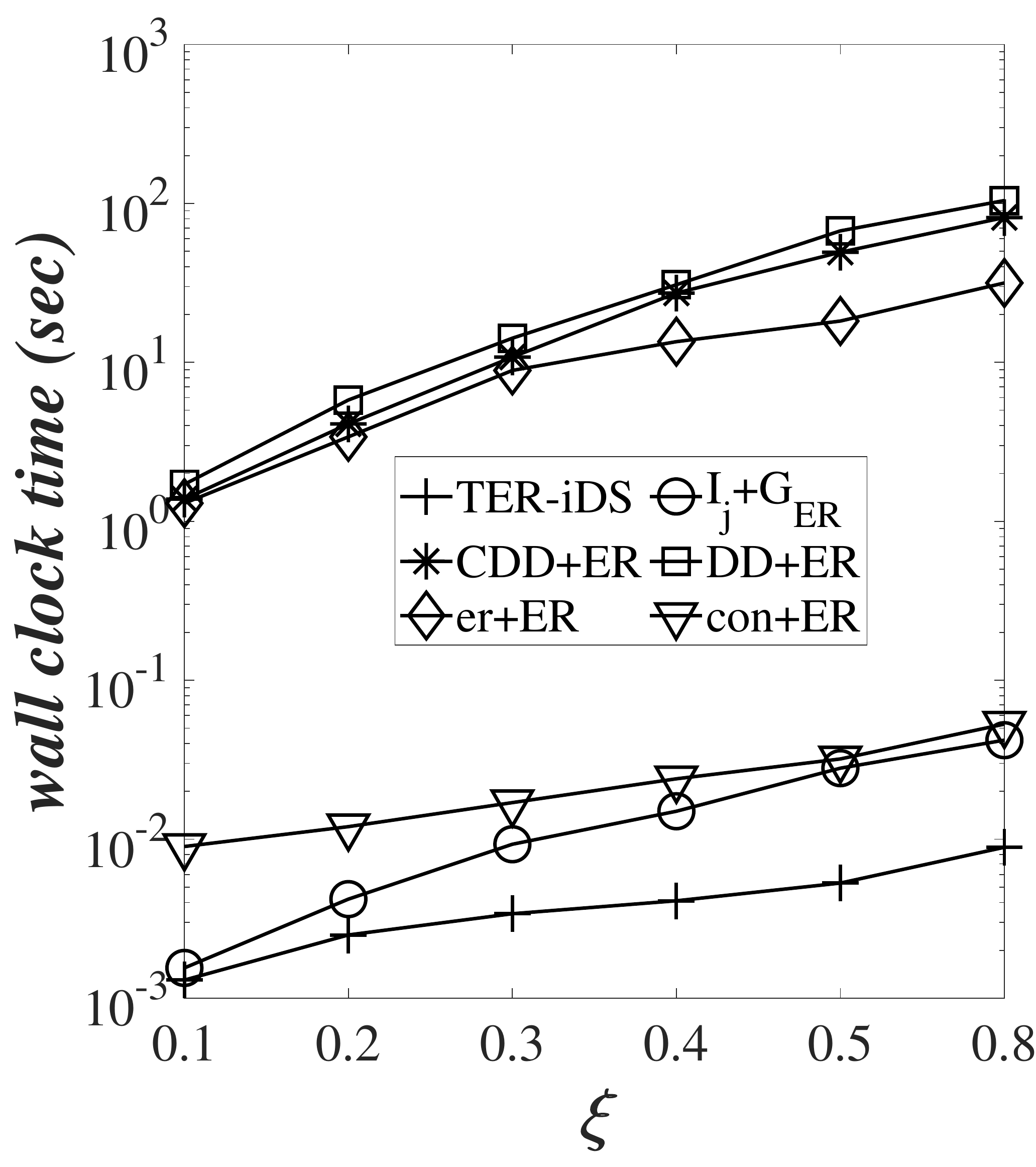}}
\label{subfig:xi_citations}
}
\subfigure[][$Anime$]{
\scalebox{0.165}[0.13]{\includegraphics{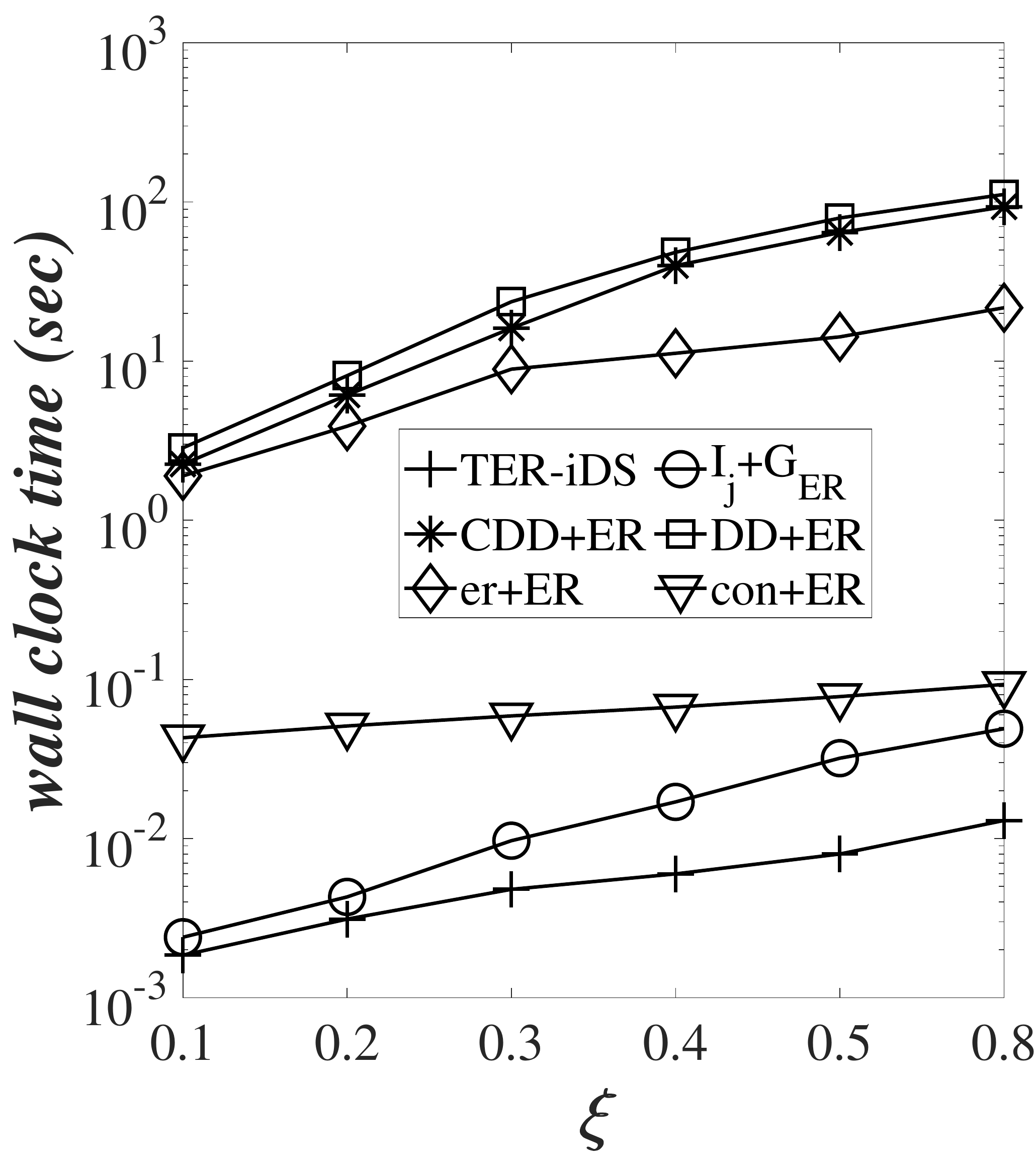}}
\label{subfig:xi_Anime}
}
\subfigure[][$Bikes$]{
\scalebox{0.165}[0.13]{\includegraphics{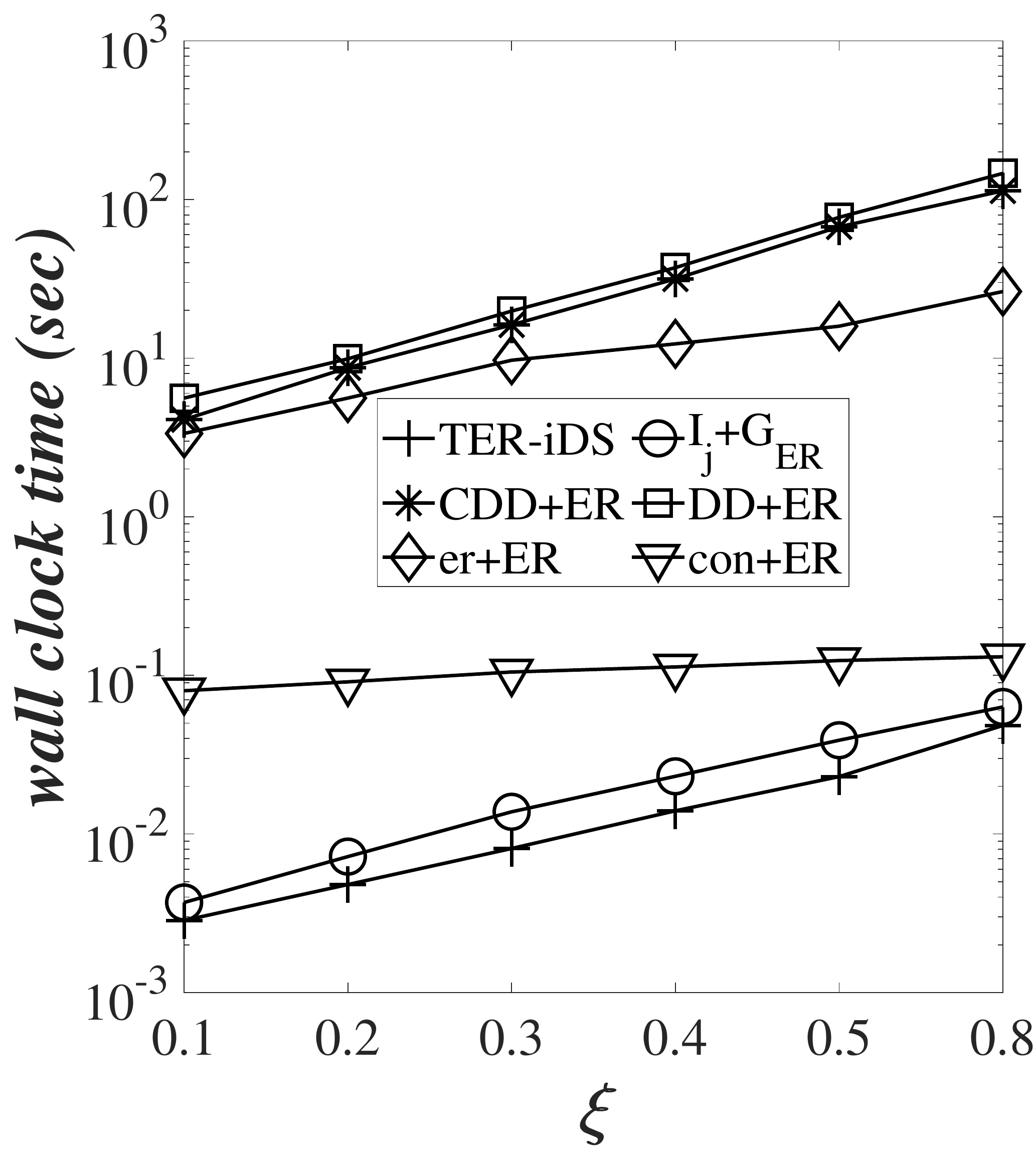}}
\label{subfig:xi_Bikes}
}
\subfigure[][$EBooks$]{
\scalebox{0.165}[0.13]{\includegraphics{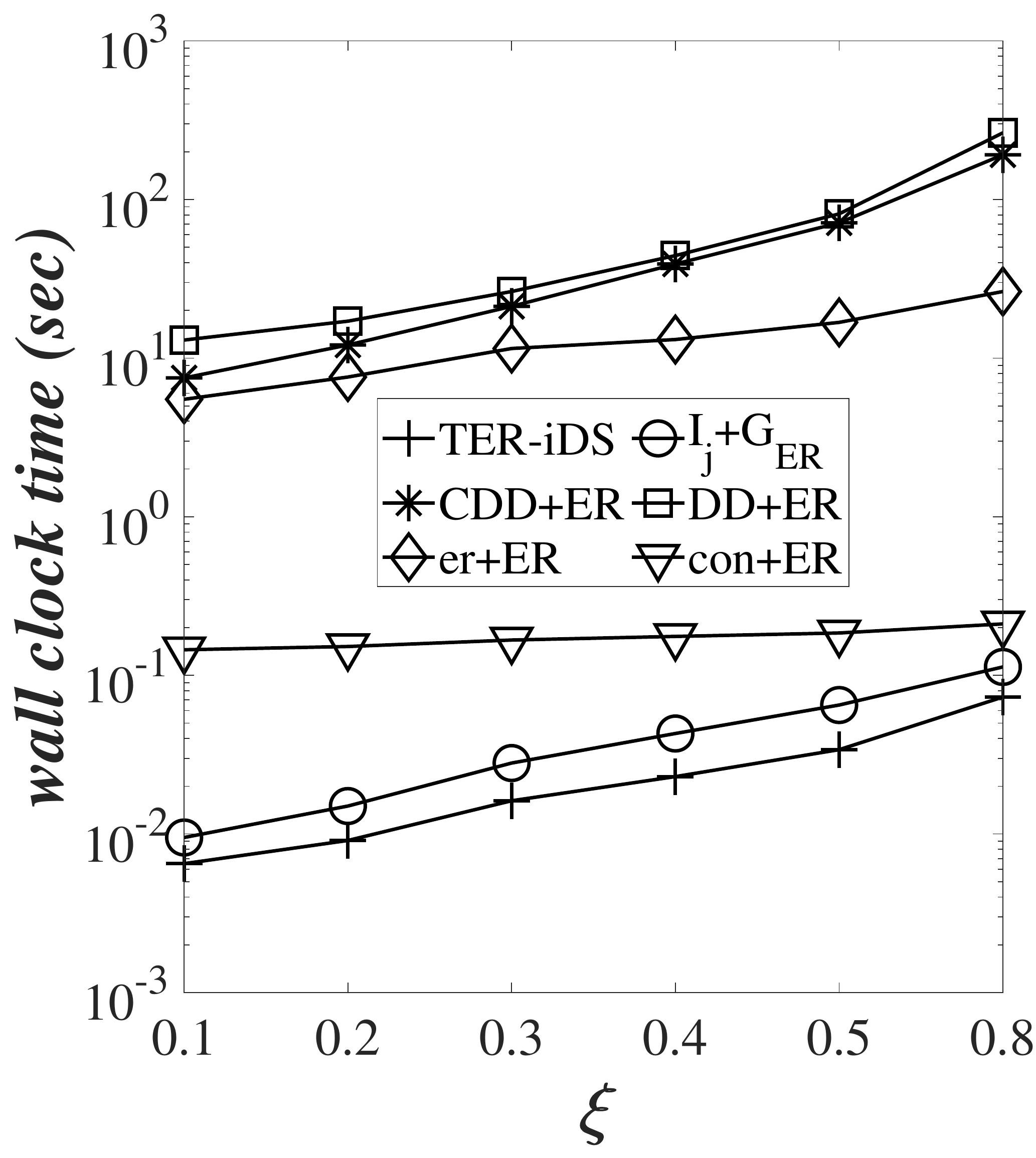}}
\label{subfig:xi_EBooks}
}
\subfigure[][$Songs$]{\hspace{-1ex}
\scalebox{0.165}[0.13]{\includegraphics{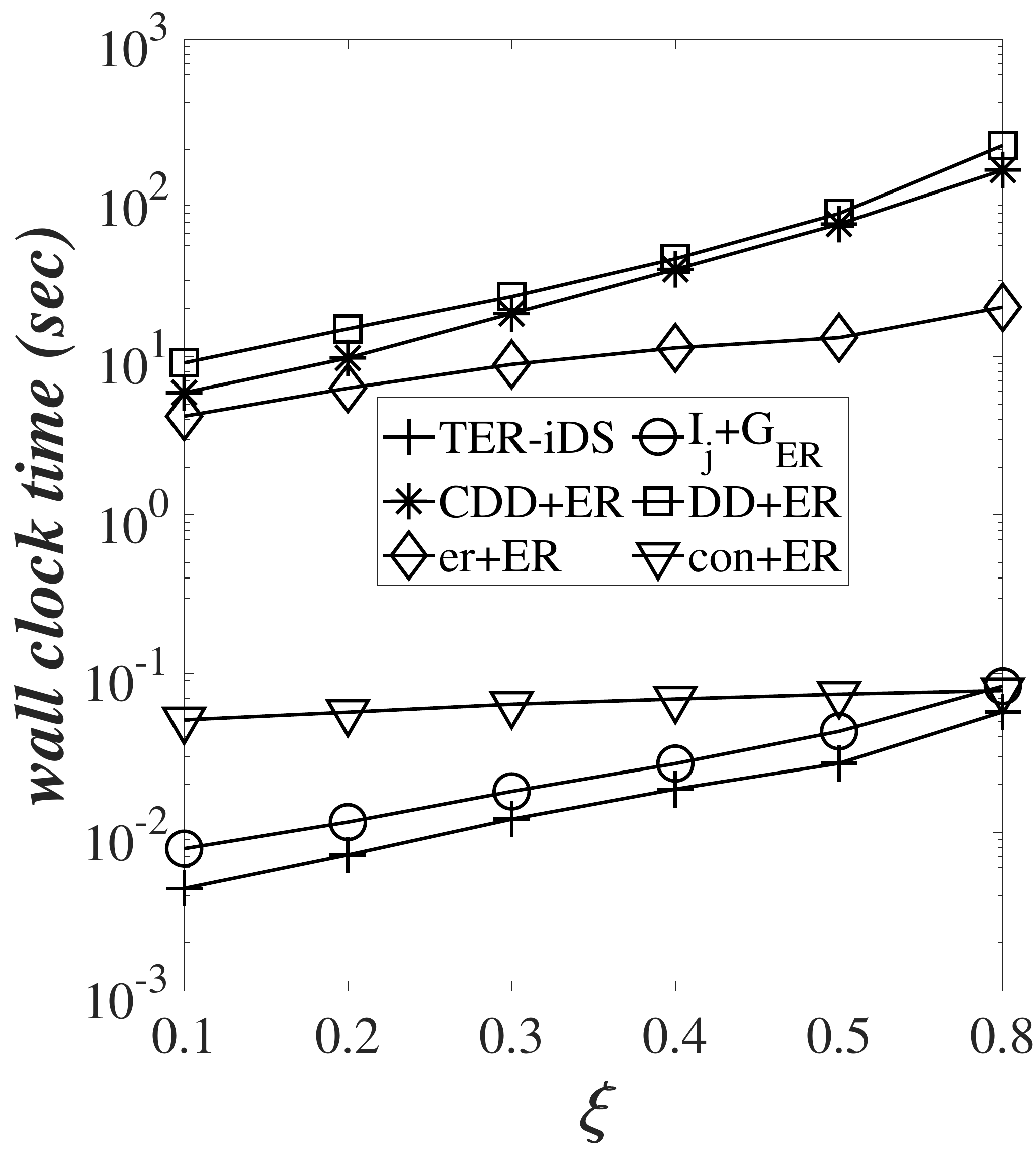}}
\label{subfig:xi_Songs}
}\vspace{-4ex}
\caption{\small The TER-iDS efficiency vs. the missing rate, $\xi$, of incomplete tuples in data streams $iDS_i$.} 
\label{exper:xi} \vspace{-2ex}
\end{figure*} 

\begin{figure*}[ht]
\centering \vspace{-1ex}\hspace{-2ex}
\subfigure[][{\small $Citations$}]{\hspace{-1ex}                  
\scalebox{0.17}[0.16]{\includegraphics{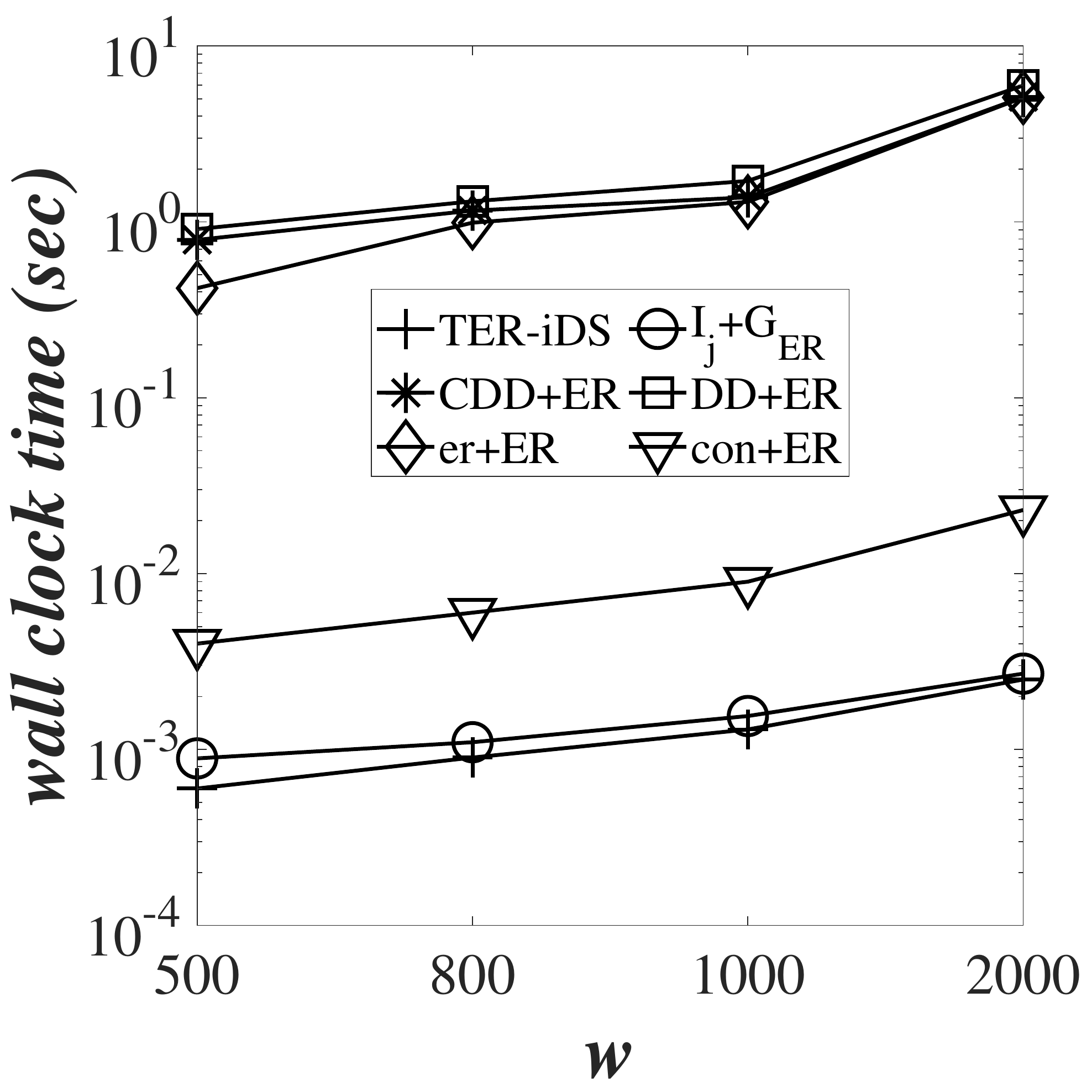}}
\label{subfig:w_citations}
}
\subfigure[][$Anime$]{
\scalebox{0.17}[0.16]{\includegraphics{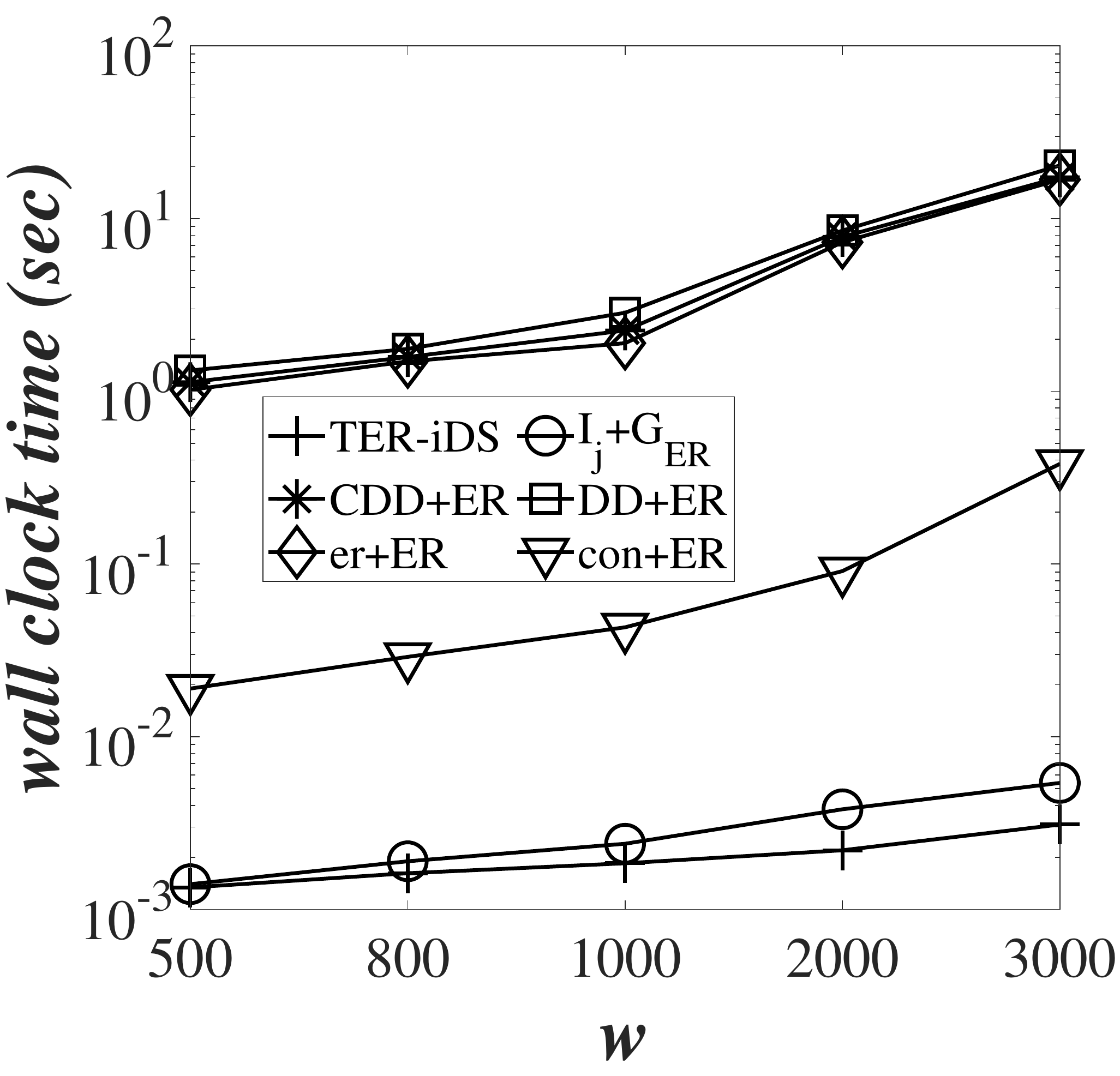}}
\label{subfig:w_Anime}
}
\subfigure[][$Bikes$]{
\scalebox{0.17}[0.16]{\includegraphics{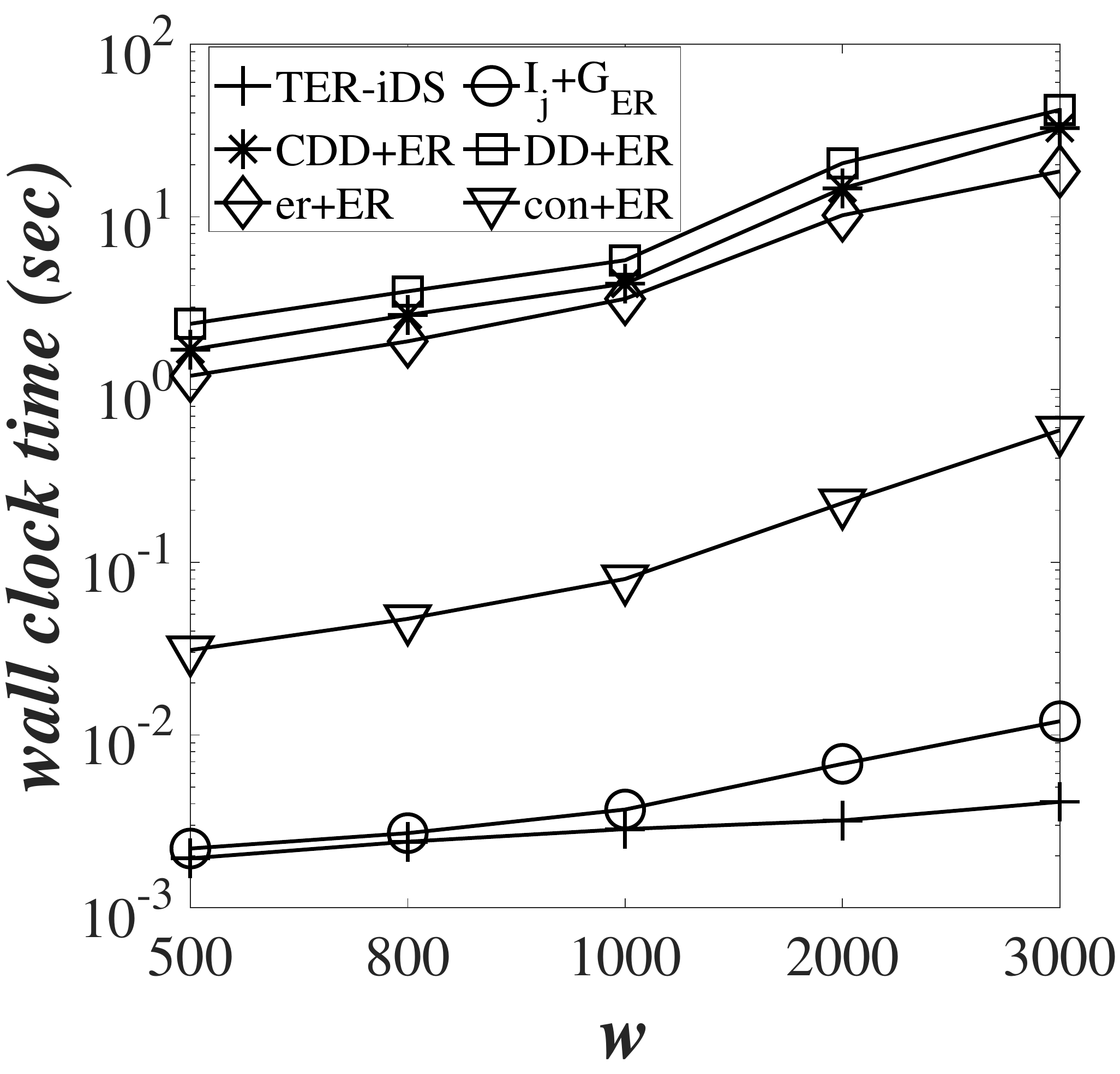}}
\label{subfig:w_Bikes}
}
\subfigure[][$EBooks$]{
\scalebox{0.17}[0.16]{\includegraphics{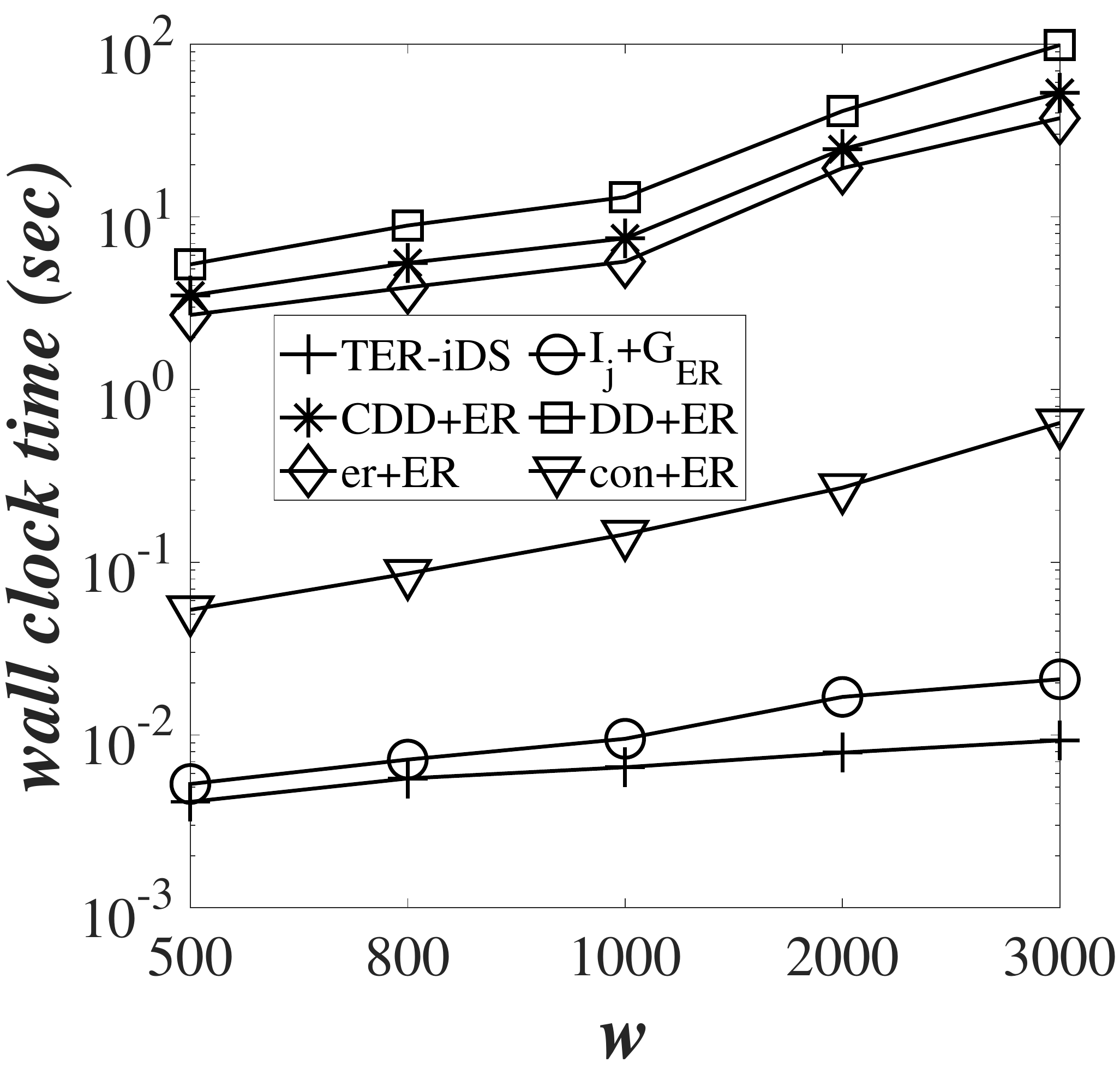}}
\label{subfig:w_EBooks}
}
\subfigure[][$Songs$]{\hspace{-1ex}
\scalebox{0.17}[0.16]{\includegraphics{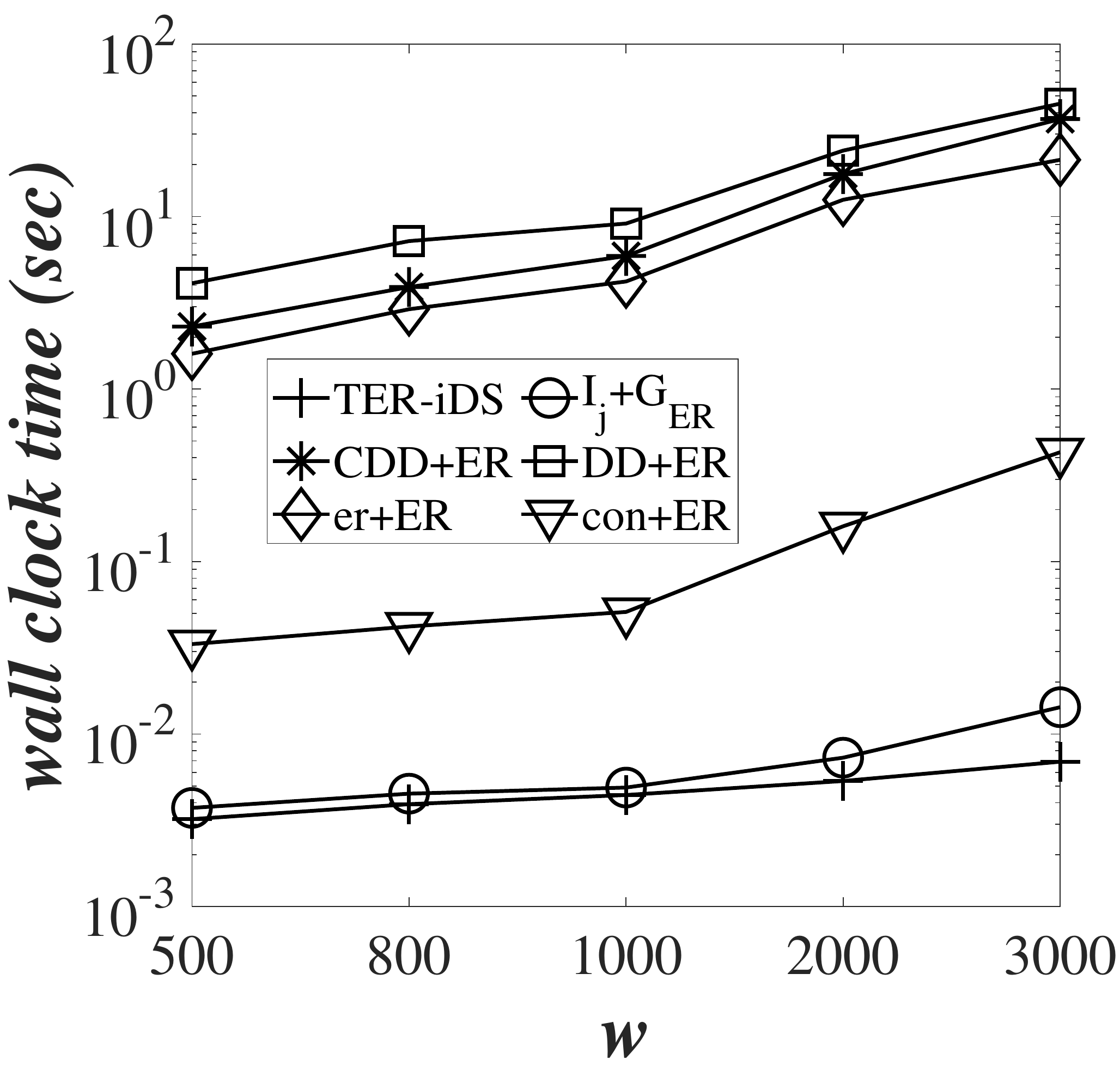}}
\label{subfig:w_Songs}
}
\vspace{-4ex}
\caption{\small The TER-iDS efficiency vs. the size, $w$, of sliding window $W_t$.} 
\label{exper:w} \vspace{-2ex}
\end{figure*}

\vspace{1ex}\noindent {\bf Parameter Settings.} Table \ref{table:exp_parameter_setting} depicts the parameter settings of our experiments, where default parameter values are in bold. In each set of experiments, we vary one parameter, while setting other parameters to their default values. We ran our experiments on a PC with Intel(R) Core(TM) i7-6600U CPU 2.70 GHz and 32 GB memory. All algorithms were implemented in C++. All the code and real data sets are available at: \textit{\url{http://www.cs.kent.edu/~wren/TER-iDS/}}.

\subsection{Evaluation of TER-iDS Pruning Strategies}
\label{subsec:pruning_evaluation}
Figure \ref{fig:pruning_power} shows the pruning power of our proposed pruning strategies (in Section \ref{fig:pruning_power}) over 5 real data sets, where all parameters are set to their default values (as depicted in Table \ref{table:exp_parameter_setting}). Specifically, we apply pruning theorems, in the order of topic keyword pruning, similarity upper bound pruning, probability upper bound pruning, and instance-pair-level pruning. From the graph, we can see that the topic keyword pruning can prune majority of tuple pairs (i.e., 77.51\%$\sim$86.51\%). Then, the similarity upper bound pruning can further prune the remaining unpruned tuple pairs (i.e., 5.59\%$\sim$14.23\%), followed by the probability upper bound pruning (i.e., 2.15\%$\sim$3.64\%) and instance-pair-level pruning (i.e., 1.54\%$\sim$4.35\%). Overall, all the 4 pruning methods can together prune 98.32\%$\sim$99.43\% of tuple pairs, which confirms the effectiveness of our proposed pruning strategies.

\subsection{The TER-iDS Effectiveness and Efficiency}
\label{subsec:exp_results}
\vspace{1ex}\noindent {\bf The TER-iDS effectiveness vs. real data sets.} Figure \ref{subfig:Fscore} compares the accuracy of our $TER\text{-}iDS$ approach with that of three baselines (i.e., $DD+ER$, $er+ER$, and $con+ER$) on 5 real-world data sets, in terms of $F\text{-}score$, where all parameters are set to default values (as depicted in Table \ref{table:exp_parameter_setting}). Note that, we do not report the accuracy of two baselines, $I_j+G_{ER}$ and $CDD+ER$, since they have the same $F\text{-}score$ as our $TER\text{-}iDS$ approach by using the same CDD-based imputation method. From the figure, we can see that our $TER\text{-}iDS$ approach achieves the highest accuracy (i.e., 94.62\%$\sim$97.34\%), and $DD+ER$ has the second highest accuracy, followed by $er+ER$ and $con+ER$. This is because, given a limited size of data repository, CDD and DD have a higher chance to obtain suitable samples for imputation (since they can tolerate differential differences among attribute values) and thus have higher imputation accuracy than \textit{editing rule}. Moreover, compared to DD, CDD has tighter constraints and more accurate imputation accuracy. For $con+ER$, it achieves the worst accuracy, since the constraint-based imputation method \cite{zhang2016sequential} does not adequately consider the semantic association among textual attribute values.

\vspace{1ex}\noindent {\bf The TER-iDS efficiency vs. real data sets.} Figure \ref{subfig:efficiency} illustrates the \textit{wall clock time} of our $TER\text{-}iDS$ approach and 5 baselines, $I_j+G_{ER}$, $CDD+ER$, $DD+ER$, $er+ER$, $con+ER$, over 5 real data sets, where all parameters follow their default values (as depicted in Table \ref{table:exp_parameter_setting}. From the experimental results, our $TER\text{-}iDS$ approach outperforms $CDD+ER$, $DD+ER$, and $er+ER$, by 3-4 orders of magnitude, performs better than $con+ER$ by 1-2 orders of magnitude, and has lower cost than $I_j+G_{ER}$. This confirms the efficiency of our index-join idea (i.e., imputation and ER at the same time) in $TER\text{-}iDS$. Meanwhile, $I_j+G_{ER}$ (applying indexes without join) has the second lowest time cost, which shows the efficiency of our proposed indexes/synopsis over CDD rules, data repository, and incomplete data streams. Moreover, $con+ER$ achieves the third lowest time cost, since it does not need to access the data repository $R$. However, it has the worst ER accuracy (as shown in Figure \ref{subfig:Fscore}). Furthermore, $DD+ER$ has the highest time cost, since DD retrieves more samples for imputation (due to its constraint intervals), and leads to more possible instances of incomplete tuples. Note that, all the approaches surprisingly achieve the highest time costs on $EBooks$ data (instead of $Songs$). After we carefully check the data sets, we find that $EBooks$ has significantly larger token sizes on some attributes (e.g., description) than that of other data sets, which requires a higher time cost for the checking of tuple pairs.

\underline{A break-up cost analysis of TER-iDS}. Figure \ref{fig:break_up_cost} illustrates the break-up cost of our TER-iDS approach over 5 real data sets, which includes online CDD selection cost, online imputation cost (based on selected CDDs), and online ER cost (based on pruning strategies in Section \ref{sec:pruning_strategy}). Note that, our TER-iDS method online obtains suitable CDD rules, imputes missing attribute values, and conducts the ER operator at the same time by joining indexes/synopsis (as discussed in Section \ref{subsec:Algorithm_TER_iDS}). Thus, we obtain the break-up cost in Figure \ref{fig:break_up_cost} by accumulating their costs. From the graph, we can see that the cost of processing the ER operator takes the majority of the TER-iDS cost over data sets except $Songs$, due to the intrinsic quadratic complexity of ER operator. Our TER-iDS approach spends more time over $Songs$ data (with a large size, $300K$, of data repository) for selecting suitable CDDs and retrieving samples for imputation (from data repository) than that of other data sets. Moreover, our TER-iDS approach has the highest time cost for conducting the ER operator over $EBooks$, due to large token set sizes of some attributes (e.g., description) in $EBooks$.

In the sequel, we will test the robustness of our $TER\text{-}iDS$ approach over 5 real data sets, by varying different parameters in Table \ref{table:exp_parameter_setting}. Moreover, we see the similar trend of the break-up cost of our TER-iDS method in following experiments, and thus we will not seperately report the break-up wall clock time.

\vspace{1ex}\noindent {\bf The TER-iDS efficiency vs. probabilistic threshold $\alpha$.} Figure \ref{exper:alpha} shows the effect of the probabilistic threshold $\alpha$ on our $TER\text{-}iDS$ approach and five competitors over 5 real data sets, where $\alpha$ varies from $0.1$ to $0.9$ and other parameters are the default. From the graphs we can see that the time cost of $TER\text{-}iDS$ decreases as $\alpha$ increases. This is reasonable, since fewer candidates of matching pairs need to be checked for larger $\alpha$. Moreover, $TER\text{-}iDS$ has the lowest time cost (i.e., 0.0008 $sec\sim$ 0.0175 $sec$) for all $\alpha$ values, which shows good efficiency of our $TER\text{-}iDS$ approach for different $\alpha$ values. 

\vspace{1ex}\noindent {\bf The TER-iDS efficiency vs. the ratio, $\rho$, of similarity threshold $\gamma$ w.r.t. dimensionality $d$.} Figure \ref{exper:rho} reports the performance of $TER\text{-}iDS$ and 5 baselines, by varying the ratio, $\rho = \gamma/d$, from 0.3 to 0.7, where default values are used for other parameters. From figures, when $\rho$ increases, the time cost decreases smoothly for $TER\text{-}iDS$ and 5 baselines. This is because, for larger $\rho$, there will be fewer candidate ER pairs in data streams. $TER\text{-}iDS$ still has the lowest time cost among all methods (i.e., 0.0007 $sec\sim$ 0.007 $sec$), which confirms the efficiency of our $TER\text{-}iDS$ approach.

\vspace{1ex}\noindent {\bf The TER-iDS efficiency vs. the missing rate, $\xi$, of incomplete tuples in $iDS_i$.} Figure \ref{exper:xi} illustrates the effect of the missing rate, $\xi$, of incomplete tuples in streams on the $TER\text{-}iDS$ performance, compared with 5 baselines, where $\xi=0.1, 0.2, 0.3, 0.4$, $0.5$, and $0.8$, and default values are used for other parameters. From figures, with higher missing rate $\xi$, the time cost increases for all the approaches, since we need to impute more incomplete data. Nevertheless, the wall clock time of our $TER\text{-}iDS$ approach (i.e., 0.0013 $sec\sim$ 0.073 $sec$) outperforms that of baselines, which confirms the $TER\text{-}iDS$ efficiency.

\vspace{1ex}\noindent {\bf The TER-iDS efficiency vs. the size, $w$, of sliding window $W_t$.} Figure \ref{exper:w} demonstrates the performance of $TER\text{-}iDS$ and 5 baselines, where the window size $w$ of incomplete data streams varies from $500$ to $3000$, and other parameters are set to default values. Specifically, for $Citations$ data, we vary the $w$ from $500$ to $2000$, since the size of $Citations$ cannot reach $3000$. From figures, for larger $w$, the time cost increases for all the methods, since there are more tuples in sliding windows to impute and perform ER. Similar to previous experimental results, $TER\text{-}iDS$ has the lowest time cost (i.e., 0.0006 $sec\sim$ 0.0093 $sec$), which indicates the efficiency of our $TER\text{-}iDS$ approach.

We also did experiments on other parameters (e.g., the number, $m$, of missing attributes, and the size ratio, $\eta$, of data repository $R$ w.r.t. data streams $iDS_i$). Please refer to Appendix \ref{sec:more_exp_results} for more experimental results. In summary, our $TER\text{-}iDS$ approach can achieve robust and efficient performance under various parameter settings.

\section{Related Work}
\label{sec:related}
\vspace{1ex}\noindent {\bf Entity Resolution.} Entity resolution (ER) is a key task for data cleaning and data integration. There are many existing works (e.g., \cite{papadakis2014meta,li2015linking,shen2014probabilistic,ebraheem2018distributed,firmani2016online,dragut2015query}) for resolving data records that refer to the same entities. Shen et al. \cite{shen2014probabilistic} proposed a probabilistic model for dealing with entity linking with a heterogeneous information network. Papadakis et al. \cite{papadakis2014meta} introduced meta-blocking that can be combined with any blocking method to further improve the efficiency. Li et al. \cite{li2015linking} built an up-to-date history for real-world entities by linking temporal records from different sources. Dragut et al. \cite{dragut2015query} proposed a general framework for online record linkage over static Web databases. Firmani et al. \cite{firmani2016online} leveraged crowdsourcing platforms to improve the accuracy of online ER tasks over static data. Ebraheem et al. \cite{ebraheem2018distributed} solved the ER problem based on deep learning techniques. Papadakis et al. \cite{papadakis2019survey} did a comprehensive survey for existing ER techniques. These works usually focused on ER tasks over static and complete data. In contrast, in this paper, we consider online topic-related ER problem over incomplete data streams, which is more challenging to tackle effectively and efficiently. 

\vspace{0.5ex}\noindent {\bf Conditional Differential Dependency.} \textit{Conditional differential dependency} (CDD) \cite{kwashie2015conditional,wang2017discovering} is an extension and refinement of the \textit{differential dependency} (DD) \cite{song2011differential}. CDDs can be applied to all scenarios that DDs are applicable. In particular, DDs can be used for data imputation \cite{song2015enriching}, data cleaning \cite{prokoshyna2015combining}, data repairing \cite{song2014repairing}, and so on. Song et al. \cite{song2014repairing,song2015enriching} used DDs to repair vertex labels in network graphs, or impute missing attributes on static databases. Prokoshyna et al. \cite{prokoshyna2015combining} utilized DDs to clean inconsistent records that violate DD rules. These works were applied to static databases. In contrast, in this paper, we not only consider the data imputation over data streams via CDDs, but also conduct entity resolution at the same time.

\vspace{0.5ex}\noindent {\bf Stream Processing.} Stream processing is a hot yet challenging task, due to limited memory consumption and fast processing speeds. Besides entity resolution, previous works studied various query types over data streams, such as join \cite{das2003approximate}, nearest neighbor query \cite{koudas2004approximate}, top-$k$ query \cite{choudhury2017monitoring}, skyline query \cite{tao2006maintaining}, event detection \cite{zhou2014event}, and so on. These works were designed for handling complete data streams, and thus their proposed techniques cannot be directly adopted to our TER-iDS problem in the scenario of incomplete data streams.

\vspace{0.5ex}\noindent {\bf Incomplete Databases.} In this paper, we consider the \textit{missing at random} (MAR) model \cite{graham2012missing} for incomplete data. Under the MAR model, we can classify the existing imputation methods of incomplete data into categories such as statistical-based \cite{mayfield2010eracer}, rule-based \cite{fan2010towards}, constraint-based \cite{song2015screen,zhang2017time}, and pattern-based \cite{liu2016adaptive} imputation methods. Due to textual property and sparseness of ER data sets, these works may fail to impute incomplete data, when there are only a few or even no samples for imputing missing attributes. To overcome this drawback, the differential dependency (DD) \cite{song2011differential} was proposed for increasing the imputation accuracy of incomplete data. Moreover, the conditional differential dependency (CDD) \cite{kwashie2015conditional,wang2017discovering} was proposed to further refine and improve the imputation accuracy of DD rules. In this paper, based on complete data repositories, we adopt CDDs as our imputation technique for imputing missing attributes. Moreover, 
there are some works on queries over incomplete data streams, such as join, skyline, and top-$k$ operators \cite{ren2019efficient,ren2019skyline,ren2021effective}. However, their works have different query semantics from our TER-iDS problem (i.e., topic-based entity resolution), thus, we cannot directly adopt their methods to solve our TER-iDS problem. Note that, data may be missing systematically, and such an absence may provide additional information \cite{newman2003longitudinal,graham2012missing}, which we would like to leave as our future work.

\section{Conclusions}
\label{sec:conclusion}

In this paper, we formulate and tackle the TER-iDS problem, which performs online imputation and topic-based ER processes among incomplete data streams. In order to effectively and efficiently process the TER-iDS operator, we design effective imputation, pruning, and indexing mechanisms to facilitate the TER-iDS processing, and develop efficient algorithms via index joins to enable online imputation and ER at the same time. We demonstrate through extensive experiments the performance of our proposed TER-iDS approaches over real data sets.

\begin{acks}
Xiang Lian is supported by NSF OAC No. 1739491 and Lian Startup
No. 220981, Kent State University. Specifically, we would like to thank Dr. Shaoxu Song from Tsinghua University for the fruitful discussions on this work. We also thank anonymous reviewers for their useful suggestions.
\end{acks}

\balance

\small
\bgroup
\bibliographystyle{abbrv}
\let\xxx=\bibitem\def\bibitem{\par\vspace{-0mm}\xxx}
\bibliography{TER-iDS.bib}
\egroup

\newpage
\appendix 

\noindent {\bf \LARGE Appendix}


\section{Proofs of Theorems/Lemmas for Pruning Strategies}
\label{sec:all_proofs}

\subsection{Proof of Theorem~\ref{lem:lem1}}
\label{subsec:proof_keyword_pruning}
\begin{proof}
Since $\varpi(r_{i,m}, \mathcal{K})=false$ and $\varpi(r_{j,m'}, \mathcal{K})=false$ hold for any imputed tuple instances $r_{i,m}$ and $r_{j,m'}$, respectively, we have $\chi(\varpi(r_{i,m},\mathcal{K}) \vee \varpi(r_{j,m'},\mathcal{K}))=0$, where function $\chi(\cdot)$ is given by our problem statement in Section \ref{subsec:TER-iDS}.

Based on the equation of the TER-iDS probability (in Inequality~(\ref{eq:eq2})), we can obtain:
\begin{eqnarray}
&&Pr_{TER\text{-}iDS}(r_i, r_j)\notag\\ 
&=& \sum_{\forall r_{i,m}}\sum_{\forall r_{j,m'}} r_{i,m}.p \cdot r_{j,m'}.p \cdot \chi((\varpi(r_{i,m},\mathcal{K}) \vee \varpi(r_{j,m'},\mathcal{K}))\notag\\ 
&& \wedge sim(r_{i,m}, r_{j,m'}) > \gamma)\notag\\ 
&\leq& \sum_{\forall r_{i,m}}\sum_{\forall r_{j,m'}} r_{i,m}.p \cdot r_{j,m'}.p \cdot \chi(\varpi(r_{i,m},\mathcal{K}) \vee \varpi(r_{j,m'},\mathcal{K})) \notag\\
&=& \sum_{\forall r_{i,m}}\sum_{\forall r_{j,m'}} r_{i,m}.p \cdot r_{j,m'}.p \cdot 0\notag\\
&=& 0 \le \alpha, \notag
\end{eqnarray}
which violates the condition of Inequality~(\ref{eq:eq2}).

Thus, the tuple pair $(r_i, r_j)$ cannot be the ER result and can be safely pruned, which completes the proof.
\end{proof}

\subsection{Proof of Theorem \ref{lem:lem2}}
\label{subsec:proof_similarity_pruning}
\begin{proof}
From the theorem assumption that $ub\_sim(r_{i,m}, r_{j,m'})$ $\le \gamma$ holds for all instance pairs $(r_{i,m}, r_{j,m'})$, based on the inequality transition, we have $sim(r_{i,m}, r_{j,m'})\le ub\_sim(r_{i,m}, r_{j,m'})\le \gamma$. Thus, it holds that $\chi(sim(r_{i,m}, r_{j,m'})>\gamma) = false$.

Moreover, we have:
\begin{eqnarray}
&&Pr_{TER\text{-}iDS}(r_i, r_j)\notag\\ 
&=& \sum_{\forall r_{i,m}}\sum_{\forall r_{j,m'}} r_{i,m}.p \cdot r_{j,m'}.p \cdot \chi((\varpi(r_{i,m},\mathcal{K}) \vee \varpi(r_{j,m'},\mathcal{K}))\notag\\ 
&& \wedge sim(r_{i,m}, r_{j,m'}) > \gamma)\notag\\ 
&\leq& \sum_{\forall r_{i,m}}\sum_{\forall r_{j,m'}} r_{i,m}.p \cdot r_{j,m'}.p \cdot \chi(sim(r_{i,m}, r_{j,m'}) > \gamma) \notag\\
&\leq& \sum_{\forall r_{i,m}}\sum_{\forall r_{j,m'}} r_{i,m}.p \cdot r_{j,m'}.p \cdot 0\notag\\
&=& 0 \le \alpha \notag
\end{eqnarray}
Thus, tuples $r_i$ and $r_j$ cannot be matched and can be safely pruned, which completes the proof.
\end{proof}

\subsection{Proof of Lemma~\ref{cor:cor1}}
\label{subsec:proof_lemma_1}
\begin{proof}
From Equation~(\ref{eq:SF}), we have: $$sim(r_i[A_k], r_j[A_k]) = \frac{|T(r_i[A_k])\cap T(r_j[A_k])|}{|T(r_i[A_k])\cup T(r_j[A_k])|}.$$

When $|T^-(r_i^p[A_k])| > |T^+(r_j^p[A_k])|$ holds, we have $|T(r_i[A_k])\cap T(r_j[A_k])| \leq \min\{|T(r_i[A_k])|, |T(r_j[A_k])|\} = |T(r_j[A_k])|\\ \leq |T^+(r_j^p[A_k])|$. Moreover, it holds that: $|T(r_i[A_k])\cup T(r_j[A_k])| \geq \max\{|T(r_i[A_k])|, |T(r_j[A_k])|\} = |T(r_i[A_k])|\geq |T^-(r_i^p[A_k])|$. Therefore, we can obtain $sim(r_i[A_k], r_j[A_k]) \leq \frac{|T^+(r_j^p[A_k])|}{|T^-(r_i^p[A_k])|}$, RHS of which is $ub\_sim(r_i[A_k], r_j[A_k])$.

The proof of the case where $|T^+(r_i^p[A_k])| < |T^-(r_j^p[A_k])|$ holds is symmetric and thus omitted.

For the third case (i.e., none of the two cases above hold), the maximum possible value for Jaccard similarity, $sim(r_i[A_k], r_j[A_k])$, is given by 1, since $|T(r_i[A_k])\cap T(r_j[A_k])| \leq |T(r_i[A_k])\cup T(r_j[A_k])|$ holds. Thus, we have $ub\_sim(r_i[A_k], r_j[A_k]) = 1$.
\end{proof}

\subsection{Proof of Lemma~\ref{cor:cor2}}
\label{subsec:proof_lemma_2}
\begin{proof}
Based on the triangle inequality, we have: $$\vspace{-4ex}sim(r_i, r_j) = d - dist(r_i, r_j) = d - \sum_{k=1}^d dist(r_i[A_k], r_j[A_k])$$ $$\vspace{-3ex}\hspace{-4ex}\leq d - \sum_{k=1}^d |dist(r_i[A_k], piv[A_k]) - dist(r_j[A_k], piv[A_k])|$$
$$\hspace{-35ex}= \hspace{4ex}d - \sum_{k=1}^d |X_k - Y_k|$$

When $lb\_X_k > ub\_Y_k$ holds, we have $|X_k - Y_k| \geq lb\_X_k - ub\_Y_k$, which is exactly $min\_dist(r_i[A_k], r_j[A_k])$. 

Similarly, the case that $lb\_Y_k > ub\_X_k$ holds is symmetric and we have $|X_k - Y_k| \geq lb\_Y_k - ub\_X_k = min\_dist(r_i[A_k], r_j[A_k])$. 

When none of above two cases hold (i.e., the third case), we have $|X_k-Y_k| \geq 0 = min\_dist(r_i[A_k], r_j[A_k])$.

Hence, we can derive that: $sim(r_i, r_j) \leq d - \sum_{k=1}^d |X_k - Y_k| \leq d - \sum_{k=1}^d min\_dist(r_i[A_k], r_j[A_k])$, the RHS of which is exactly a similarity upper bound $ub\_sim(r_i, r_j)$.
\end{proof}
\subsection{Proof of Theorem~\ref{lem:lem3}}
\label{subsec:proof_probability_pruning}
\begin{proof} From the theorem assumption that $UB\_Pr_{TER\text{-}iDS}(r_i, r_j)$ $\le \alpha$ holds, we have: $Pr_{TER\text{-}iDS}(r_i, r_j)$ $\le UB\_Pr_{TER\text{-}iDS}(r_i, r_j)$ $\le \alpha$, which violates Inequality~(\ref{eq:eq2}) in our problem statement in Section \ref{subsec:TER-iDS}. Thus, tuple pair $(r_i, r_j)$ cannot be the TER-iDS result and can be safely pruned, which completes the proof.
\end{proof}

\subsection{Proof of Lemma~\ref{cor:corPZ}}
\label{proof_lemma_3}
\begin{proof}
We derive the probability upper bound, $UB\_Pr_{TER\text{-}iDS}(r_i,$ $r_j)$, below:
\begin{eqnarray}
&&Pr_{TER\text{-}iDS}(r_i, r_j)\notag\\ 
&=& \sum_{\forall r_{i,m}}\sum_{\forall r_{j,m'}} r_{i,m}.p \cdot r_{j,m'}.p \cdot \chi((\varpi(r_{i,m},\mathcal{K}) \vee \varpi(r_{j,m'},\mathcal{K}))\notag\\ 
&& \text{ }\qquad\qquad \wedge sim(r_{i,m}, r_{j,m'}) > \gamma)\notag\\ 
&\leq& \sum_{\forall r_{i,m}}\sum_{\forall r_{j,m'}} r_{i,m}.p\cdot r_{j,m'}.p \cdot \chi(sim(r_{i,m}, r_{j,m'}) >\gamma)\notag\\
&=& Pr\{sim(r_i, r_j\}>\gamma\} = Pr\{dist(r_i, r_j)<d-\gamma\}\notag\\
&=& Pr\{\sum_{k=1}^d dist(r_i[A_k], r_j[A_k])<d-\gamma\}.\notag
\end{eqnarray}
Since the Jaccard distance function $dist(\cdot, \cdot)$ follows the triangle inequality, we can relax the distance $dist(r_i[A_k], r_j[A_k])$ by utilizing the pivot tuple $piv$, that is, $|dist(piv[A_k], r_i[A_k]) - dist(piv[A_k],$ $r_j[A_k])| \leq dist(r_i[A_k], r_j[A_k])$. Thus, we can rewrite the formula above as follows.
\begin{eqnarray}
&&Pr_{TER\text{-}iDS}(r_i, r_j)\notag\\ 
&\leq& Pr\{\sum_{k=1}^d |dist(piv[A_k], r_i[A_k]) - dist(piv[A_k], r_j[A_k])| <d-\gamma\}\notag\\
&=& Pr\{\sum_{k=1}^d |X_k - Y_k| < d - \gamma\}.\notag
\end{eqnarray}
Since we have $X_k - Y_k\leq |X_k - Y_k|$ and $Y_k - X_k\leq |X_k - Y_k|$, it holds that:
\begin{eqnarray}
&&\hspace{-6ex}Pr_{TER\text{-}iDS}(r_i, r_j)\notag\\ 
&\hspace{-8ex}\leq&\hspace{-6ex} Pr\{\max\{\sum_{k=1}^d (X_k - Y_k), \sum_{k=1}^d (Y_k - X_k)\} <d-\gamma\}\notag\\
&\hspace{-8ex}=&\hspace{-6ex} \min\{Pr\{\sum_{k=1}^d (X_k - Y_k) <d-\gamma\}, Pr\{\sum_{k=1}^d (Y_k - X_k) <d-\gamma\}\}\notag\\
&\hspace{-8ex}=&\hspace{-6ex} \min\{Pr\{X - Y <d-\gamma\}, Pr\{Y - X <d-\gamma\}\}\notag\\
&\hspace{-8ex}=&\hspace{-6ex} \min\{1-Pr\{X - Y \geq d-\gamma\}, 1- Pr\{Y - X \geq d-\gamma\}\}\notag\\
&\hspace{-8ex}=&\hspace{-6ex} \min\left\{1-Pr\{X - Y\geq \frac{d-\gamma}{E(X - Y)} \cdot E(X - Y)\},\right.\notag\\
&&\left. 1-Pr\{Y - X\geq \frac{d-\gamma}{E(Y - X)} \cdot E(Y - X)\}\right\}.\notag
\end{eqnarray}
Based on the Paley-Zygmund Inequality \cite{paley1932some}, that is, $Pr(Z>\theta\cdot E(Z))\geq (1-\theta)^2 \cdot \frac{E^2(Z)}{E(Z^2)}$ (for $Z\geq 0$ and $0\leq \theta\leq 1$), we can further relax the inequality above (i.e., $\theta = \frac{d-\gamma}{E(X - Y)}$ or $\frac{d-\gamma}{E(Y - X)}$).
\begin{eqnarray}
&&Pr_{TER\text{-}iDS}(r_i, r_j)\notag\\ 
&\leq& \min\left\{1 - (1 - \frac{d-\gamma}{E(X - Y)})^2 \cdot \frac{E^2(X - Y)}{E((X - Y)^2)}, \right.\notag\\
&&\left. \qquad\qquad 1 - (1 - \frac{d-\gamma}{E(Y - X)})^2 \cdot \frac{E^2(Y - X)}{E((Y - X)^2)}\right\} \notag\\
&&\textit{ \qquad\qquad\qquad\qquad // Paley-Zygmund Inequality \cite{paley1932some}}\notag
\end{eqnarray}

Finally, given an upper bound, $ub\_Z$, of a variable $Z$ ($=X-Y$ or $Y-X$), we have $ub\_Z\geq Z$, where $ub\_Z$ can be given by $(ub\_X - lb\_Y)$ or $(ub\_Y - lb\_X)$, respectively. Then, it holds that: $E(Z^2)\leq ub\_Z \cdot E(Z)$. Thus, we can rewrite the inequality above as:
\begin{eqnarray}
&&Pr_{TER\text{-}iDS}(r_i, r_j)\notag\\ 
&\leq& \min\left\{1-(1 - \frac{d-\gamma}{E(X) - E(Y)})^2 \cdot \frac{(E(X) - E(Y))^2}{E((ub\_X - lb\_Y) \cdot (X-Y))},\right.\notag\\
&&\left. \qquad\qquad 1- (1 - \frac{d-\gamma}{E(Y) - E(X)})^2 \cdot \frac{(E(Y) - E(X))^2}{E((ub\_Y - lb\_X) \cdot (Y-X))}\right\}\notag\\
&=& \min\left\{1-(1 - \frac{d-\gamma}{E(X) - E(Y)})^2 \cdot \frac{E(X) - E(Y)}{ub\_X - lb\_Y},\right.\notag\\
&&\left. \qquad\qquad 1- (1 - \frac{d-\gamma}{E(Y) - E(X)})^2 \cdot \frac{E(Y) - E(X)}{ub\_Y - lb\_X}\right\}\notag\\
&=& UB\_Pr_{TER\text{-}iDS}(r_i, r_j),\notag
\end{eqnarray}

Note that, for the derivation above, we used the Paley-Zygmund Inequality, which is under the condition that either (1) $lb\_X\geq ub\_Y$ (i.e., $X-Y\geq0$) and $0\leq \frac{d-\gamma}{E(X)-E(Y)}$ $\leq 1$ (i.e., $0\leq \theta\leq 1$) hold, or (2) $lb\_Y$ $\geq ub\_X$ (i.e., $Y-X\geq0$) and $0\leq \frac{d-\gamma}{E(Y)-E(X)}$ $\leq 1$ (i.e., $0\leq \theta\leq 1$) hold. If both conditions do not hold, then the probability upper bound, $UB\_Pr_{TER\text{-}iDS}(r_i, r_j)$, is given by 1. Therefore, the probability upper bound is correct in Lemma \ref{cor:corPZ}.
\end{proof}

\subsection{Proof of Theorem~\ref{lem:lem5}}
\label{proof_instance_pruning}
\begin{proof}
Given the set, $S$, of instance pairs we have processed so far (for the calculation of the TER-iDS probability), we can classify all combinations of instance pairs $(r_{i, m}, r_{j, m'})$ (for $(r_i^p, r_j^p)$) into two categories, those in $S$ and those not in $S$. Thus, we can rewrite the TER-iDS probability, $Pr_{TER\text{-}iDS}(r_i, r_j)$, by considering these two portions as follows: 

\begin{eqnarray}
&&Pr_{TER\text{-}iDS}(r_i, r_j)\notag\\ 
&=& \sum_{\forall r_{i,m}}\sum_{\forall r_{j,m'}} r_{i,m}.p \cdot r_{j,m'}.p \cdot \chi((\varpi(r_{i,m},\mathcal{K}) \vee \varpi(r_{j,m'},\mathcal{K}))\notag\\ 
&& \text{ }\qquad\qquad \wedge sim(r_{i,m}, r_{j,m'}) > \gamma)\notag\\ 
&=& \sum_{\forall (r_{i, m}, r_{j, m'}) \in (r_i^p, r_j^p)} Pr(r_{i, m}, r_{j, m'}) \notag\\
&=& \sum_{\forall (r_{i, m}, r_{j, m'}) \in S} Pr(r_{i, m}, r_{j, m'}) + \sum_{\forall (r_{i, m}, r_{j, m'}) \notin S} Pr(r_{i, m}, r_{j, m'}) \notag\\
&\leq& \sum_{\forall (r_{i, m}, r_{j, m'}) \in S} Pr(r_{i, m}, r_{j, m'}) + \sum_{\forall (r_{i, m}, r_{j, m'}) \notin S} r_{i, m}.p\cdot r_{j, m'}.p \notag\\
&=& \sum_{\forall (r_{i, m}, r_{j, m'}) \in S} Pr(r_{i, m}, r_{j, m'}) + (1 - \sum_{\forall (r_{i, m}, r_{j, m'})\in S} r_{i, m}.p \cdot r_{j, m'}.p) \notag
\end{eqnarray}
From the theorem assumption that $\sum_{\forall (r_{i, m}, r_{j, m'}) \in S} Pr(r_{i, m}, r_{j, m'}) + (1 - \sum_{\forall (r_{i, m}, r_{j, m'})\in S} r_{i, m}.p \cdot r_{j, m'}.p)\leq \alpha$, we obtain $Pr_{TER\text{-}iDS}(r_i, r_j)$ $\leq \alpha$. Thus, tuple pair $(r_i, r_j)$ cannot match with each other, and can be safely pruned, which completes the proof.
\end{proof}

\section{The Cost-Model-Based Algorithm for Selecting Pivot Tuples}
\label{sec:cost_model_for_pivots}
For each attribute $A_x$ ($1\leq x \leq d$), we propose a cost-model-based algorithm to select $n_x$ ($n_x\geq 1$) attribute pivots, $att\_piv_a$ ($1\leq a\leq n_x$), from the domain of of attribute $A_x$ in data repository $R$. As a result, we can obtain at most $\prod n_x$ pivot tuples, by integrating attribute pivots on $d$ attributes. We call the attribute pivots $att\_piv_1$ and $att\_piv_a$ (for $a > 2$) as \textit{main} and \textit{auxiliary pivots}, respectively. We also denote parameters, $eMin$ and $cntMax$, as a minimal entropy threshold and maximal allowed number of attribute pivots, respectively. Then, we can compute the number, $n_x$, of attribute pivots for attribute $A_x$, by selecting the least number ($\leq cntMax$) of attribute pivots and having the Shannon entropy (Equation (\ref{eq:entropy})) no smaller than $eMin$.

Below, we illustrate how to select $n_x$ ($1\leq n_x\leq cntMax$) attribute pivots $att\_piv_a$ for each attribute $A_x$. As given in Equation~(\ref{eq:entropy}), larger Shannon entropy indicates better converting quality of the pivot, i.e., evenly distributing the converted attribute values in the converted space. Therefore, for each attribute $A_x$, we select the best (main) attribute pivot $att\_piv_1$ (i.e., $piv_1[A_x]$) among all values $val\in dom(A_x)$ of $R$ with the maximal Shannon entropy, that is, $att\_piv_1 = arg \max\limits_{val}\sum_{\forall val\in dom(A_x)} H(val)$. If $H(att\_piv_1) \ge eMin$, it indicates that the selected main pivot $att\_piv_1$ is good enough for evenly distributing the converted values on attributes $A_x$; otherwise, the selected $att\_piv_1$ itself cannot evenly distribute the converted values of $s[A_x]$ in the converted space on attribute $A_x$. In this case, we will choose more (i.e., $n_x-1$) auxiliary pivots $att\_piv_a$ ($a > 1$) on attribute $A_x$ for a larger Shannon entropy (Equation (\ref{eq:entropy})), which can divide the converted space on attribute $A_x$ into more sub-intervals.

\begin{figure}[t!]
\centering
\subfigure[][{\small wall clock time}]{\hspace{-3ex}                  
\scalebox{0.19}[0.18]{\includegraphics{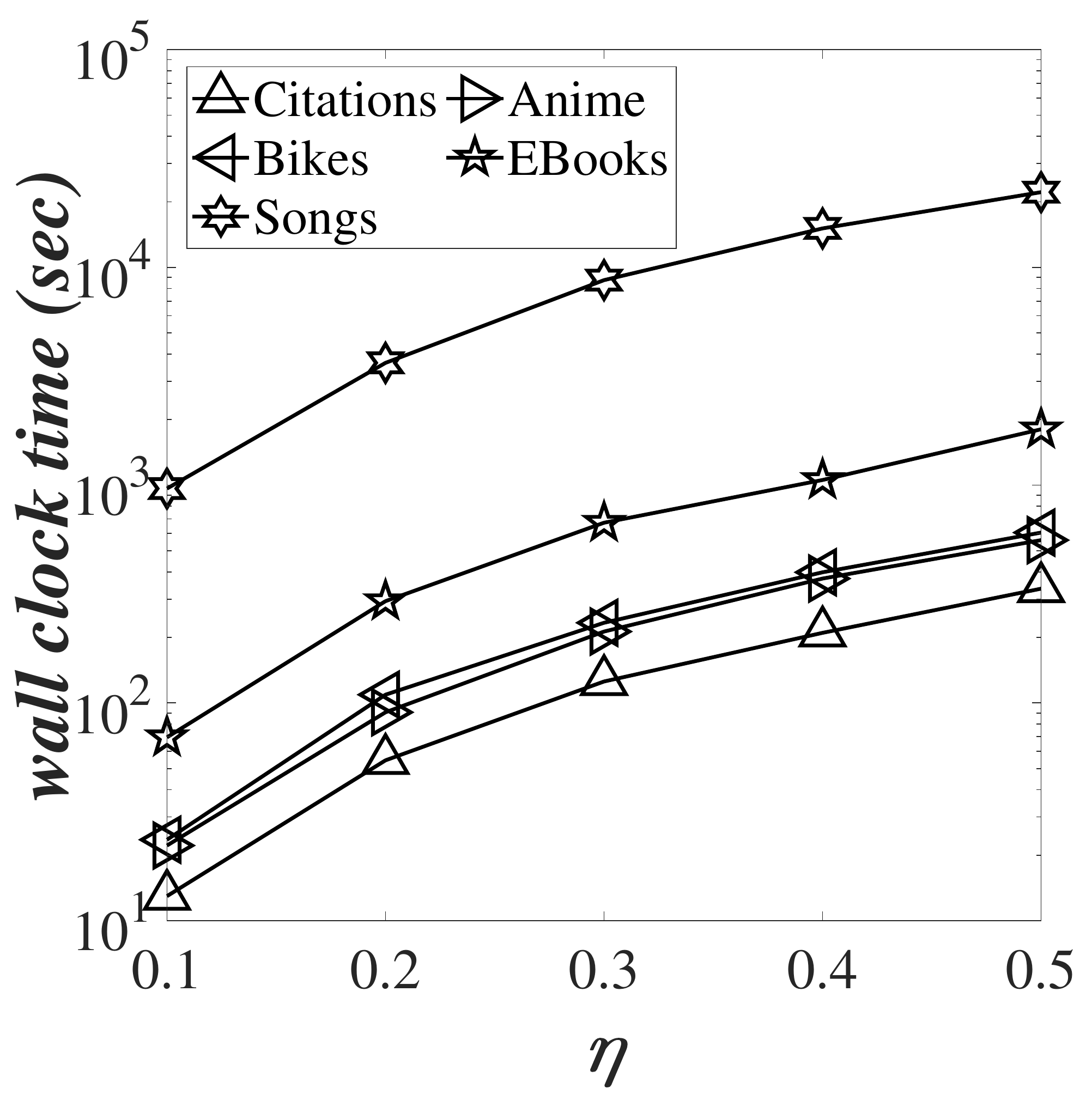}}
\label{subfig:pivot_eta}
}\quad
\subfigure[][{\small wall clock time}]{\hspace{-3ex}                  
\scalebox{0.19}[0.185]{\includegraphics{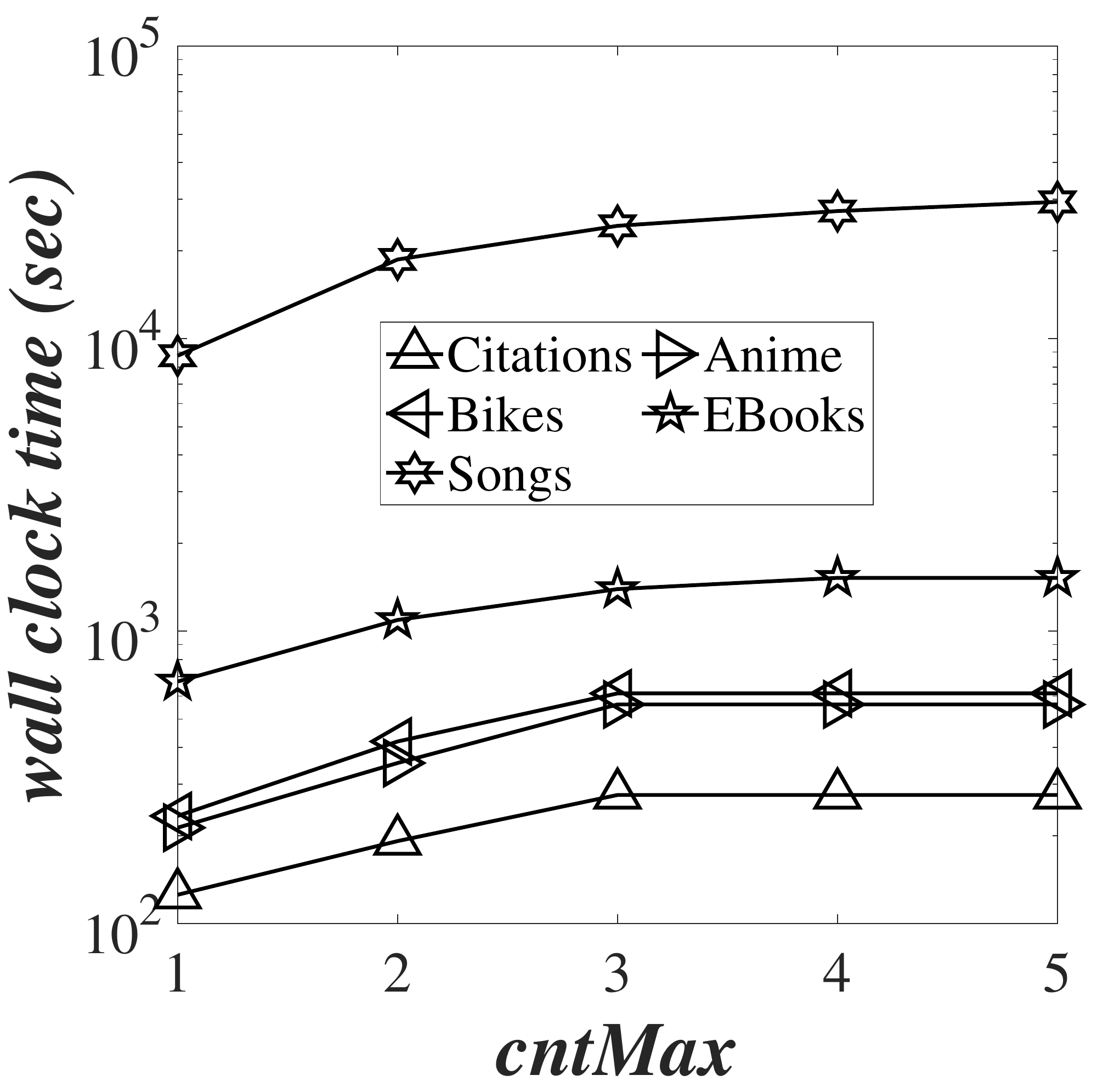}}
\label{subfig:pivot_cnt}
}\vspace{-3ex}
\caption{\small The evaluation of the cost-model-based algorithm.}
\label{fig:cost_model_evaluation}\vspace{0ex}
\end{figure}

\begin{figure}[t!]\vspace{-1ex}
\scalebox{0.2}[0.2]{\hspace{0ex}\includegraphics{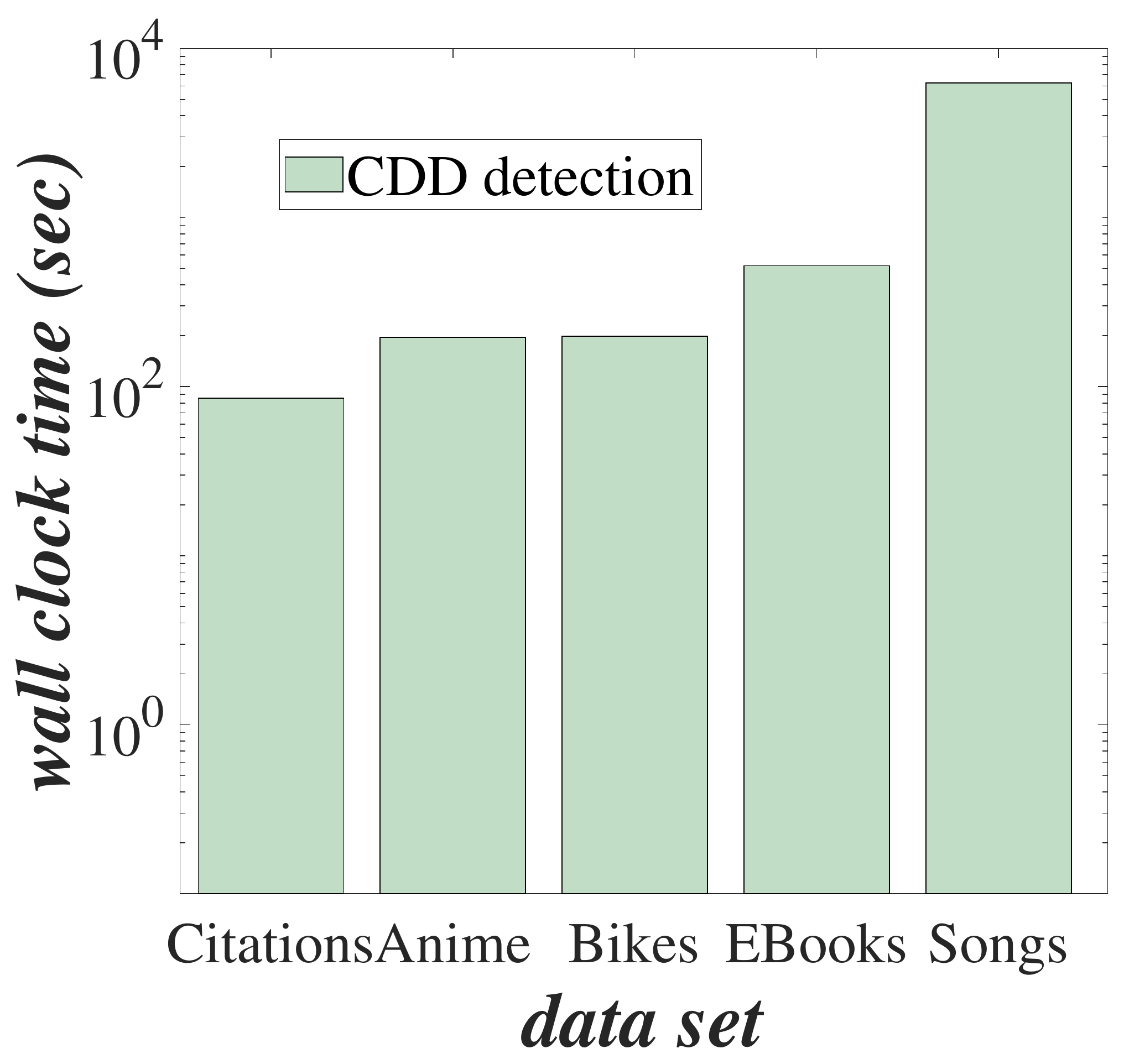}}\vspace{-2ex}
\caption{\small {The time cost of offline CDD detection vs. data sets}.}
\label{fig:CDD_detect_cost}
\end{figure}

\begin{figure*}[ht]
\centering \vspace{-1ex}
\subfigure[][{\small $Citations$}]{\hspace{-3ex}   

\scalebox{0.17}[0.17]{\includegraphics{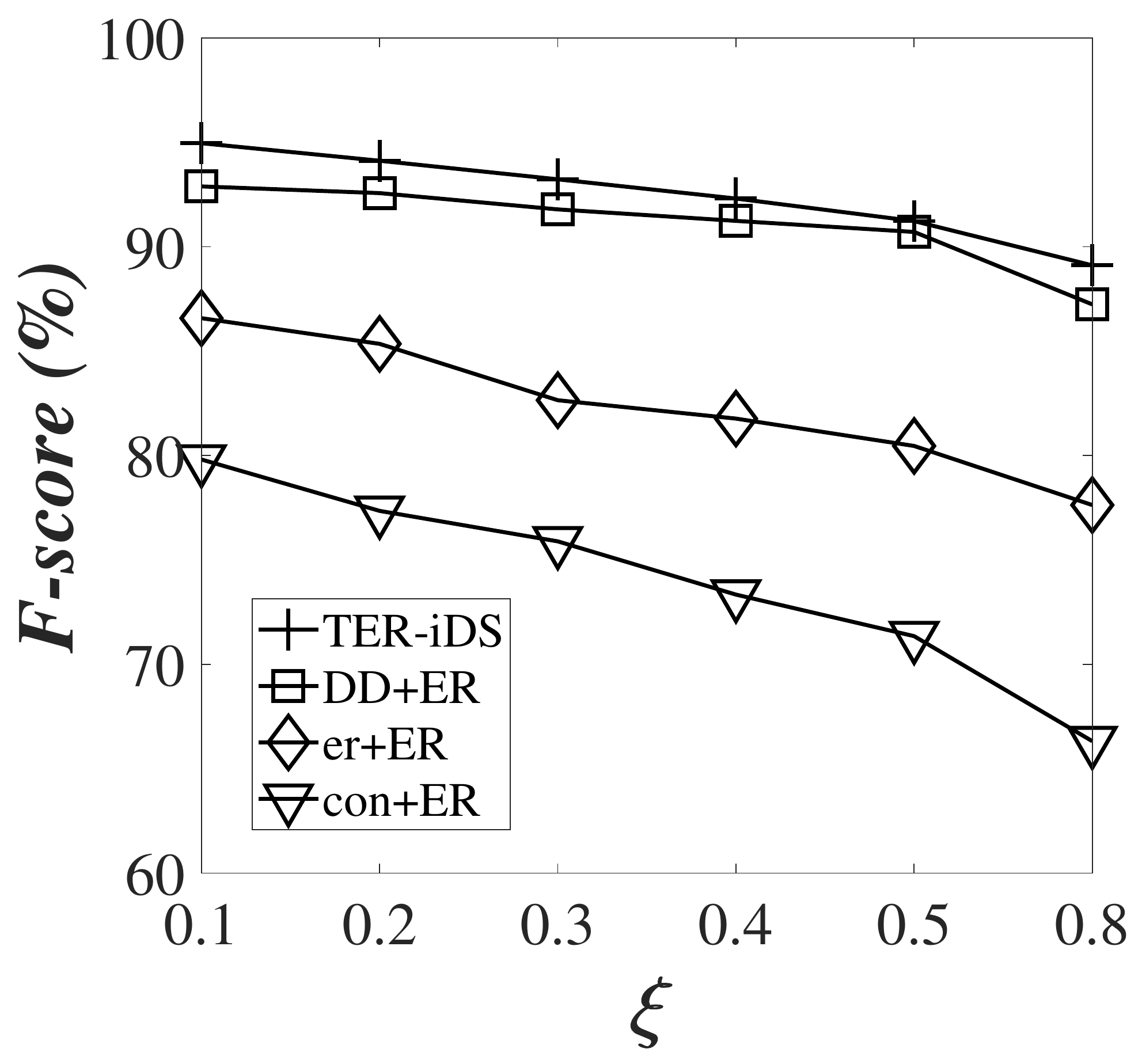}}
\label{subfig:xi_citations_F1}
}
\subfigure[][$Anime$]{
\scalebox{0.17}[0.17]{\includegraphics{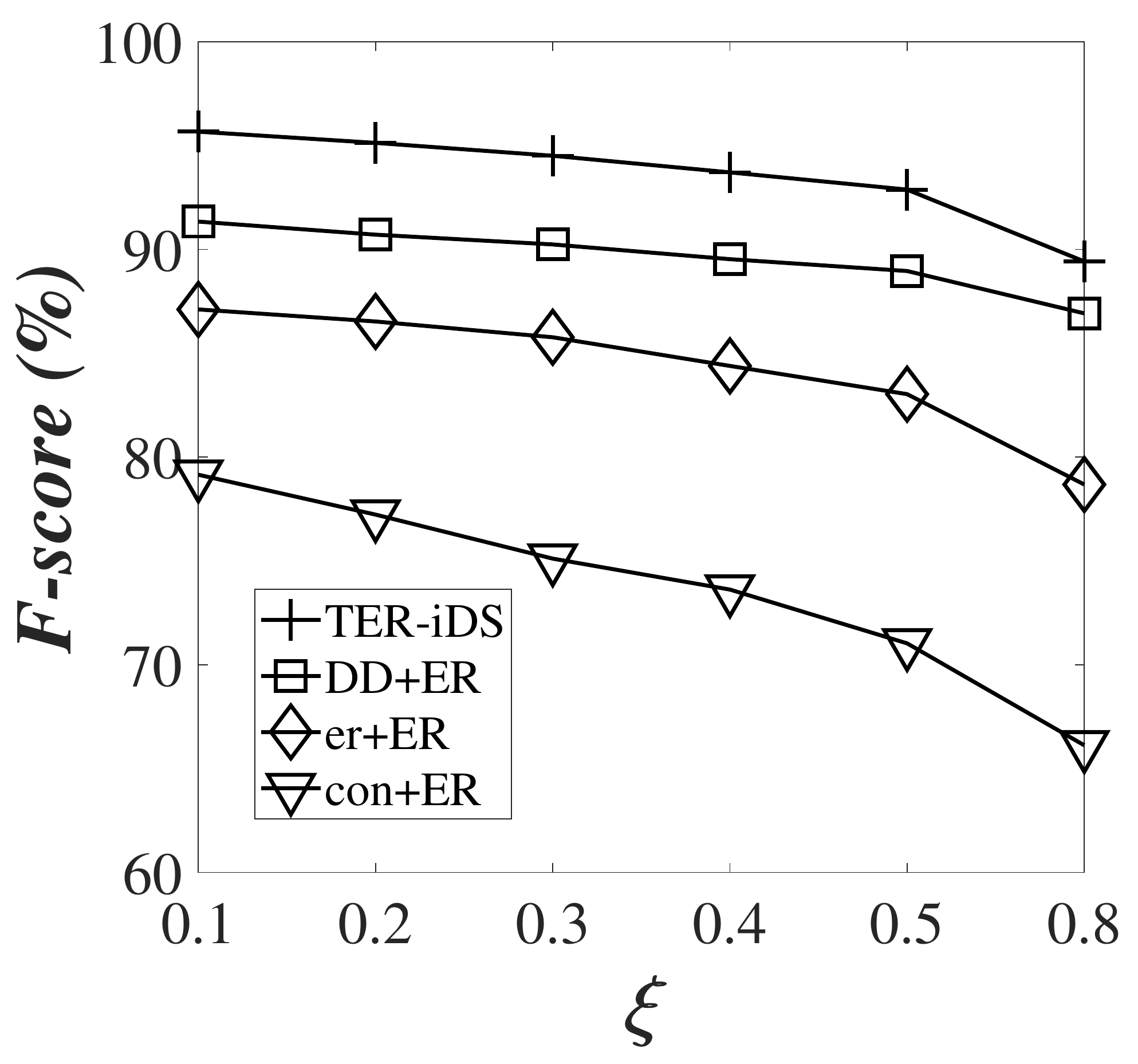}}
\label{subfig:xi_Anime_F1}
}
\subfigure[][$Bikes$]{
\scalebox{0.17}[0.17]{\includegraphics{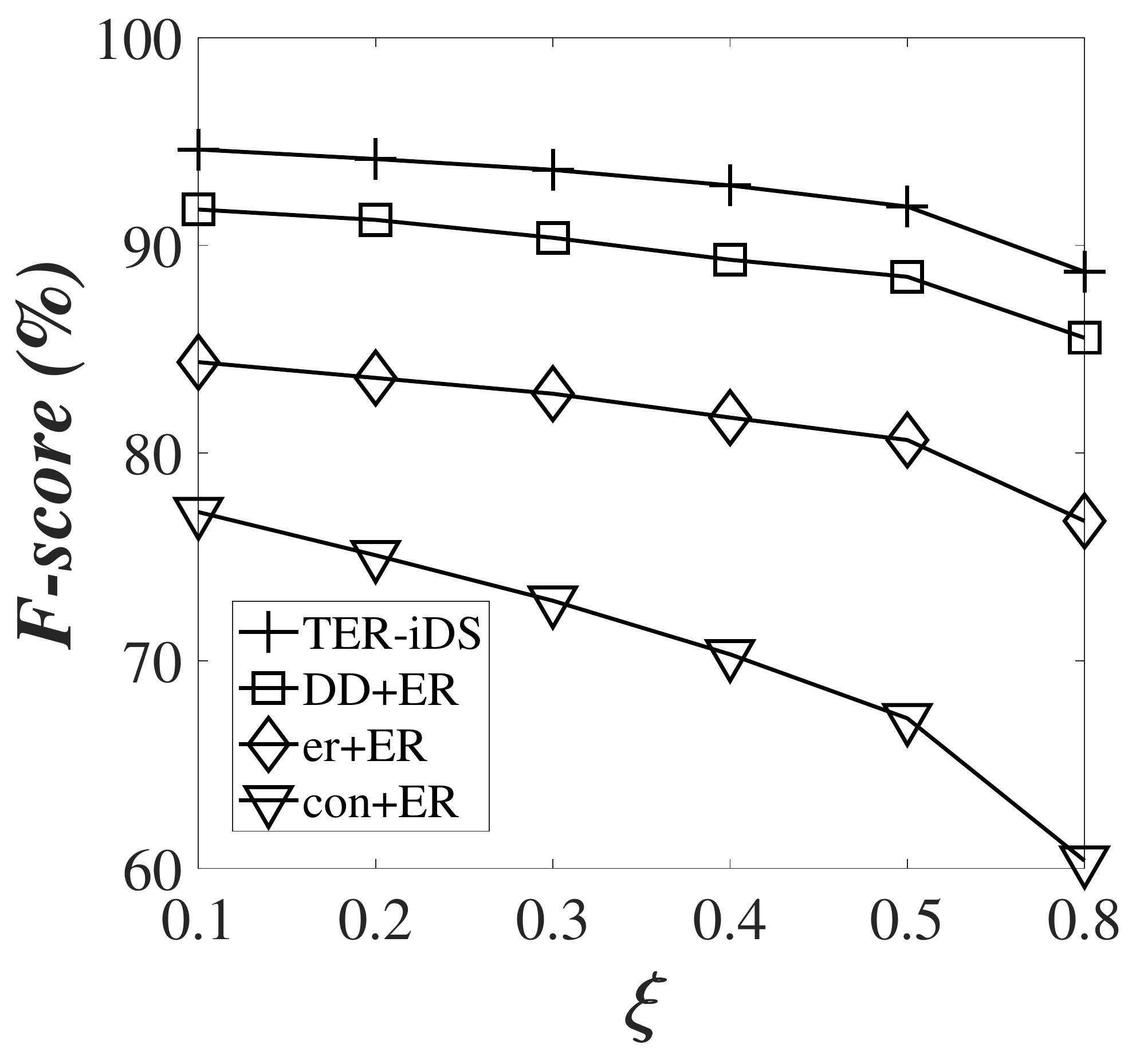}}
\label{subfig:xi_Bikes_F1}
}
\subfigure[][$EBooks$]{
\scalebox{0.17}[0.17]{\includegraphics{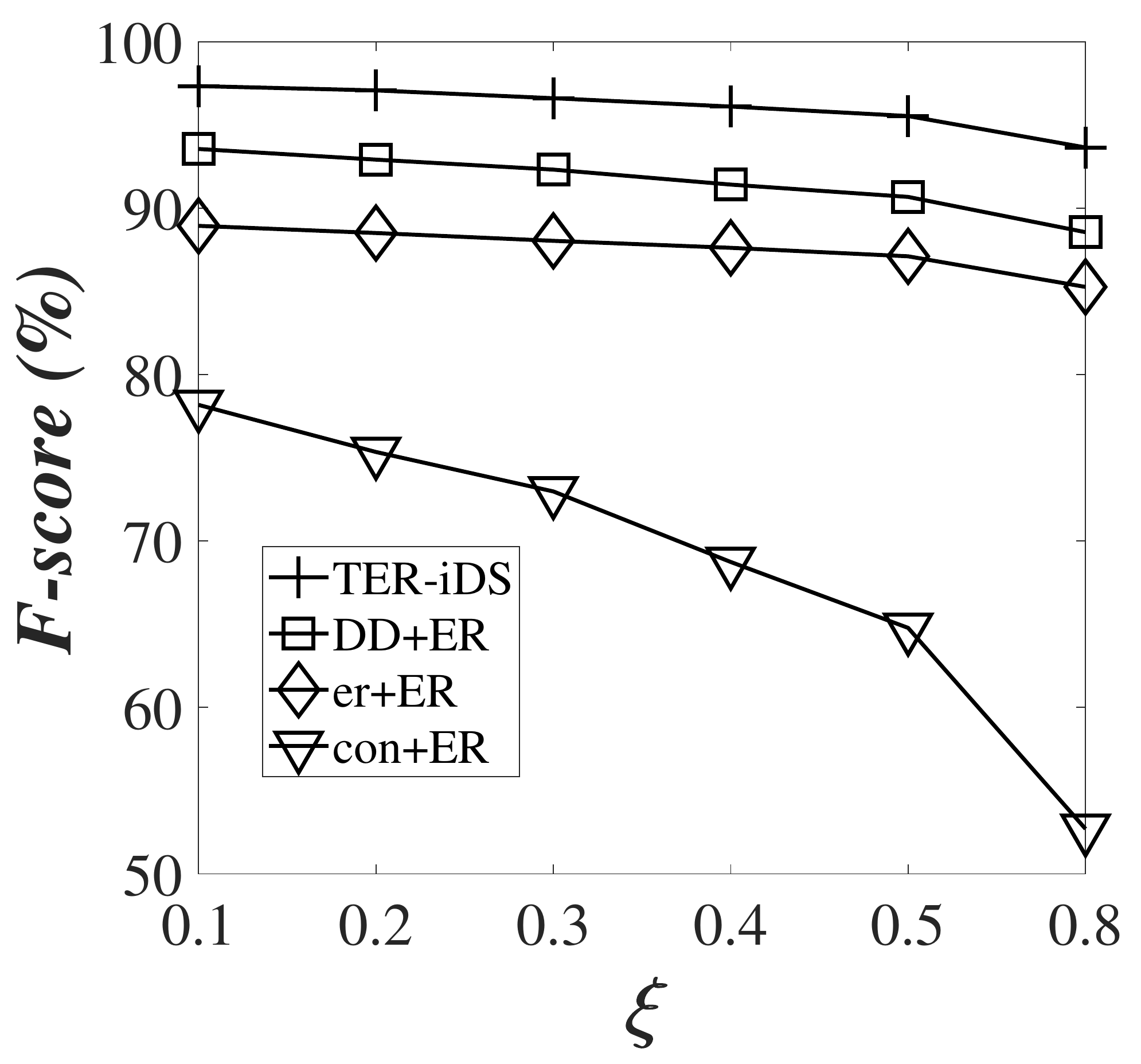}}
\label{subfig:xi_EBooks_F1}
}
\subfigure[][$Songs$]{\hspace{-1ex}
\scalebox{0.17}[0.17]{\includegraphics{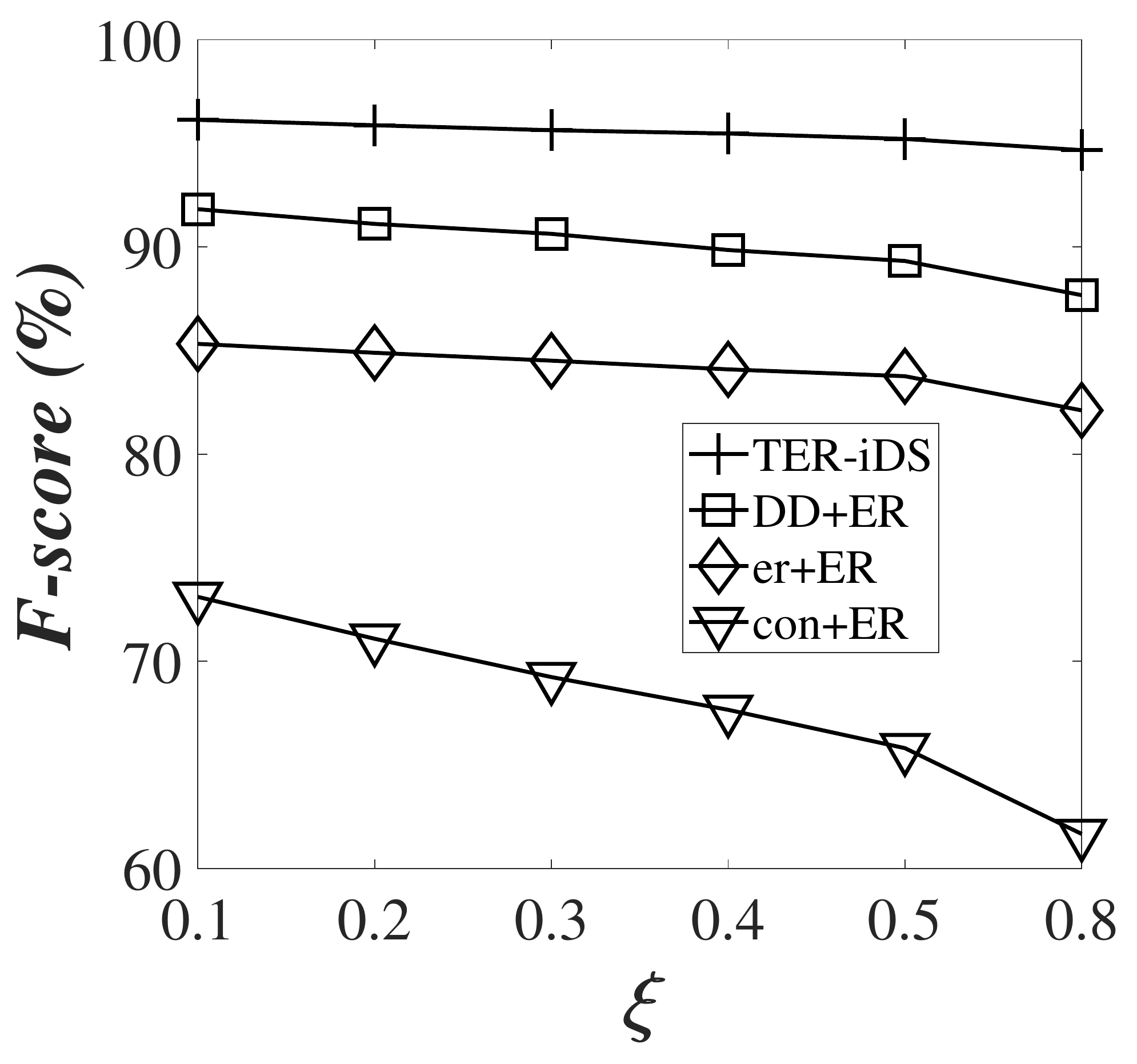}}
\label{subfig:xi_Songs_F1}
}\vspace{-4ex}
\caption{\small The TER-iDS effectiveness vs. the missing rate, $\xi$, of incomplete tuples in data streams $iDS_i$.} 
\label{exper:xi_F1} \vspace{-2ex}
\end{figure*} 

\begin{figure*}[ht]
\centering \vspace{-1ex}
\subfigure[][{\small $Citations$}]{\hspace{-3ex}   

\scalebox{0.17}[0.16]{\includegraphics{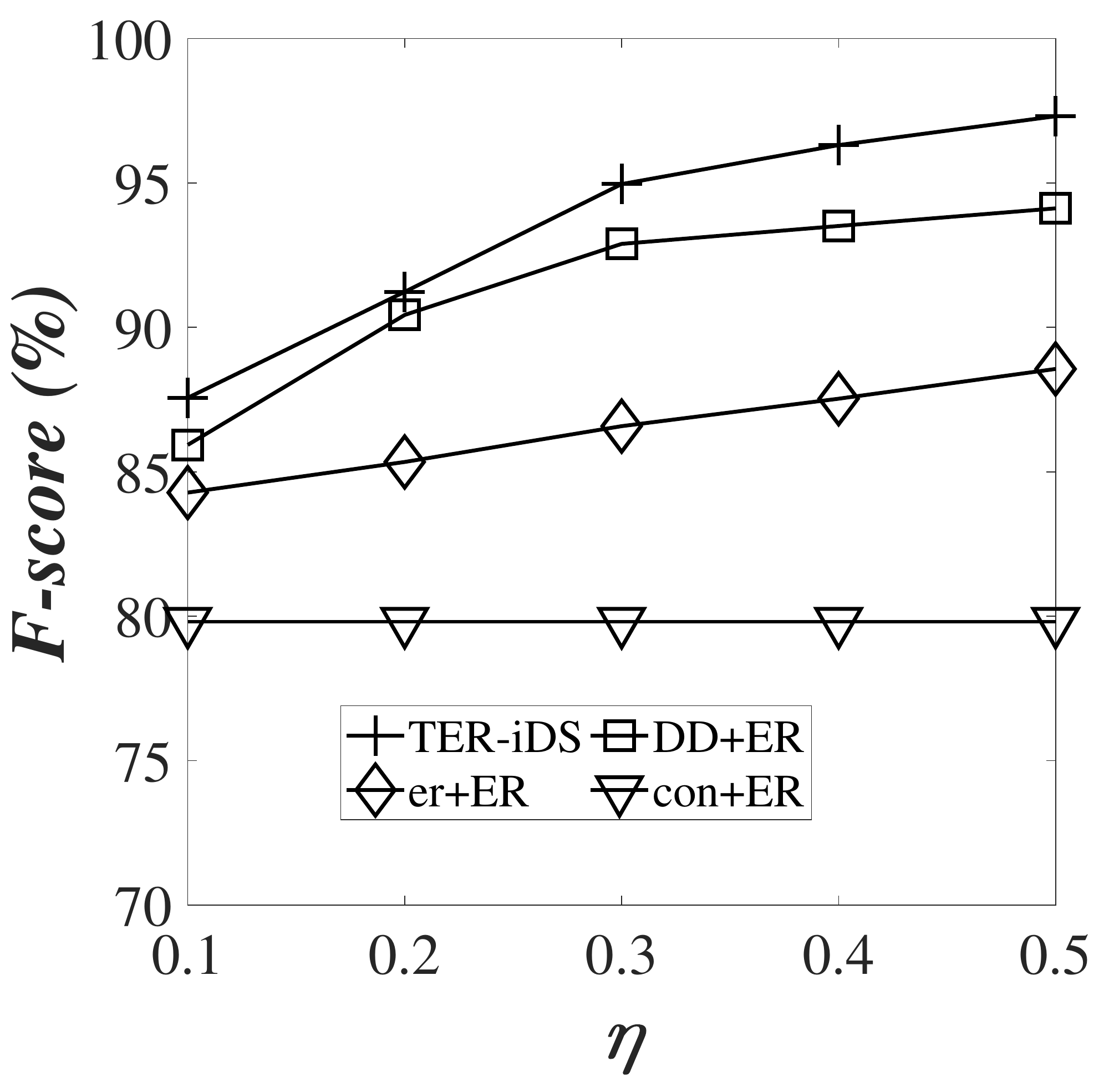}}
\label{subfig:eta_citations_F1}
}
\subfigure[][$Anime$]{
\scalebox{0.17}[0.16]{\includegraphics{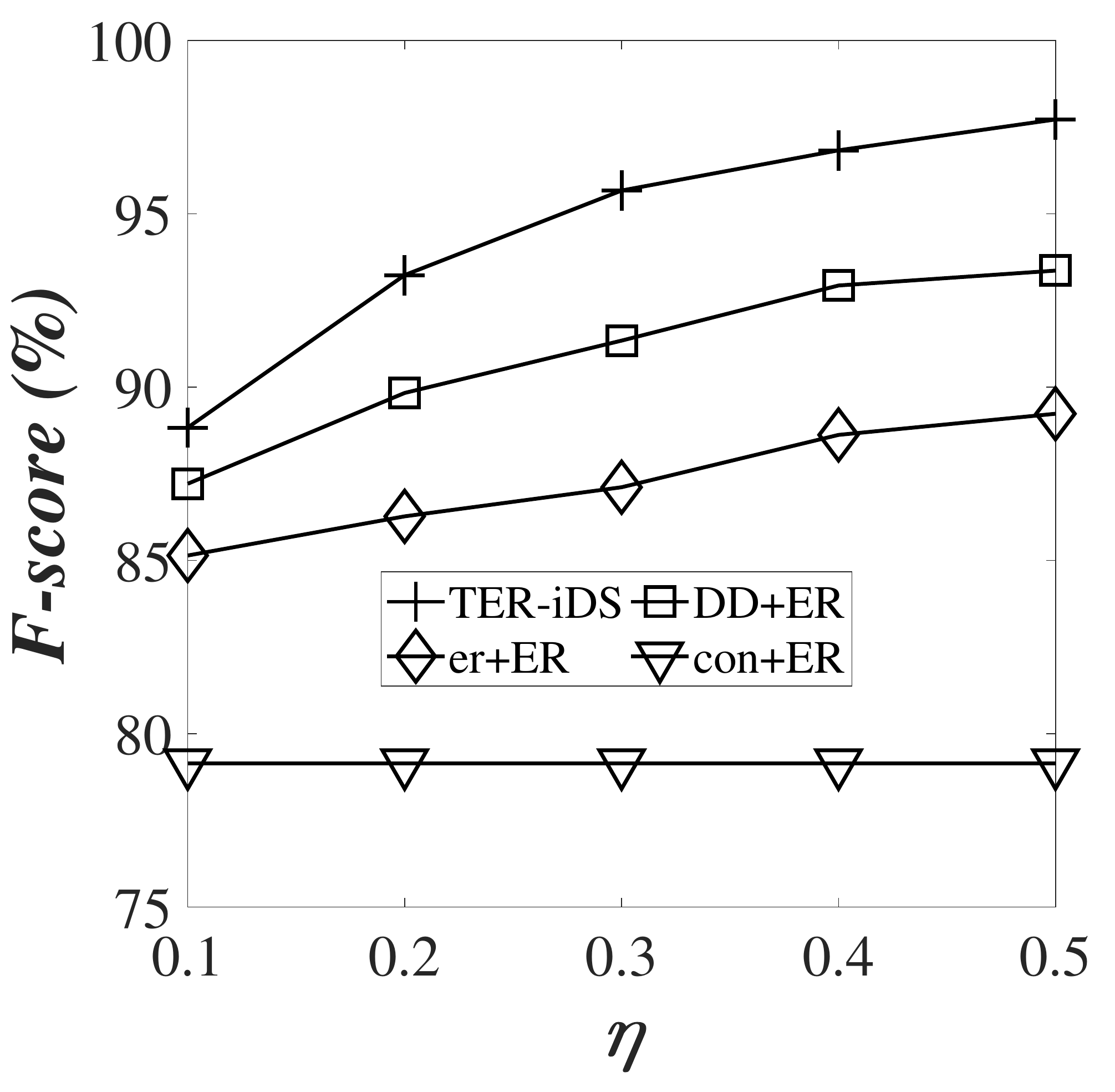}}
\label{subfig:eta_Anime_F1}
}
\subfigure[][$Bikes$]{
\scalebox{0.17}[0.16]{\includegraphics{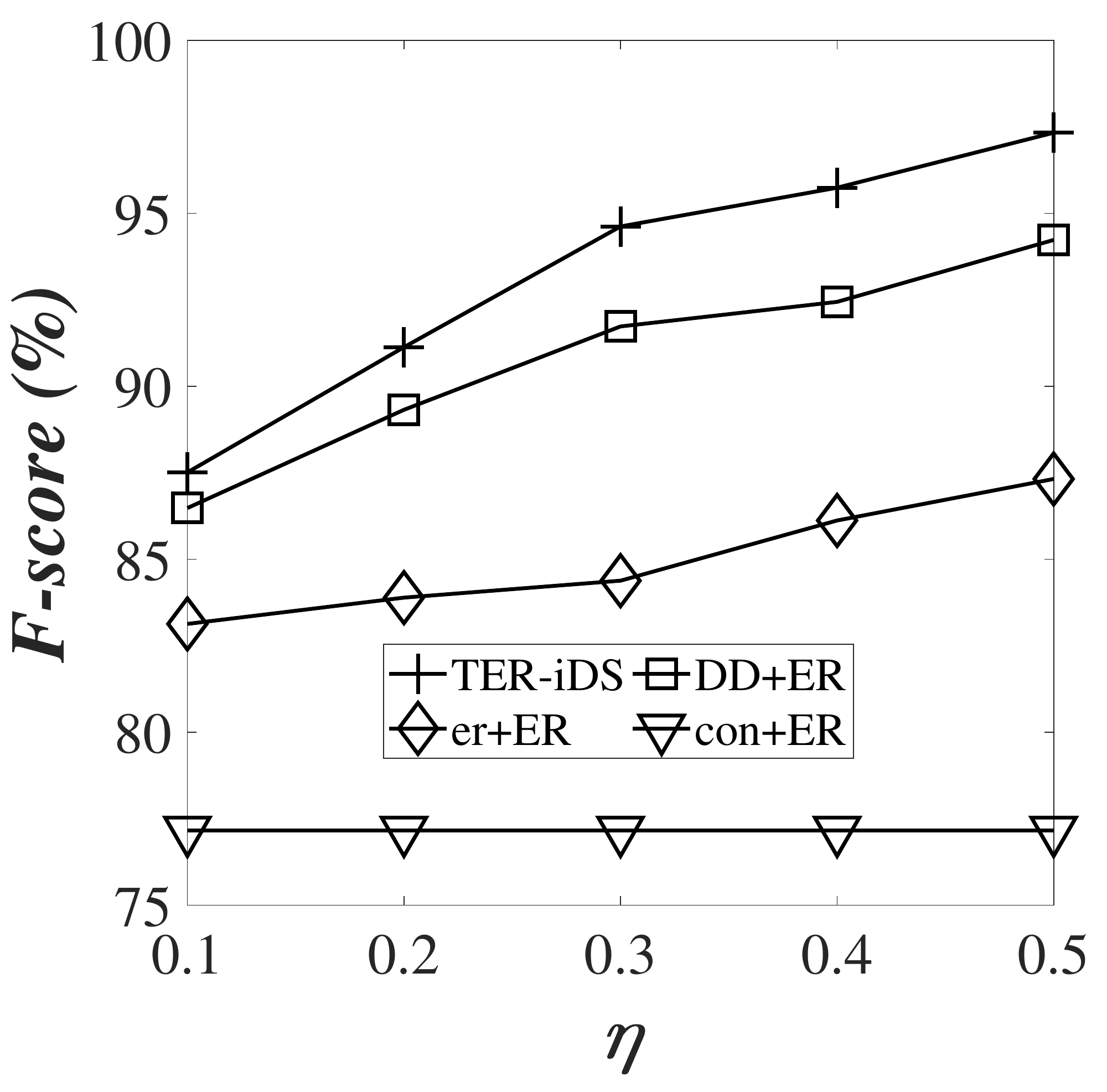}}
\label{subfig:eta_Bikes_F1}
}
\subfigure[][$EBooks$]{
\scalebox{0.17}[0.16]{\includegraphics{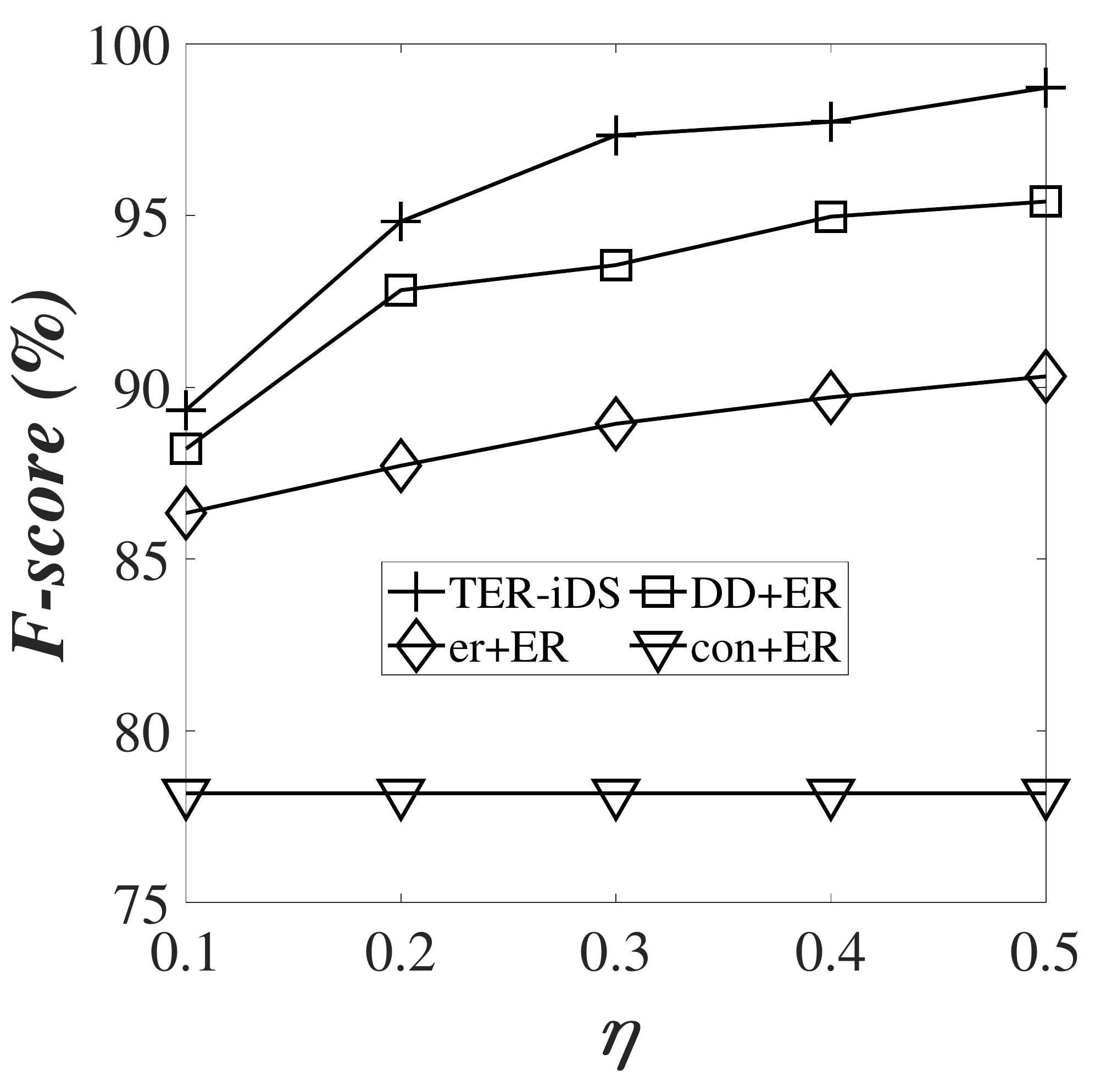}}
\label{subfig:eta_EBooks_F1}
}
\subfigure[][$Songs$]{\hspace{-1ex}
\scalebox{0.17}[0.16]{\includegraphics{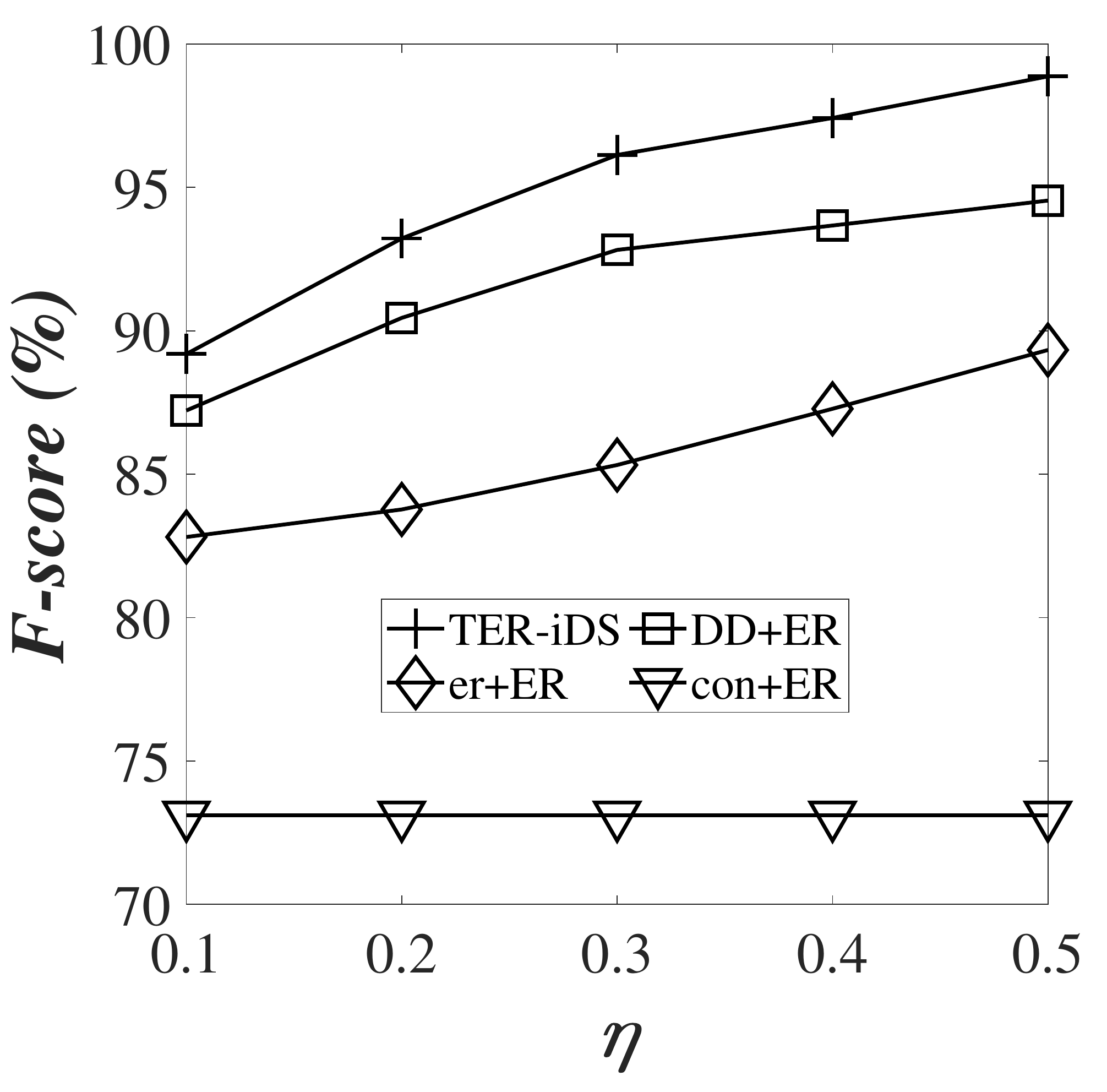}}
\label{subfig:eta_Songs_F1}
}\vspace{-4ex}
\caption{\small The TER-iDS effectiveness vs. the size ratio, $\eta$, of data repository $R$ w.r.t. data stream $iDS$.} 
\label{exper:eta_F1} \vspace{-2ex}
\end{figure*} 

\begin{figure*}[ht]
\centering \vspace{-1ex}
\subfigure[][{\small $Citations$}]{\hspace{-3ex}   

\scalebox{0.18}[0.18]{\includegraphics{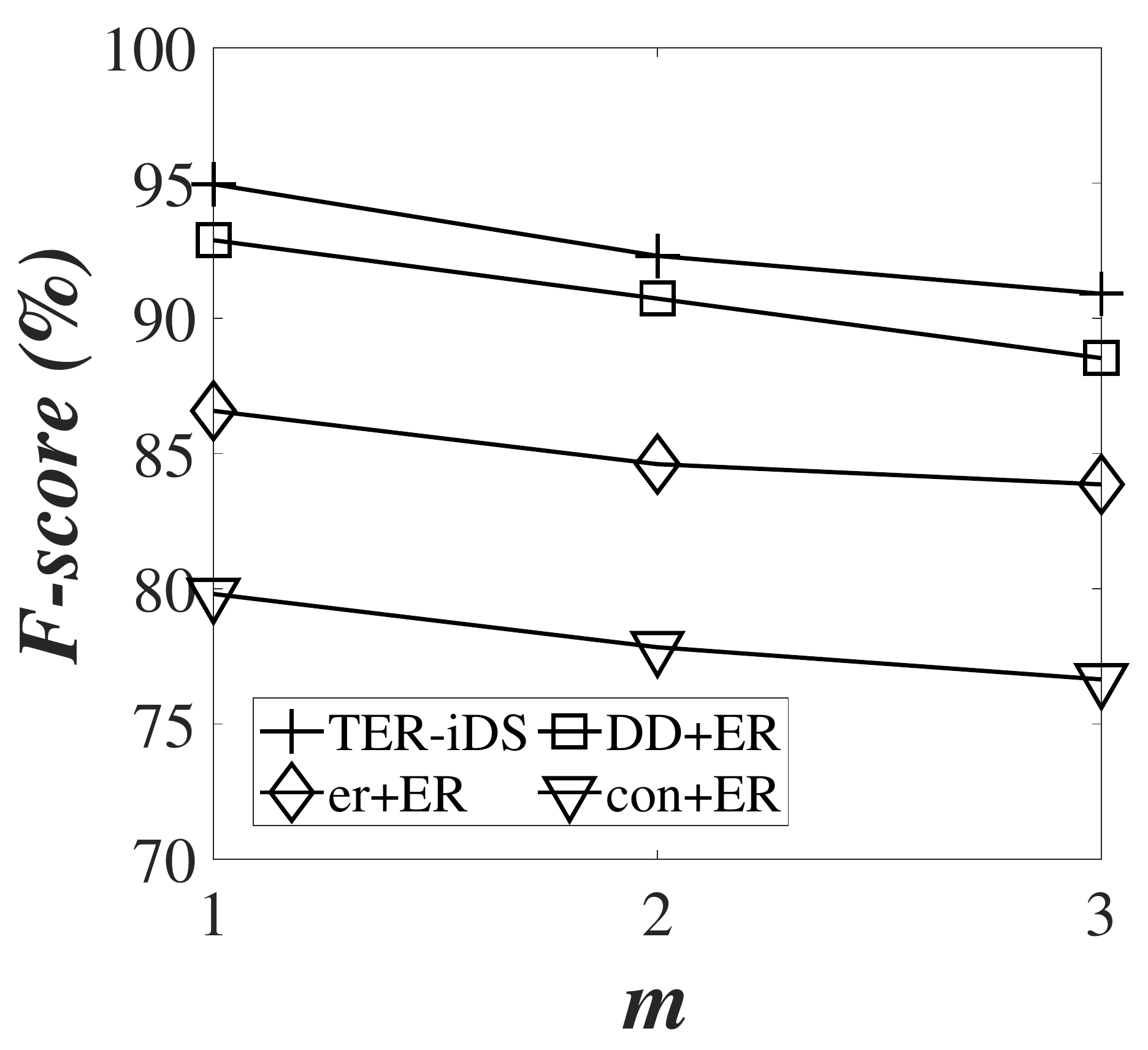}}
\label{subfig:m_citations_F1}
}
\subfigure[][$Anime$]{
\scalebox{0.18}[0.18]{\includegraphics{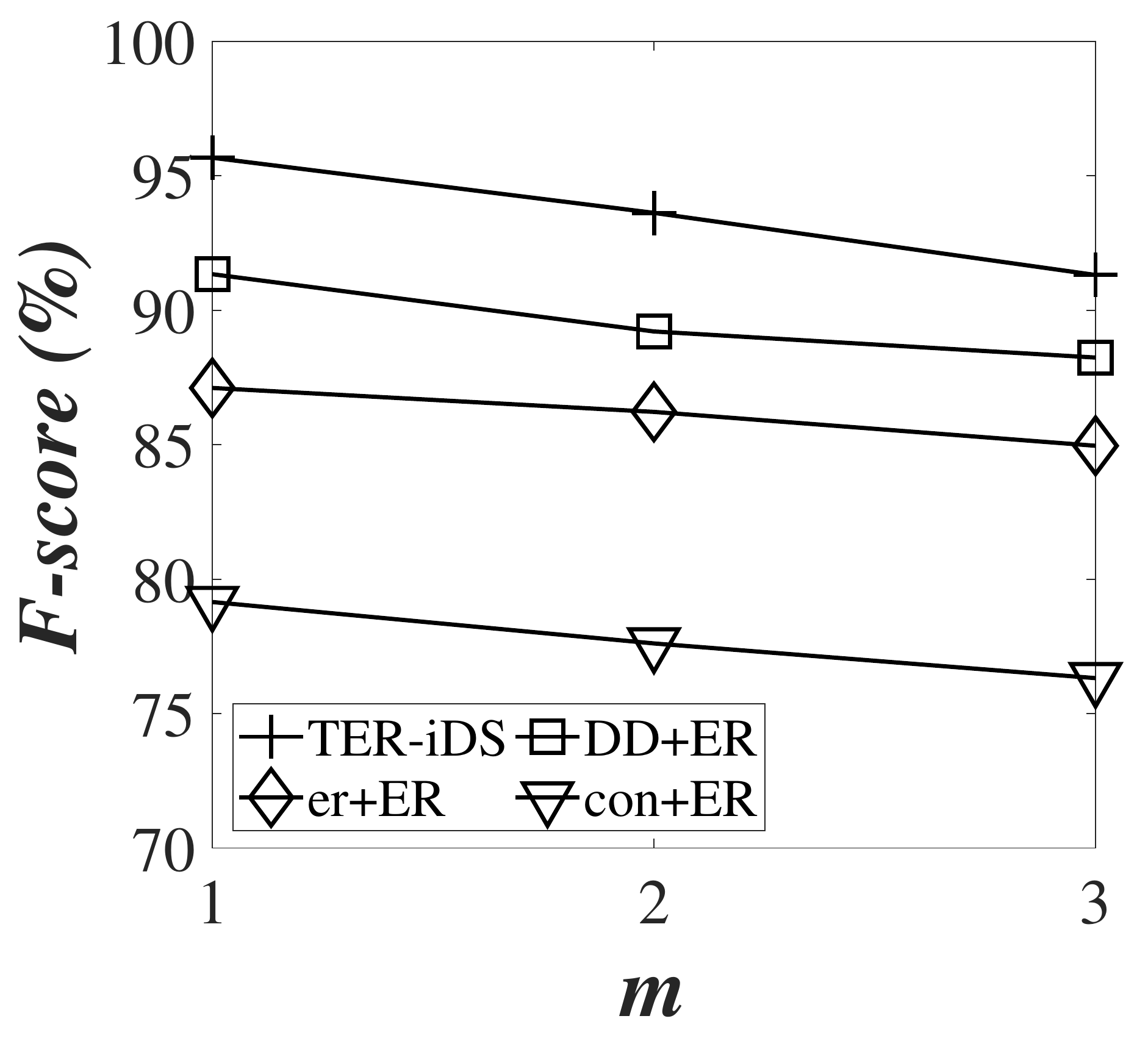}}
\label{subfig:m_Anime_F1}
}
\subfigure[][$Bikes$]{
\scalebox{0.18}[0.18]{\includegraphics{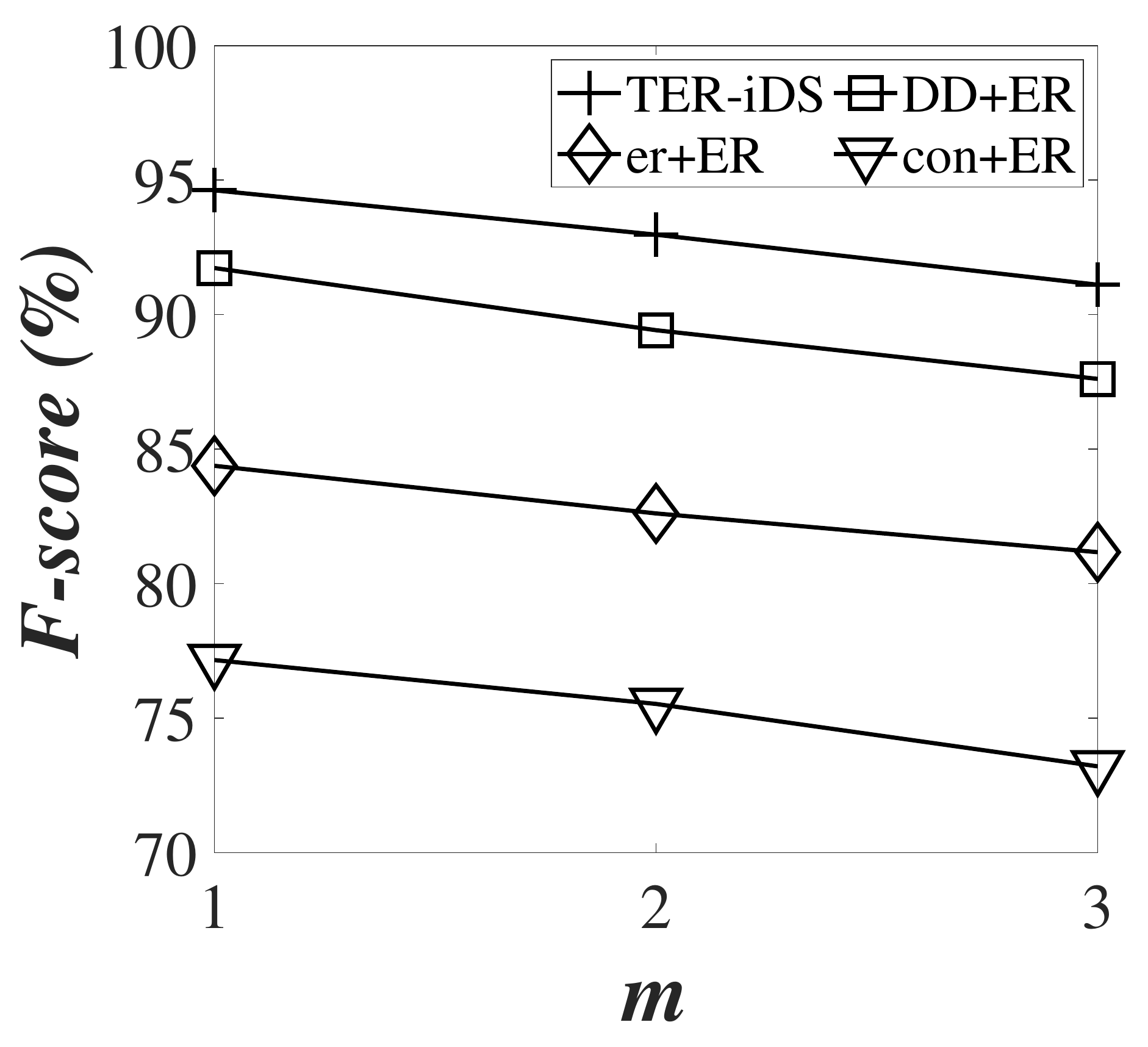}}
\label{subfig:m_Bikes_F1}
}
\subfigure[][$EBooks$]{
\scalebox{0.18}[0.18]{\includegraphics{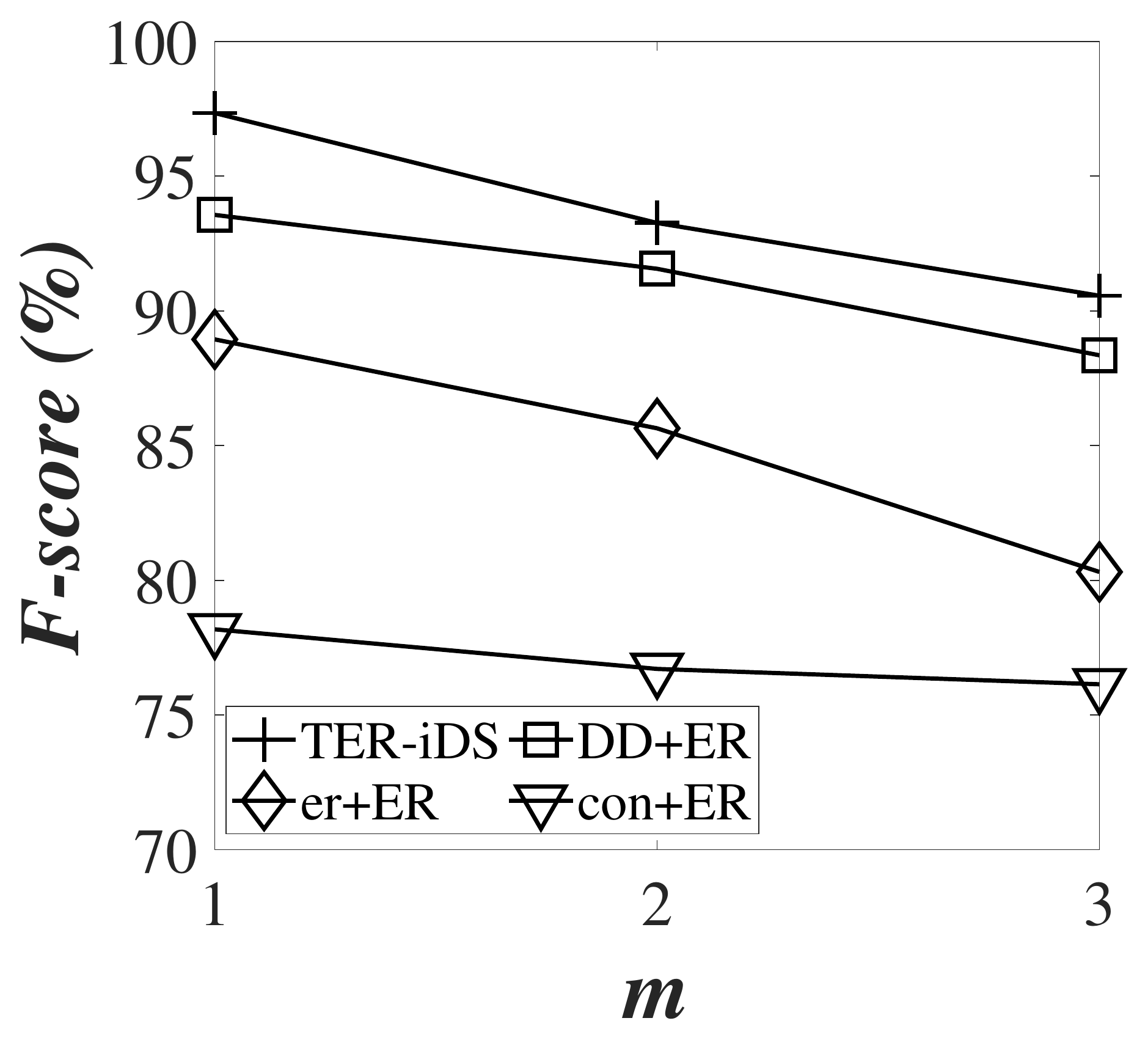}}
\label{subfig:m_EBooks_F1}
}
\subfigure[][$Songs$]{\hspace{-1ex}
\scalebox{0.18}[0.18]{\includegraphics{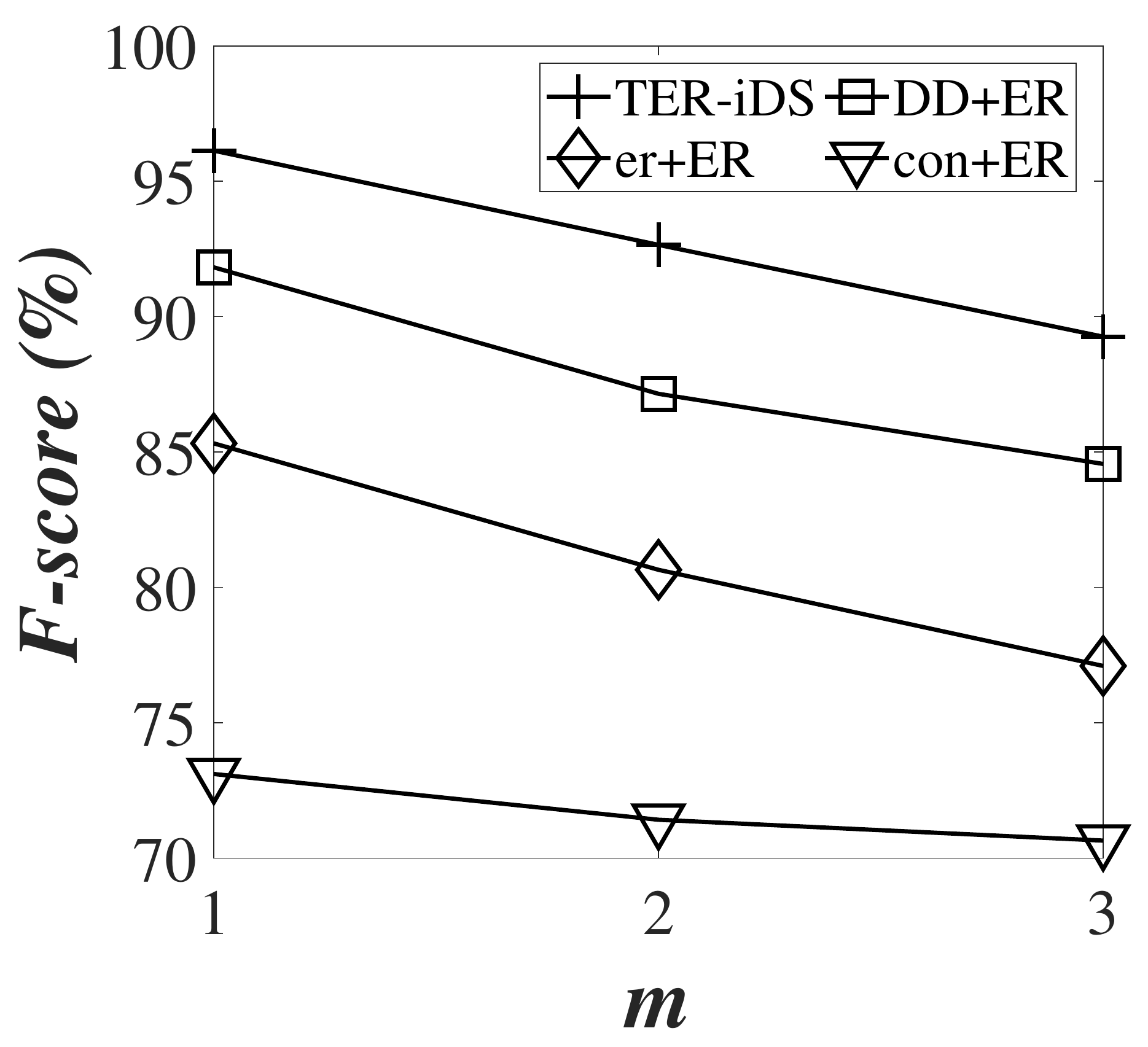}}
\label{subfig:m_Songs_F1}
}\vspace{-4ex}
\caption{\small The TER-iDS effectiveness vs. the number, $m$, of missing attributes.} 
\label{exper:m_F1} \vspace{-2ex}
\end{figure*}

\begin{figure*}[ht]
\centering \vspace{-1ex}
\subfigure[][{\small $Citations$}]{\hspace{-3ex}   

\scalebox{0.17}[0.16]{\includegraphics{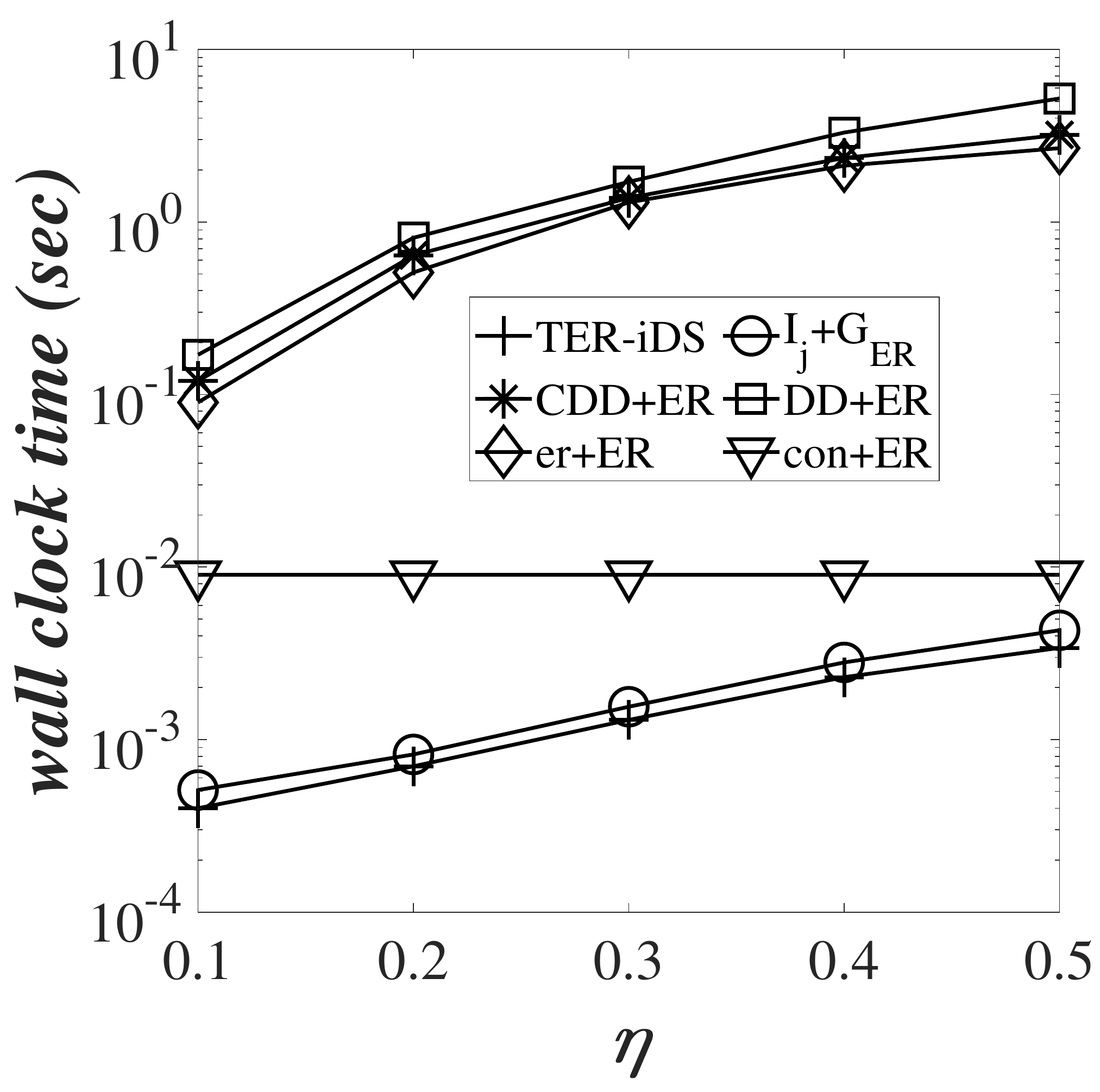}}
\label{subfig:eta_citations}
}
\subfigure[][$Anime$]{
\scalebox{0.17}[0.16]{\includegraphics{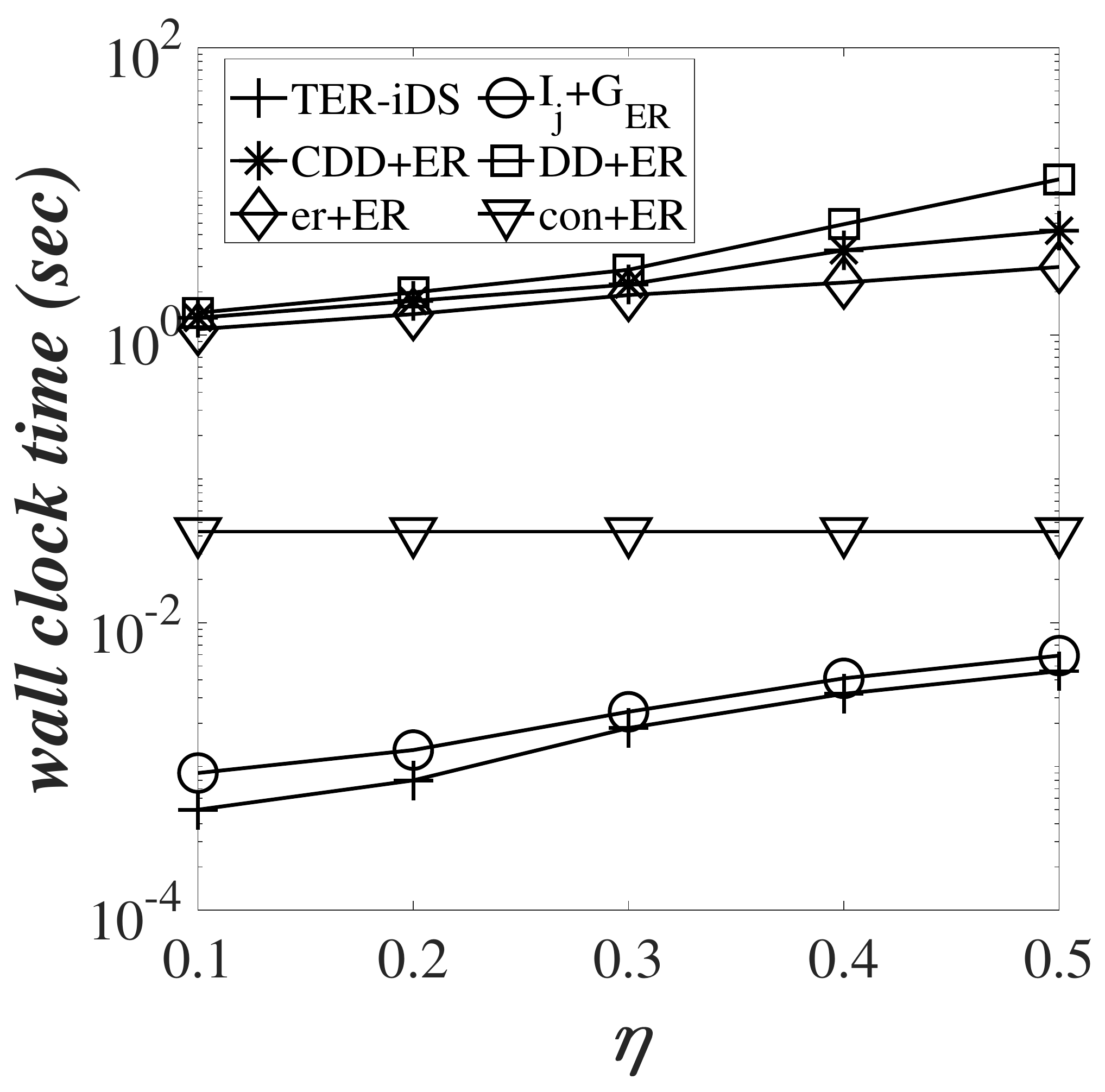}}
\label{subfig:eta_Anime}
}
\subfigure[][$Bikes$]{
\scalebox{0.17}[0.16]{\includegraphics{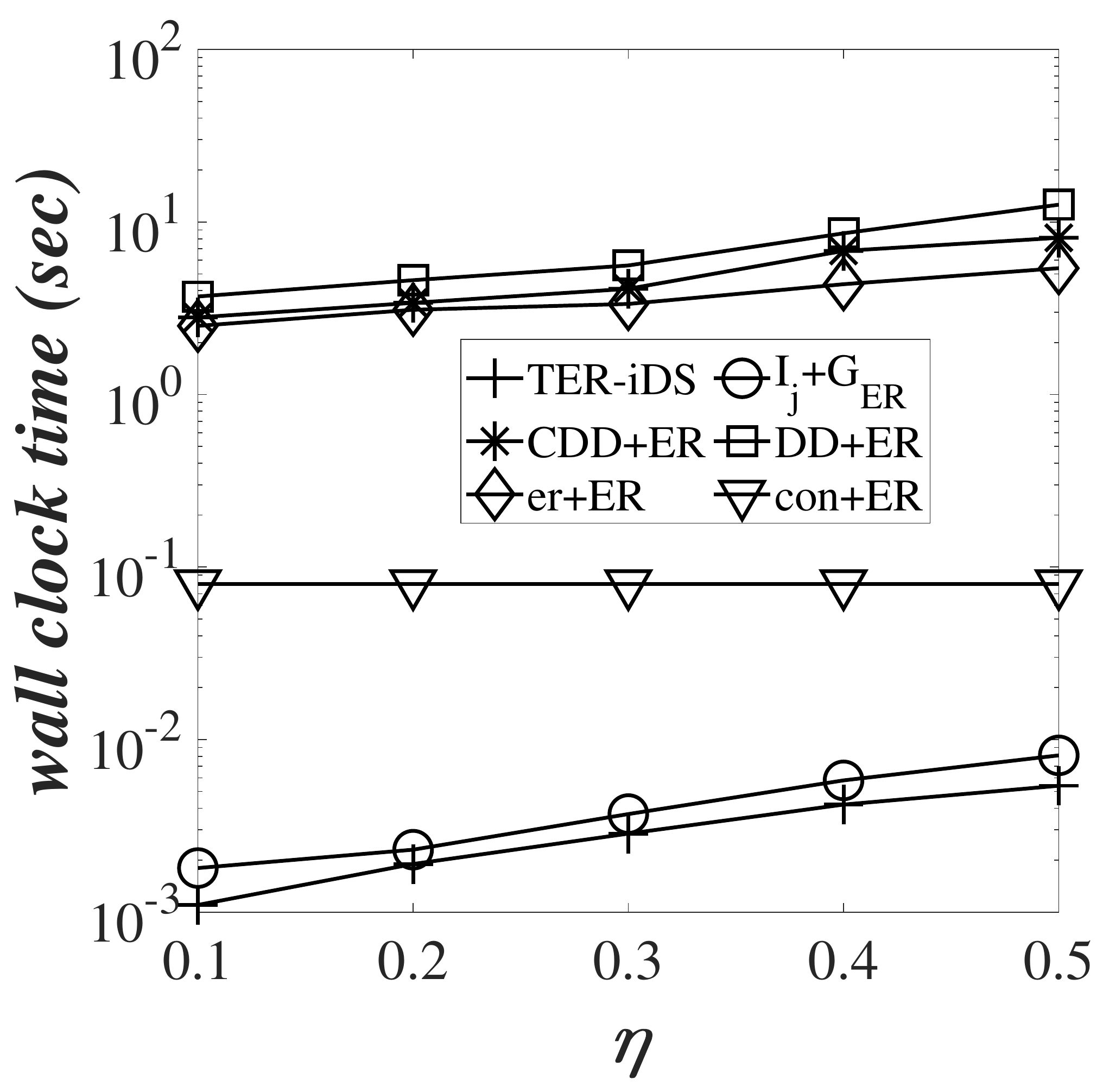}}
\label{subfig:eta_Bikes}
}
\subfigure[][$EBooks$]{
\scalebox{0.17}[0.16]{\includegraphics{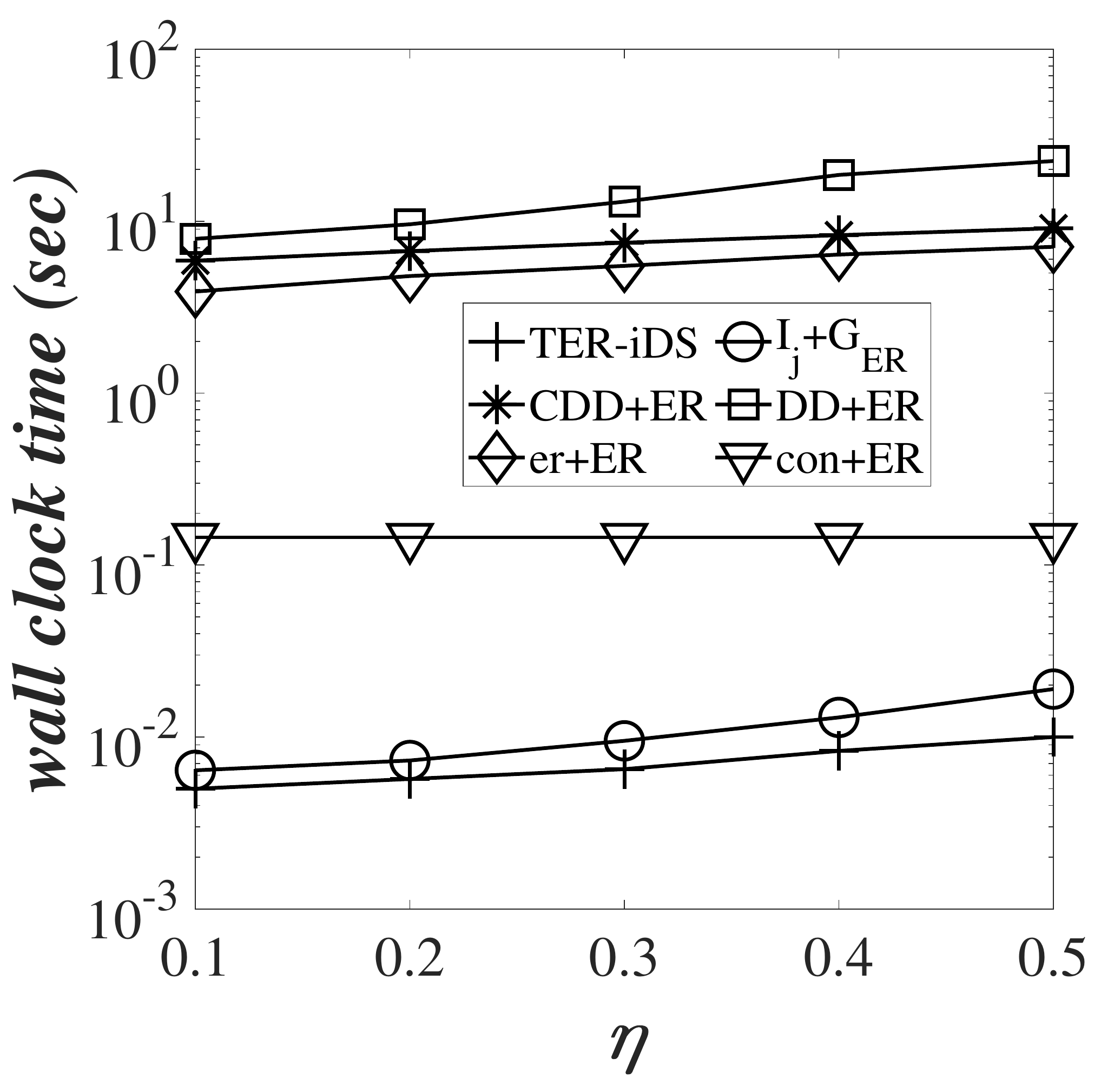}}
\label{subfig:eta_EBooks}
}
\subfigure[][$Songs$]{\hspace{-1ex}
\scalebox{0.17}[0.16]{\includegraphics{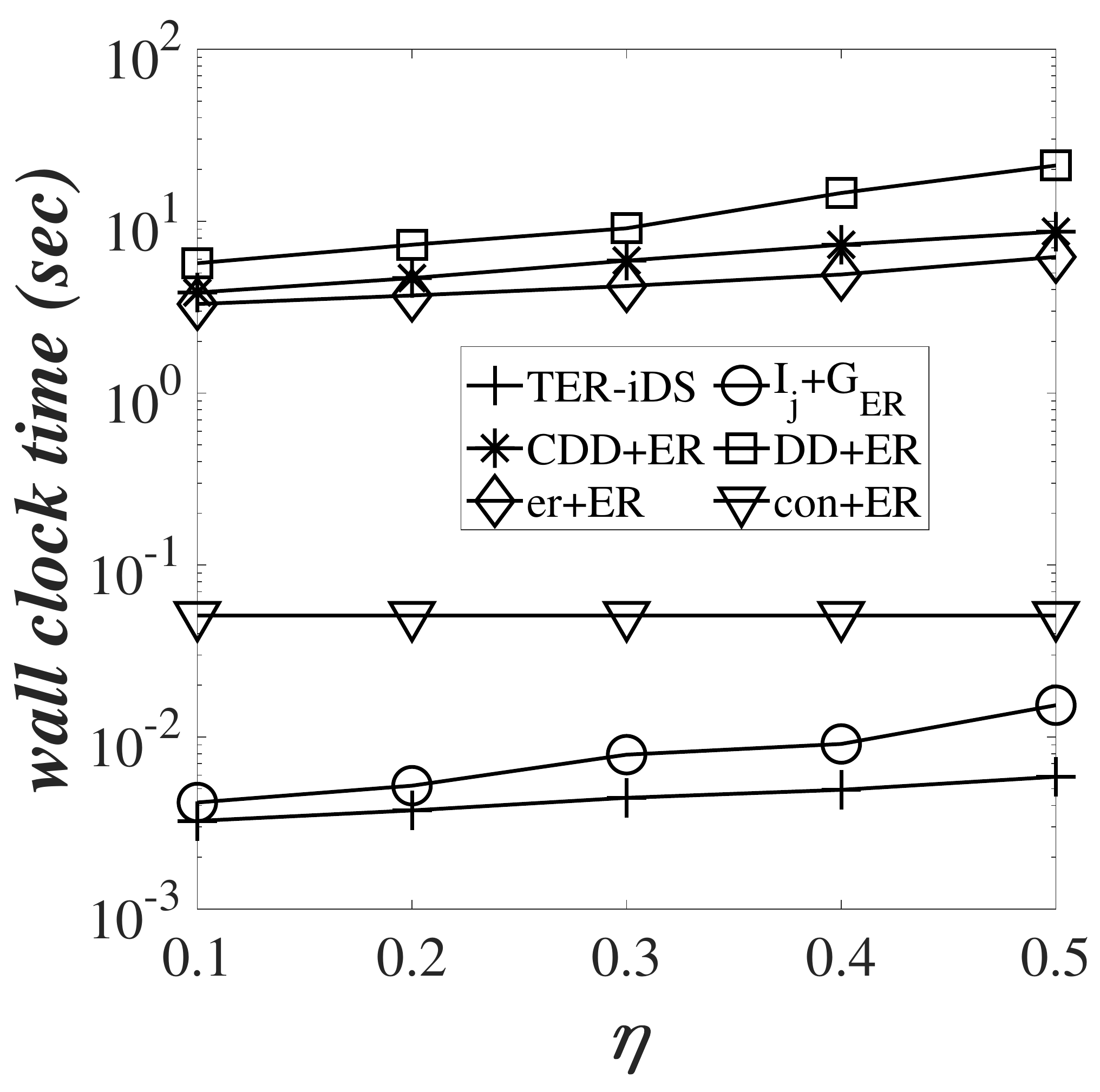}}
\label{subfig:eta_Songs}
}\vspace{-4ex}
\caption{\small The TER-iDS efficiency vs. the size ratio, $\eta$, of data repository $R$ w.r.t. data stream $iDS$.} 
\label{exper:eta} \vspace{-2ex}
\end{figure*} 

\begin{figure*}[ht]
\centering \vspace{-1ex}
\subfigure[][{\small $Citations$}]{\hspace{-3ex}   

\scalebox{0.17}[0.16]{\includegraphics{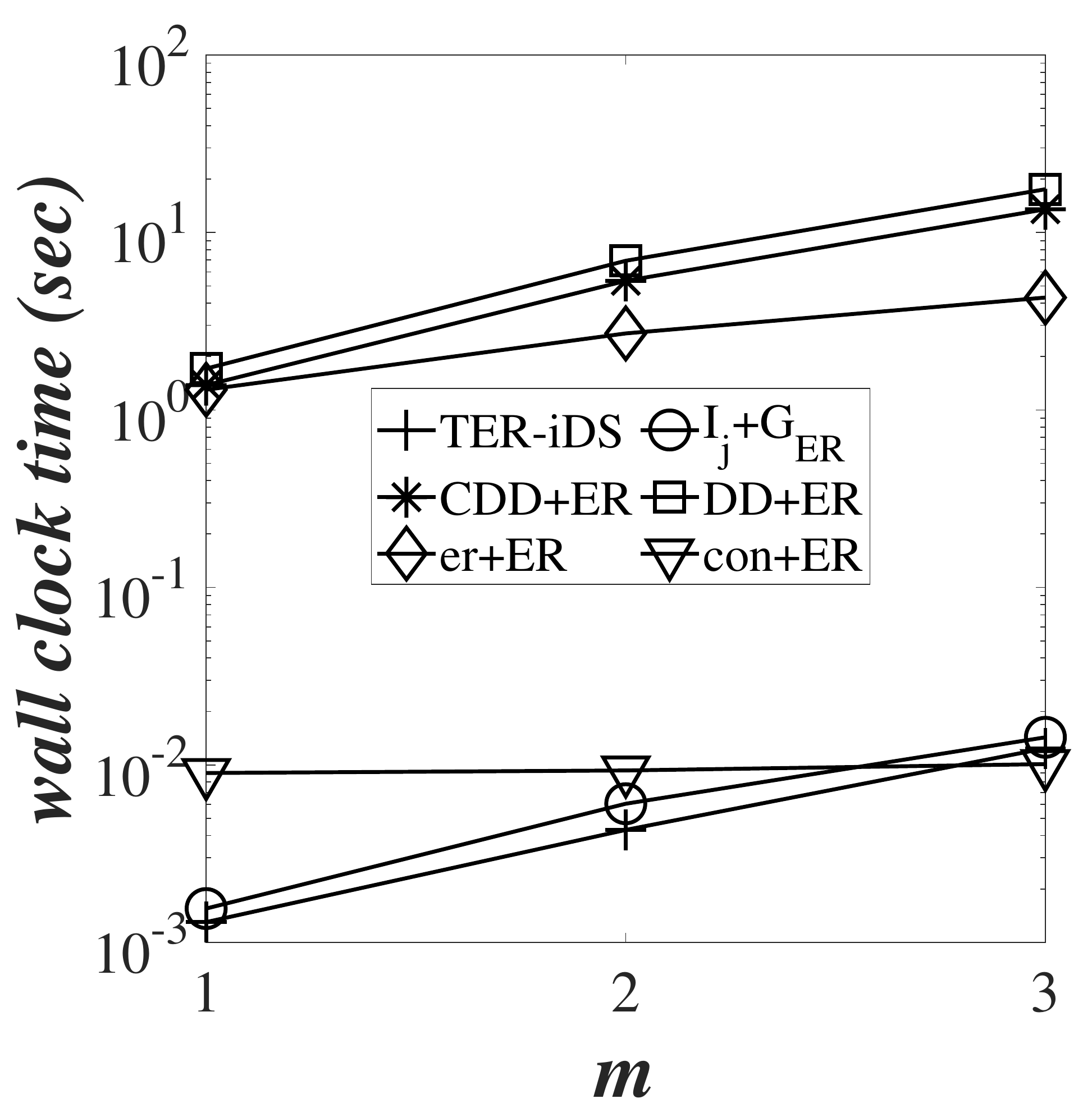}}
\label{subfig:m_citations}
}
\subfigure[][$Anime$]{
\scalebox{0.17}[0.16]{\includegraphics{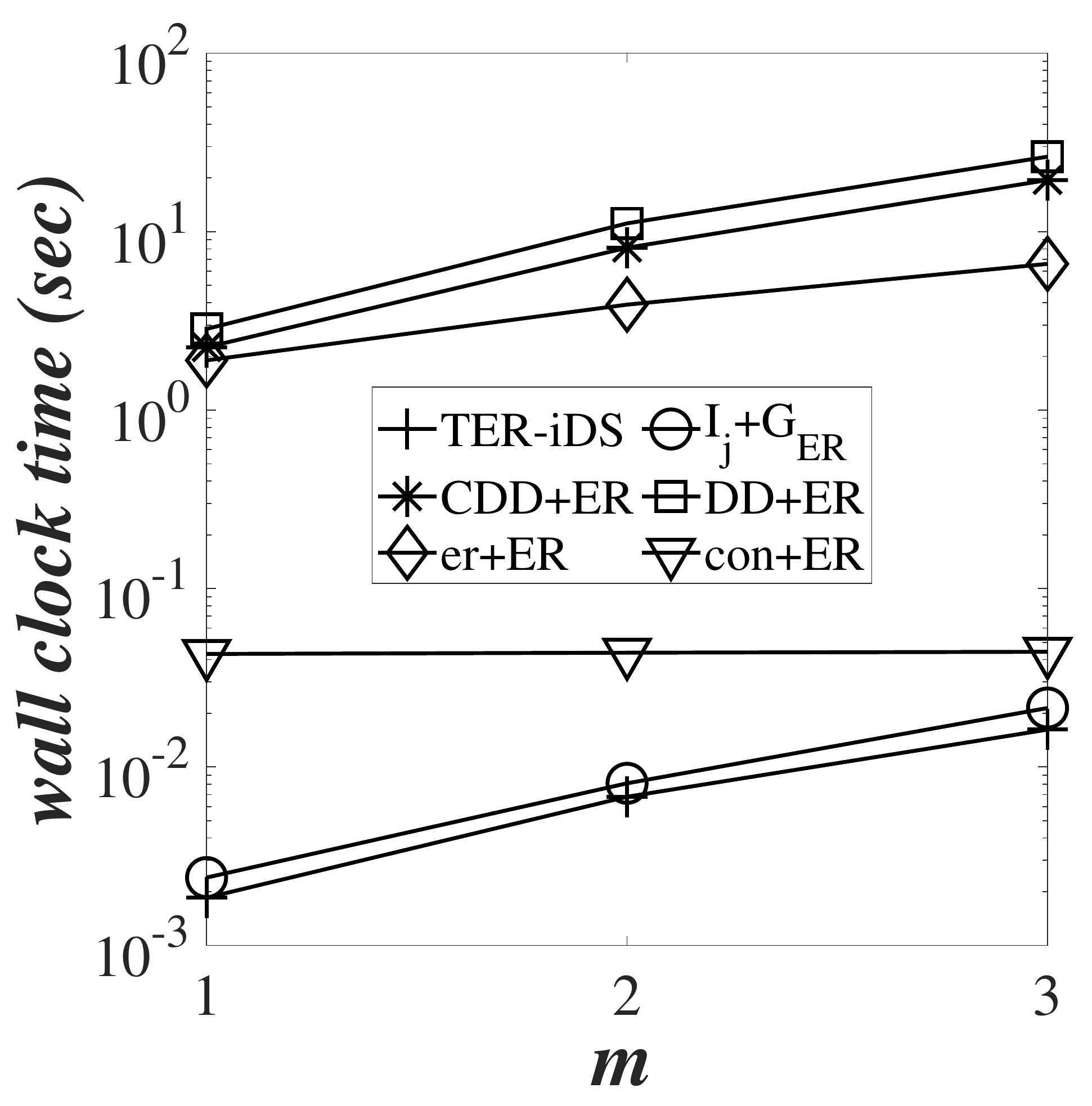}}
\label{subfig:m_Anime}
}
\subfigure[][$Bikes$]{
\scalebox{0.17}[0.16]{\includegraphics{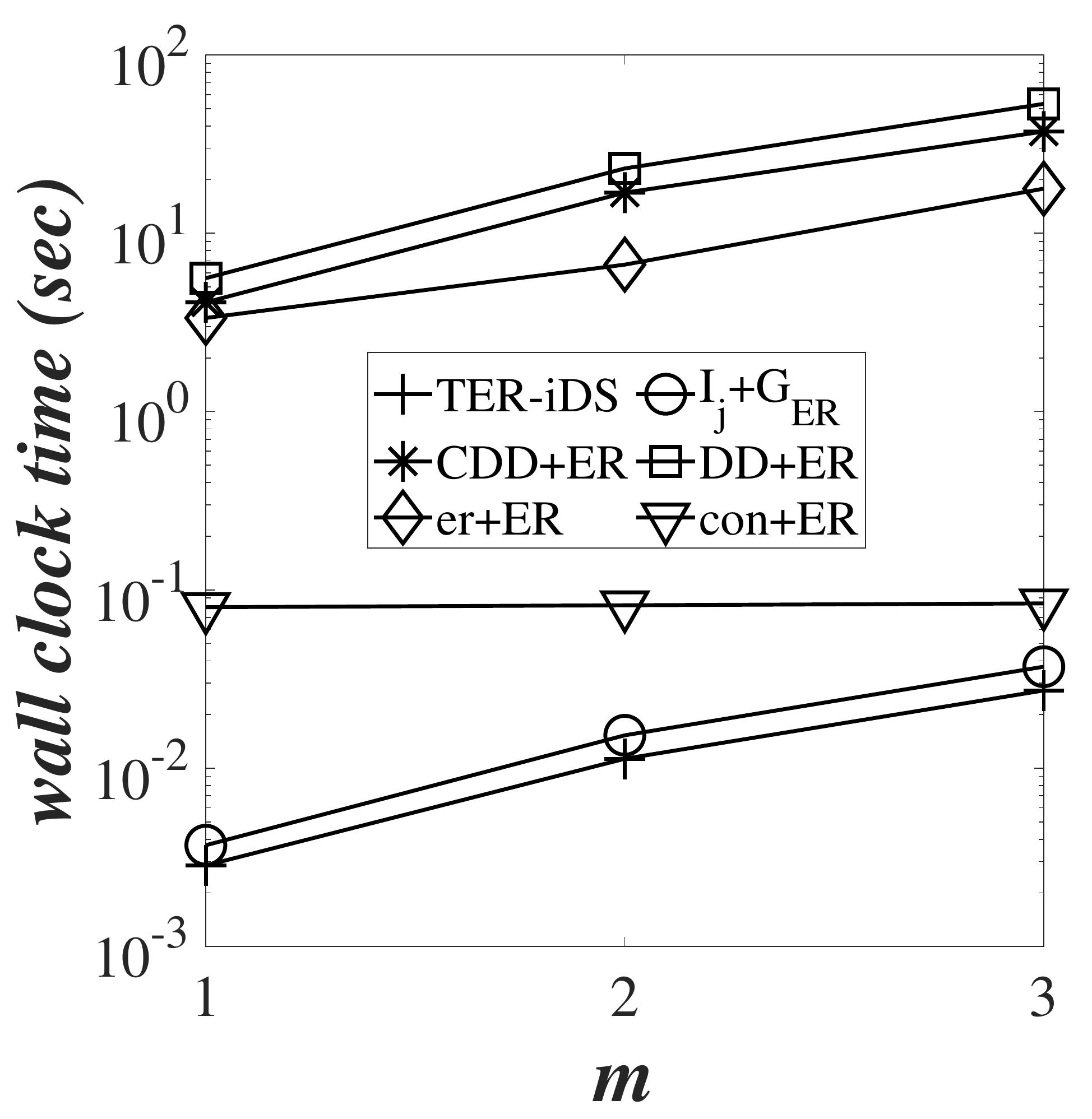}}
\label{subfig:m_Bikes}
}
\subfigure[][$EBooks$]{
\scalebox{0.17}[0.16]{\includegraphics{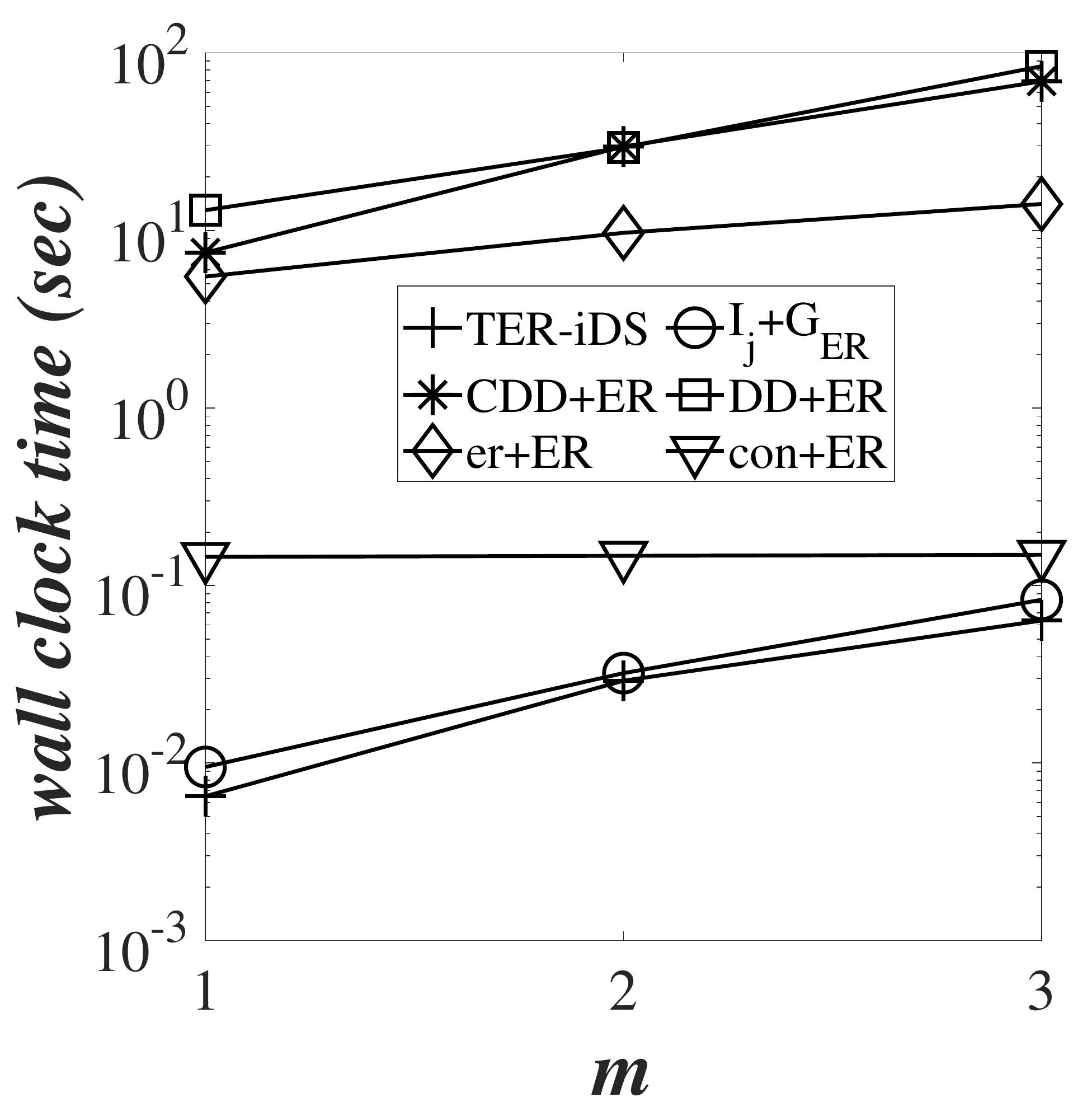}}
\label{subfig:m_EBooks}
}
\subfigure[][$Songs$]{\hspace{-1ex}
\scalebox{0.17}[0.16]{\includegraphics{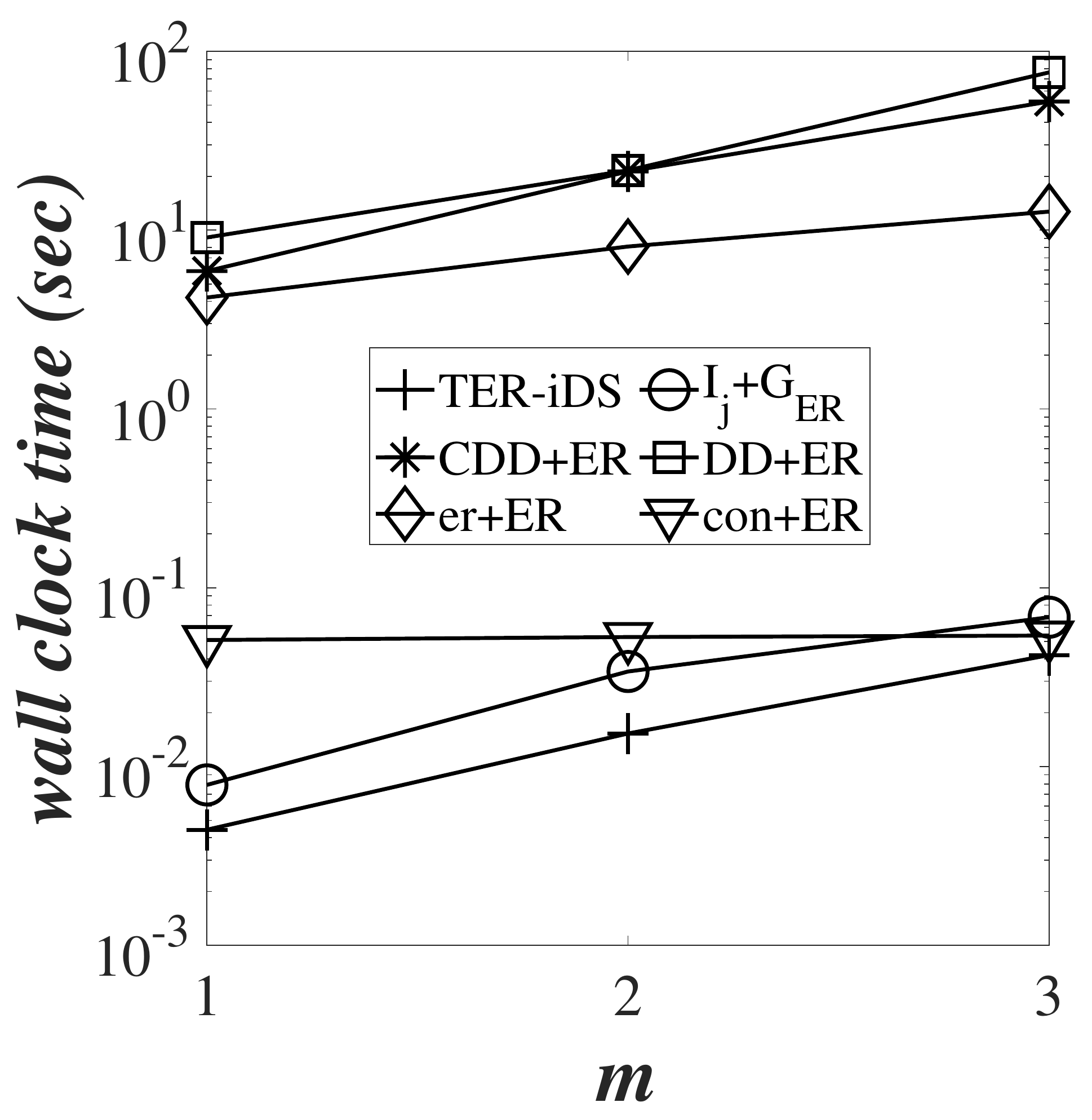}}
\label{subfig:m_Songs}
}\vspace{-4ex}
\caption{\small The TER-iDS efficiency vs. the number, $m$, of missing attributes.} 
\label{exper:m} \vspace{-2ex}
\end{figure*} 

\section{More Experimental Results}
\label{sec:more_exp_results}

\subsection{Verification of the Cost-Model-Based Algorithm}
\label{subsec:cost_model_verification}
For the evaluation of our cost-model-based algorithm in Appendix \ref{sec:cost_model_for_pivots}, we divide the converted space into 10 sub-intervals on each attribute (i.e., $P=10$), and set the minimal entropy threshold $eMin$ to $1.5$. Note that, the pivot selection is processed offline prior to our online TER-iDS approach.

\vspace{1ex}\noindent {\bf The time cost of the cost-model-based algorithm vs. the size ratio, $\eta$, of data repository $R$ w.r.t. data stream $iDS$.} Figure \ref{subfig:pivot_eta} shows the time cost of choosing pivots from data repository $R$, where $\eta$ varies from $0.1$ to $0.5$, $P=10$, $eMin=1.5$, and other parameters are set to default values. From the figure, for larger size ratio $\eta$, the time cost becomes larger over all the data sets, since our cost model needs to check more samples in larger data repositories for choosing the suitable pivots. Similarly, we can see that it takes more time for our cost model to select suitable pivots for data sets with large size. Although our cost-model-based algorithm offline chooses pivots, it achieves reasonable time cost for data sets with large size (e.g., 22,161.65 $sec$ for $Songs$ with $|R|=500K$).

\vspace{1ex}\noindent {\bf The time cost of the cost-model-based algorithm vs. the maximal allowed number, $cntMax$, of attribute pivots on each attribute.} Figure \ref{subfig:pivot_cnt} demonstrates the CPU time of obtaining suitable pivot tuples from data repository $R$, where $cntMax$ is set within $[1,5]$, $P=10$, $eMin=1.5$, and other parameters are by default. From the graph, when $cntMax$ increases, the time cost (i.e., 125.49 $sec\sim$ 29,355.12 $sec$) smoothly increases for all the data sets. Specifically, for data sets such as $Citations$, the time cost remains almost the same when when $cntMax$ reaches some thresholds (e.g., time cost is fixed as 275.32 $sec$ for $Citations$ when $cntMax\geq 3$). This is because, our cost-model-based algorithm will stop choosing more auxiliary attribute pivots when the selected $n_x$ ($\leq cntMax$) pivots can together achieve a Shannon entropy (Equation (\ref{eq:entropy})) no smaller than $eMin$ ($=1.5$). 

\subsection{Evaluation of CDD Detection}
\label{subsec:CDD_detect_eva}
Figure \ref{fig:CDD_detect_cost} shows the time cost (85.59 $sec \sim$ 6,260.5 $sec$) of offline detecting (creating) CDD rules from data repositories of 5 real data sets, where all parameters are set to their default values (as depicted in Table \ref{table:exp_parameter_setting}). From the figure, we can see that data sets with larger sizes of data repository need more time (e.g., 6,260.5 $sec$ for $Songs$ of size $300K$) to detect CDD rules. This is reasonable, since we need to check more data for detecting and validating a valid CDD rule. Moreover, the CDD detection cost of $EBooks$ data (i.e., 519.06 $sec$) is much higher than that of $Citations$ (i.e., 85.59 $sec$), $Anime$ (i.e., 195.21 $sec$), and $Bikes$ (i.e., 198.53 $sec$), since $EBooks$ has larger sizes of token sets on some attributes. Please refer to \cite{kwashie2015conditional,wang2017discovering} for more evaluation of the CDD detection. 

\subsection{More TER-iDS Effectiveness Evaluation}

In this subsection, we test and report the topic-related ER accuracy of our $TER\text{-}iDS$ approaches and its 3 baselines (i.e., $DD+ER$, $er+ER$, and $con+ER$) over 5 real data sets, by varying 3 parameters, which are the missing rate, $\xi$, of incomplete tuples in data streams $iDS_i$, the size ratio, $\eta$, of data repository $R$ w.r.t. data stream $iDS$, and the number, $m$, of missing attributes, respectively. Note that, we do not compare the accuracy of our $TER\text{-}iDS$ method with that of $I_j+G_{ER}$ and $CDD+ER$, since they both adopt CDDs as imputation methods and thus have the same accuracy.

\vspace{1ex}\noindent {\bf The TER-iDS effectiveness vs. the missing rate, $\xi$, of incomplete tuples in $iDS_i$.} Figure \ref{exper:xi_F1} shows the effect of the missing rate, $\xi$, of incomplete tuples in streams on the effectiveness performance of $TER\text{-}iDS$ and 3 baselines, $DD+ER$, $er+ER$, and $con+ER$, where $\xi=0.1$, $0.2$, $0.3$, $0.4$, $0.5$, and $0.8$, and other parameters are set to their default values. From figures, when $\xi$ increases, the topic-related ER accuracy decreases for all the approaches. This is reasonable, since there will be more incomplete tuples that need to be imputed for larger $\xi$. Nevertheless, $TER\text{-}iDS$ still has the highest query (imputation) accuracy (i.e., 88.73\%$\sim$97.34\%), which confirms the effectiveness of our $TER\text{-}iDS$ approach. 

\vspace{1ex}\noindent {\bf The TER-iDS effectiveness vs. the size ratio, $\eta$, of data repository $R$ w.r.t. data stream $iDS$.} Figure \ref{exper:eta_F1} reports the effectiveness of our $TER\text{-}iDS$ approach and 3 baselines, $DD+ER$, $er+ER$, and $con+ER$, where $\eta$ varies from $0.1$ to $0.5$ and default values are used for other parameters. From the figures, for all the approaches (except for $con+ER$), we can see that the accuracy increases as $\eta$ increases. This is because, for larger $\eta$, there are more sample candidates for imputing incomplete tuples, which may improve the imputation accuracy. Note that, the accuracy of $con+ER$ stays the same for all $\eta$ values, since $con+ER$ imputes missing attributes only based on incomplete data streams (rather than data repository $R$). $TER\text{-}iDS$ still has the highest imputation accuracy (i.e., 87.51\%$\sim$98.87\%) among all methods, which shows good effectiveness of our $TER\text{-}iDS$ approach. 

\vspace{1ex}\noindent {\bf The TER-iDS effectiveness vs. the number, $m$, of missing attributes.} Figure \ref{exper:m_F1} illustrates the effect of the number, $m$, of missing attributes on the effectiveness of $TER\text{-}iDS$ and 3 baselines, $DD+ER$, $er+ER$, and $con+ER$, where $m=1$, $2$, and $3$, and default values are used for other parameters. From the figures, with a larger number of missing attributes, the accuracy decreases for both $TER\text{-}iDS$ and baselines, since there are more possible imputed tuples that need to be processed and refined. Nevertheless, our $TER\text{-}iDS$ approach still has the highest accuracy among all the methods (i.e., 89.26\%$\sim$97.34\%), which verifies the effectiveness of our $TER\text{-}iDS$ approach.

\subsection{More TER-iDS Efficiency Evaluation}

In this subsection, we will evaluate the robustness of our $TER\text{-}iDS$ method over 5 real data sets, by 2 parameters, the size ratio, $\eta$, of data repository $R$ w.r.t. data stream $iDS$ and the number, $m$, of missing attributes. 

\vspace{1ex}\noindent {\bf The TER-iDS efficiency vs. the size ratio, $\eta$, of data repository $R$ w.r.t. data stream $iDS$.} Figure \ref{exper:eta} illustrates the performance of our $TER\text{-}iDS$ approach and 5 baselines over 5 real data sets, where $\eta$ varies from $0.1$ to $0.5$, and other parameters are set to their default values. From the figures, as $\eta$ increases, the time cost becomes higher for all approaches (except for $con+ER$). This is because, with larger $\eta$, we need higher computation cost to check more samples from the data repository for imputing incomplete tuples. Moreover, $con+ER$ imputes missing attributes based on data streams (instead of the data repository), which leads to almost constant time cost. Nevertheless, our $TER\text{-}iDS$ approach still achieves low time cost (i.e., 0.0004 $sec\sim$ 0.01 $sec$), and outperforms 5 baselines, which confirms the efficiency of our $TER\text{-}iDS$ approach.

\vspace{1ex}\noindent {\bf The TER-iDS efficiency vs. the number, $m$, of missing attributes.} Figure \ref{exper:m} evaluates the performance of our $TER\text{-}iDS$ approach and 5 baselines, by varying $m$ from $1$ to $3$, where other parameters are by default. From the figures, we can see that the time cost increases for $TER\text{-}iDS$ and baselines (except for $con+ER$), since larger $m$ will result in more imputed candidate tuples. For $con+ER$, its time cost is not sensitive to the $m$ values, as $con+ER$ imputes each incomplete tuple based on its near complete tuple from $iDS$ (instead of accessing data repository $R$). Nevertheless, our $TER\text{-}iDS$ approach needs the least time cost (i.e., 0.0013 $sec\sim$ 0.0635 $sec$), compared with other baselines, which confirms the efficiency of our approach.
\end{document}